\documentclass[useAMS,usenatbib]{mn2e}
\usepackage{array}
\usepackage{times}
\usepackage{graphicx}
\usepackage{mathptmx}
\usepackage{amsmath}
\usepackage{amssymb}
\usepackage[usenames]{color}

\usepackage{hyperref}
\hypersetup{colorlinks=true,citecolor=blue,linkcolor=purple,filecolor=black,runcolor=black}
\begin{document}

\newcommand{\comment}[1]{}
\definecolor{purple}{RGB}{160,32,240}
\newcommand{\peter}[1]{\textcolor{purple}{(\bf #1)}}
\newcommand{\macc}{M_\mathrm{acc}}
\newcommand{\mpeak}{M_\mathrm{peak}}
\newcommand{\mnow}{M_\mathrm{now}}
\newcommand{\vacc}{v_\mathrm{acc}} 
\newcommand{\vpeak}{v_\mathrm{peak}} 
\newcommand{\vnow}{v^\mathrm{now}_\mathrm{max}}

\newcommand{\Mnfw}{M_\mathrm{NFW}}
\newcommand{\Msun}{\mathrm{M}_{\odot}}
\newcommand{\mvir}{M_\mathrm{vir}}
\newcommand{\rvir}{R_\mathrm{vir}}
\newcommand{\vmax}{v_\mathrm{max}}
\newcommand{\vmac}{v_\mathrm{max}^\mathrm{acc}}
\newcommand{\mvac}{M_\mathrm{vir}^\mathrm{acc}}
\newcommand{\sfr}{\mathrm{SFR}}
\newcommand{\plotgrace}[1]{\includegraphics[width=\columnwidth,type=pdf,ext=.pdf,read=.pdf]{#1}}
\newcommand{\plotssgrace}[1]{\includegraphics[width=0.95\columnwidth,type=pdf,ext=.pdf,read=.pdf]{#1}}
\newcommand{\plotgraceflip}[1]{\includegraphics[width=\columnwidth,type=pdf,ext=.pdf,read=.pdf]{#1}}
\newcommand{\plotlargegrace}[1]{\includegraphics[width=2\columnwidth,type=pdf,ext=.pdf,read=.pdf]{#1}}
\newcommand{\plotlargegraceflip}[1]{\includegraphics[width=2\columnwidth,type=pdf,ext=.pdf,read=.pdf]{#1}}
\newcommand{\plotminigrace}[1]{\includegraphics[width=0.5\columnwidth,type=pdf,ext=.pdf,read=.pdf]{#1}}
\newcommand{\plotmicrograce}[1]{\includegraphics[width=0.25\columnwidth,type=pdf,ext=.pdf,read=.pdf]{#1}}
\newcommand{\plotsmallgrace}[1]{\includegraphics[width=0.66\columnwidth,type=pdf,ext=.pdf,read=.pdf]{#1}}
\newcommand{\plotappsmallgrace}[1]{\includegraphics[width=0.33\columnwidth,type=pdf,ext=.pdf,read=.pdf]{#1}}
\newcommand{\repos}[3]{\hspace{-#1}\raisebox{#2}{\llap{#3}}\hspace{#1}}

\newcommand{\hinv}{h^{-1}}
\newcommand{\mpc}{\rm{Mpc}}
\newcommand{\hmpc}{$\hinv\mpc$}

\title[Star Formation During Major Halo Mergers]{Using Galaxy Pairs to Probe Star Formation During Major Halo Mergers}

\author[Behroozi et al.]{Peter~S.~Behroozi$^{1}$,\thanks{E-mail:     behroozi@stsci.edu} Guangtun Zhu$^{2,3}$, Henry C.\ Ferguson$^{1}$, Andrew P.\ Hearin$^{4}$, \and
Jennifer Lotz$^{1}$, Joseph Silk$^{2,5,6}$, Susan Kassin$^{1}$, Yu Lu$^{7}$, Darren Croton$^{8}$, \and 
Rachel S.\ Somerville$^{9}$, Douglas F.\ Watson$^{10}$
\\
\\
$^{1}$ Space Telescope Science Institute, Baltimore, MD 21218, USA\\
$^{2}$ Department of Physics and Astronomy, Johns Hopkins University, Baltimore, MD 21218, USA\\
$^{3}$ Hubble Fellow\\
$^{4}$ Department of Physics, Yale University, New Haven, CT 06511, USA\\
$^{5}$ Institut d'Astrophysique, UMR 7095 CNRS, Universit\'e Pierre et Marie Curie, 75014 Paris, France\\
$^{6}$ Beecroft Institute of Particle Astrophysics and Cosmology, Department of Physics, University of Oxford, Oxford OX1 3RH, UK\\
$^{7}$ The Carnegie Observatories, Pasadena, CA 91101, USA\\
$^{8}$ Centre for Astrophysics \& Supercomputing, Swinburne University of Technology, Hawthorn, VIC 3122, Australia\\
$^{9}$ Department of Physics, Rutgers University, New Brunswick, NJ 08901, USA\\
$^{10}$ Kavli Institute for Cosmological Physics, University of Chicago, Chicago, IL 60637, USA
}
\date{Released \today}
%\pagerange{\pageref{firstpage}--\pageref{lastpage}} \pubyear{2015}

\maketitle

\begin{abstract}
Currently-proposed galaxy quenching mechanisms predict very different behaviours during major halo mergers, ranging from significant quenching enhancement (e.g., clump-induced gravitational heating models) to significant star formation enhancement (e.g., gas starvation models).  To test real galaxies' behaviour, we present an observational galaxy pair method for selecting galaxies whose host haloes are preferentially undergoing major mergers.  Applying the method to central $L^*$ ($10^{10} \Msun < M_* < 10^{10.5} \Msun$) galaxies in the Sloan Digital Sky Survey (SDSS) at $z<0.06$, we find that major halo mergers can at most modestly reduce the star-forming fraction, from 59\% to 47\%.  Consistent with past research, however, mergers accompany enhanced specific star formation rates for \textit{star-forming} $L^*$ centrals: $\sim$10\% when a paired galaxy is within 200 kpc (approximately the host halo's virial radius), climbing to $\sim 70\%$ when a paired galaxy is within 30 kpc.  No evidence is seen for even extremely close pairs ($<30$ kpc separation) rejuvenating star formation in quenched galaxies.  For galaxy formation models, our results suggest: (1) quenching in $L^*$ galaxies likely begins due to decoupling of the galaxy from existing hot and cold gas reservoirs, rather than a lack of available gas or gravitational heating from infalling clumps, (2) state-of-the-art semi-analytic models currently over-predict the effect of major halo mergers on quenching, and (3) major halo mergers can trigger enhanced star formation in non-quenched central galaxies.
 \end{abstract}

\section{Introduction}

\label{s:introduction}

In the context of Lambda Cold Dark Matter ($\Lambda$CDM) cosmologies, dark matter halo (gravitationally self-bound structures) masses correlate strongly with the stellar masses of the galaxies at their centres \citep[e.g.,][]{more-09,yang-09,Leauthaud12,Reddick12,Watson13,Tinker13}.  Indeed, one-to-one matching of galaxies ordered by stellar mass to haloes ordered by mass or circular velocity at fixed cumulative number density provides a remarkably close match to galaxy autocorrelation functions, conditional stellar mass functions, satellite fractions, and weak lensing measurements from $z=0$ to $z\sim5$ (see also \citealt{Nagai05,conroy:06,Watson14}).  Performing this matching at several different redshifts allows inferring average galaxy star formation rates and histories as a function of host halo mass and redshift \citep{cw-08,Firmani10,Leitner11,Bethermin12,Wang12, Moster12,Behroozi13,BWC13,Mutch13,Lu14}.  This method has shown that \textit{average} galaxy growth rates have tracked \textit{average} dark matter halo growth rates (multiplied by a halo-mass dependent efficiency) remarkably well over the past 12 Gyr \citep{Behroozi13}.

It is less clear how closely \textit{individual} galaxy growth histories track individual halo growth histories \citep{YuLu14,Genel14}, especially for central galaxies (i.e., galaxies whose host haloes are not in orbit around any more massive halo).  For example, galaxy specific star formation rates (SSFRs) at $z\sim 0$ show a clear bimodality between star-forming (SSFR $> 10^{-11}$ yr$^{-1}$)  and quenched galaxies \citep[e.g.,][]{Brinchmann04,Salim07}.  However, the fraction of star-forming versus quenched galaxies falls with increasing stellar mass, whereas the fraction of host haloes accreting (versus losing) mass rises with increasing halo mass.  As a result, quenching cannot be due to lack of cosmological accretion alone.  Yet, it is still possible that galaxy quenching correlates with halo mass accretion history.  For example, empirical models relating galaxy quenching to halo age \citep{Hearin13,Hearin13b,Watson14}---with older, earlier-forming haloes being assigned galaxies with lower star formation rates---have been very successful at matching quenched versus star-forming correlation functions, weak lensing, and radial profiles near clusters and groups.  In the simplest theoretical model, a galaxy which uses up or expels gas faster than its host halo accretes it will quench due to lack of fuel \citep[and references therein]{Feldmann15}.  Alternate scenarios could include black hole feedback which correlates with the merger history of the halo \citep{Silk98,Springel05c,Croton06,Somerville08}, or quenching due to gravitational heating from mergers or infalling clumps \citep{Cox04,Dekel08,Khochfar08,Johansson09,Johansson12,Birnboim11,Moster11}.  

These quenching models have very different behaviours during major halo mergers (here, when a halo's virial radius contains a smaller halo with a mass ratio of $>1:3$).  For a starvation model, incoming lower-mass star-forming galaxies would transfer their gas reservoir to the larger host, and therefore would rejuvenate star formation in quenched hosts.  In contrast, for a gravitational heating model, an incoming major merger (and/or associated accretion) would disrupt the flow of gas and suppress star formation.  Finally, for a merger-fed black hole feedback model, no significant change would be expected until the \textit{galaxies} themselves merge (as opposed to the smaller halo simply coming within the virial radius of the larger halo).  These differences should be especially apparent for central $L^*$ galaxies (here, galaxies with $10^{10}\,\Msun < M_* < 10^{10.5}\,\Msun$) at $z\sim0$.  Below this stellar mass range, most galaxies are star-forming, and above it, most galaxies are quenched \citep{Brinchmann04,Salim07}, so $L^*$ galaxies are important probes of the quenching process.

In this paper, we develop a galaxy pair-based selection method to preferentially identify haloes undergoing major mergers, and we examine the impact on central galaxies' star formation rates.  While many existing studies have found star formation enhancement in close pairs \citep[e.g.,][]{Lambas03,Nikolic04,Alonso04,Woods06,Perez09,Woods10,Barton07,Lin07,Li08,Ellison08,Rogers09,Robaina09,Wong11,Scudder12,Xu12b,Robotham13,Patton13,Ellison13,Scott14}, the effect size depends strongly on the selection process.  Previous studies have typically focussed on pre-merging \textit{galaxies} instead of merging \textit{haloes}, and so have usually excluded the more distant pair candidates included in this paper (see, however, \citealt{Nikolic04,Lin07,Li08,Robaina09,Patton13}).  We also design our method to avoid past selection biases.  Well-known biases include whether the pairs are in a cluster environment \citep{Barton07}, and whether the galaxies are required to be star-forming.  Many subtler selection biases also exist.  For example, the presence of a close companion can bias the distribution of host halo masses for galaxies in close pairs versus those not in close pairs; indeed, this has been exploited to constrain the mass of the Milky Way's halo from its satellite distribution \citep{Busha11b,Cautun14}.  We therefore construct several mock catalogs from simulations for the purposes of testing for and avoiding such biases in our method.

We divide the results into several sections.  The observational selection method is described in \S \ref{s:method}, and we describe the observational data sets, the mock observational catalogs, and validation tests in \S \ref{s:data}.  We present our main findings in \S \ref{s:results}, discuss the impact of these results in \S \ref{s:discussion} and conclude in \S \ref{s:conclusions}.  Throughout this work, we adopt a flat $\Lambda$CDM cosmology ($\Omega_M = 0.27, \Omega_\Lambda = 0.73, h=0.7, n_s = 0.95, \sigma_8 = 0.82$) in close agreement with recent WMAP9 cosmology constraints \citep{WMAP9}.  Stellar masses and star formation rates come from methods in \cite{Kauffmann03} and \cite{Brinchmann04}, respectively, updated for the SDSS DR7, and are renormalized to a \cite{Chabrier03} initial mass function.  Halo masses are defined according to the virial spherical overdensity criterion of \cite{mvir_conv}.

\section{Method}
\label{s:method}

\begin{figure}
\vspace{-7ex}
\plotgrace{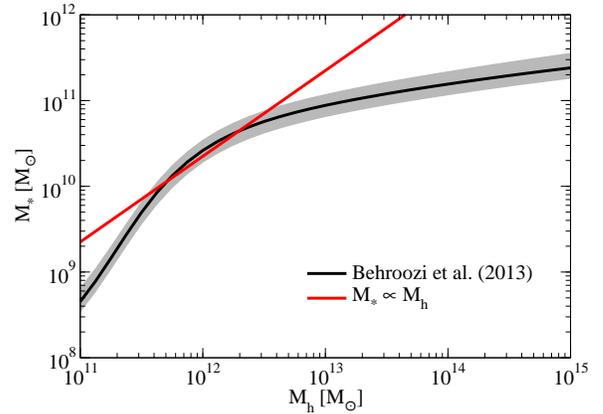}
\caption{The stellar mass---halo mass relation at $z=0.1$, from \protect\cite{BWC13}.  A given galaxy stellar mass ratio only corresponds to the same host halo mass ratio in a small window near $M_* = 10^{10.5}\,\Msun$, as seen by comparison to the \textit{red line} with $M_* \propto M_h$.  Focussing on this mass range minimizes halo mass biases when selecting pairs based on stellar mass; this mass range is also where galaxies transition to being primarily quenched.}
\label{f:smhm}
\end{figure}

Observational galaxy pair selection requires a compromise between simplicity, bias avoidance, and sample size.  Traditionally, paired galaxies are selected based on being within a specified mass or luminosity ratio as well as within a specified projected distance and redshift window.  Satellite galaxies in cluster environments often match these criteria, yet they are subject to very different physical conditions; many studies therefore also adopt an exclusion criterion, e.g., that no larger galaxy exists within a specified projected distance and redshift window of the galaxy pair.  For simplicity and readers' familiarity, we use a similar approach, but we adjust selection parameters so as to preferentially select major halo mergers and to minimise biases as compared to previous techniques (see \S\ref{s:prev} for discussion).

We find that a suitable selection exists for galaxy pairs where the larger galaxy has mass $10^{10}\,\Msun < M_\ast < 10^{10.5}\,\Msun$, and where the smaller galaxy's mass is within a ratio of 0.5 dex.  To simplify further discussion, we call the larger galaxy the \textit{host galaxy}, and the smaller galaxy the \textit{paired galaxy}.  We adopt a projected radius cut for \textit{close pairs} of 200 kpc.  While larger than typical for past studies, this distance corresponds to the virial radius of the smallest host haloes expected to be in our sample.  We also adopt a redshift window of $500$ km s$^{-1}$ for the paired galaxy, corresponding to the escape velocity of the largest host haloes expected to be in our sample.  Our results are not sensitive to these specific parameter choices, as we verify by checking many alternate choices in Appendix \ref{a:variation}. 

For the satellite exclusion criteria, we require that no galaxy more massive than the host galaxy be present within a projected radius of 500 kpc and a redshift window of 1000 km s$^{-1}$.  From tests with our mock catalogs (\S \ref{s:mocks}), we find that this cut retains 77\% of the central galaxies in our stellar mass range, with a purity of 97.4\%; this is very consistent with the expected completeness of 75\% from \cite{Liu10}.  We again explore several different choices for these criteria in Appendix \ref{a:variation} to verify that they do not affect our results.

These selection criteria are designed primarily to avoid halo mass biases.  Dark matter haloes are roughly self-similar in their subhalo mass ratio distributions; however, a given ratio in dark matter masses for close companions can imply a very different ratio in their stellar masses (Fig.\ \ref{f:smhm}).  This is why, for example, galaxies smaller than $10^{10}\,\Msun$ have few massive satellites, whereas galaxies larger than $10^{11}\,\Msun$ often have massive companions \citep[e.g.,][]{Bundy09}.  Selecting close pairs based on a fixed stellar mass ratio will therefore tend to bias the host halo masses of the close pair galaxies to be higher than the non-close pairs.  However, somewhat counteracting this effect, selecting close pairs within a fixed projected radius will probe a smaller fraction of the halo radius in more massive haloes.  In our host stellar mass window ($10^{10}\,\Msun < M_\ast < 10^{10.5}\,\Msun$), these two biases nearly cancel each other out, and stellar mass ratios of 0.5 dex correspond very nearly to 0.5 dex ratios in halo mass---i.e., major mergers (see validation tests in \S \ref{s:validation}).

\begin{figure}
\vspace{-7ex}
\plotgrace{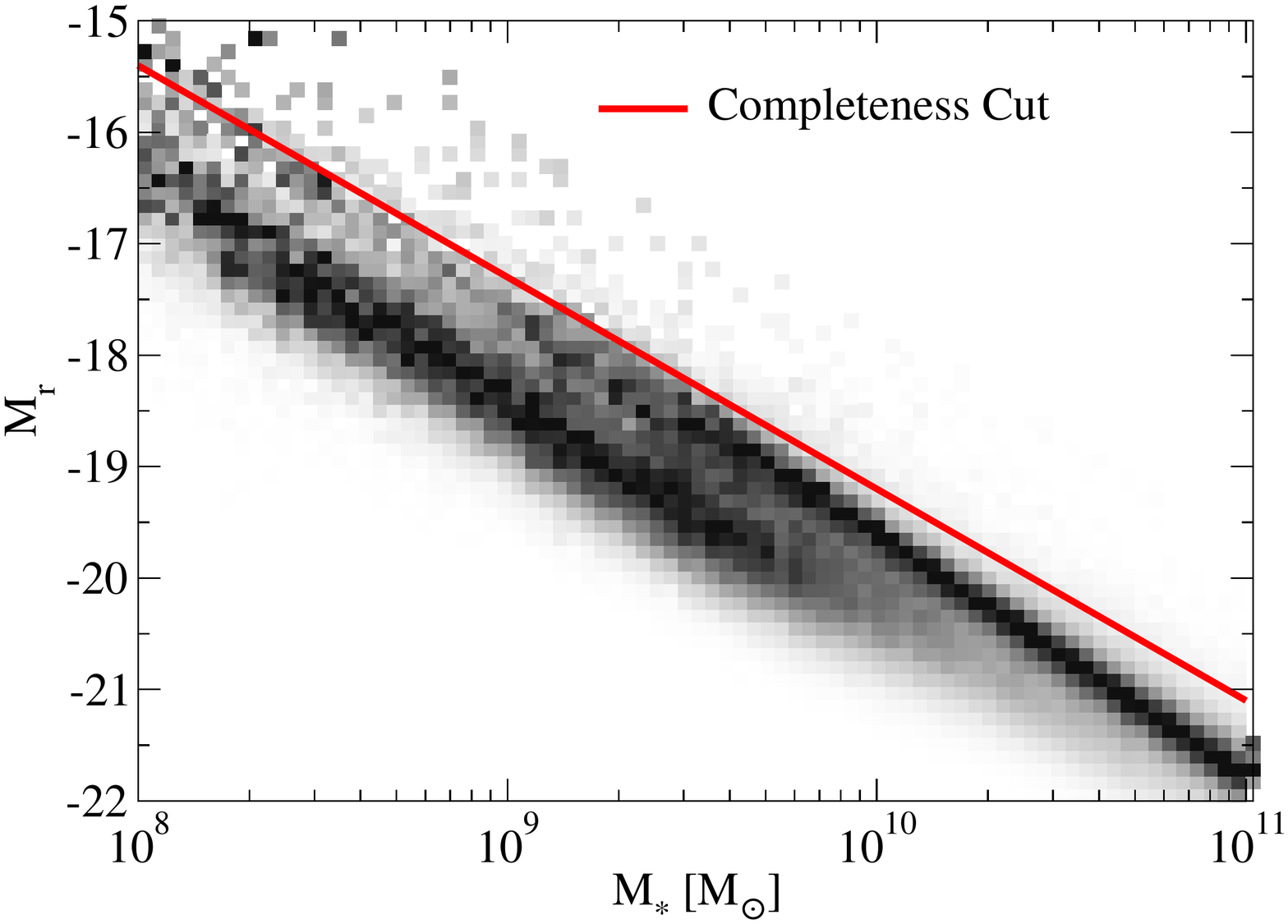}\\[-6ex]
\plotgrace{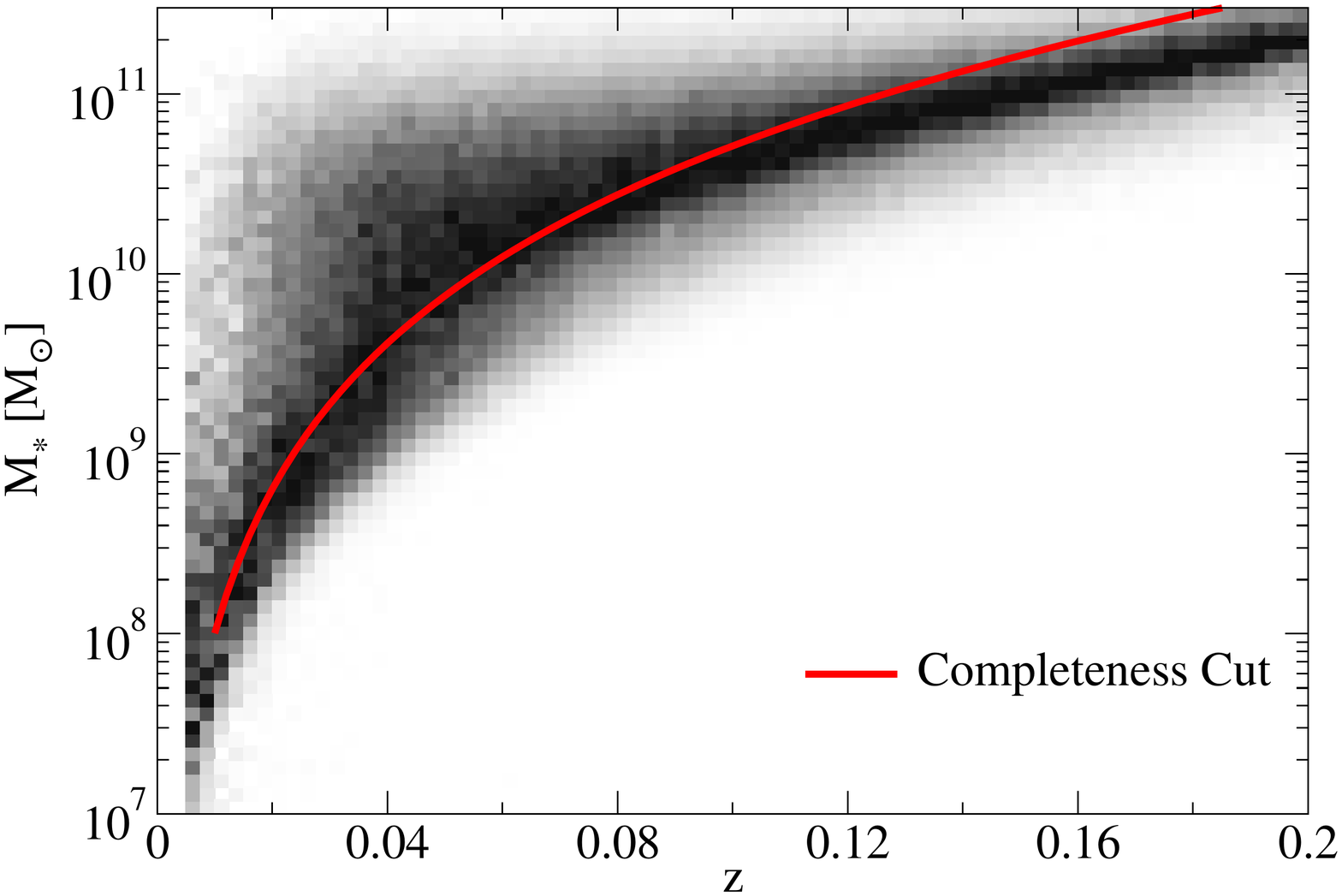}\\[-5ex]
\caption{\textbf{Top} panel: Conditional probability density for absolute $r$-band magnitude as a function of $M_*$ in the SDSS (observable volume-corrected).  The \textit{red line} shows the adopted completeness cut, which is fainter than 96 percent of spectroscopically-targeted galaxies with mass $10^{9.5}$-$10^{10}\,\Msun$.  \textbf{Bottom} panel: Conditional probability density for $M_*$ as a function of $z$ in the SDSS.  The \textit{red line} shows the adopted completeness cut from the top panel; galaxies above the red line are taken to be volume-complete.}
\label{f:completeness}
\end{figure}

\begin{figure}
\includegraphics[width=\columnwidth]{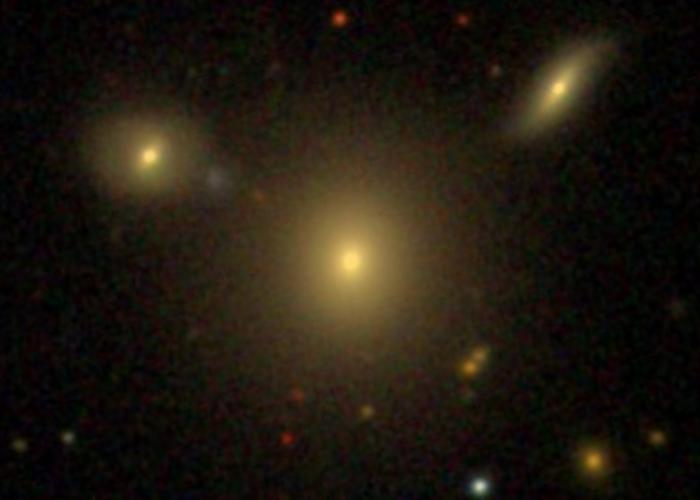}\repos{51ex}{31.5ex}{\Large$\color{white}\Box$}\repos{12ex}{38ex}{\Large$\color{white}\Box$}\repos{55.5ex}{9.3ex}{\textcolor{white}{SDSS gri}}\repos{53ex}{7ex}{\textcolor{white}{Ra: 219.597}}\repos{53ex}{4.7ex}{\textcolor{white}{Dec: 23.334}}\repos{54.75ex}{2.4ex}{\textcolor{white}{119''$\times$85''}}
\caption{SDSS gri colour image of a BCG (119''$\times$85'', corresponding to 72$\times$52 kpc at $z=0.03$).   The spectroscopic target selection algorithm for the SDSS maximises the number of non fibre-collided targets.  This selects against BCGs, whose central location would result in fibre collisions with many nearby satellites.  As in the above example, more spectroscopic targets (white squares) can be obtained in a given area if satellites are targeted instead of the BCG.  Cases where this bias impacted host galaxy isolation criteria were cleaned from our sample via visual inspection.}
\label{f:bcg_bias}
\end{figure}

Unlike many previous studies, the host and paired galaxies are drawn from a stellar mass-complete sample with no requirements on star formation activity.  We also select close pairs based on a fixed stellar mass ratio, rather than a fixed luminosity ratio.  Since star-forming galaxies at fixed stellar mass are brighter than quiescent galaxies, using a fixed luminosity ratio means that a star-forming satellite may be selected as being in a close pair, whereas a quiescent satellite would not be.  Because of galactic conformity (i.e., star-forming galaxies have larger fractions of nearby star-forming galaxies than quiescent galaxies; \citealt{Weinmann06,Kauffmann13,Phillips14,Hearin14}), using a fixed luminosity ratio would result in close pairs being artificially more star-forming than galaxies without close companions.

In summary, we make the following cuts for eligible host galaxies in our \textbf{host sample}:
\begin{enumerate}
\item Stellar mass between $10^{10}\,\Msun$ and $10^{10.5}\,\Msun$, and
\item No more massive galaxy within 500 kpc in projected distance and 1000 km s$^{-1}$ in redshift.
\end{enumerate}
For each of these host galaxies, we calculate the nearest \textbf{paired galaxy}, which is a galaxy that satisfies
\begin{enumerate}
\item Stellar mass 0 to 0.5 dex less than that of the host galaxy, and
\item Redshift separation within 500 km s$^{-1}$ of the host galaxy.
\end{enumerate}
If this paired galaxy is within 200 kpc (physical projected distance) of the host galaxy, we call the two galaxies a \textbf{close pair}.

\section{Data and Simulations}
\label{s:data}

\subsection{Observations}
\label{s:obs_data}

\begin{figure*}
\includegraphics[width=\textwidth]{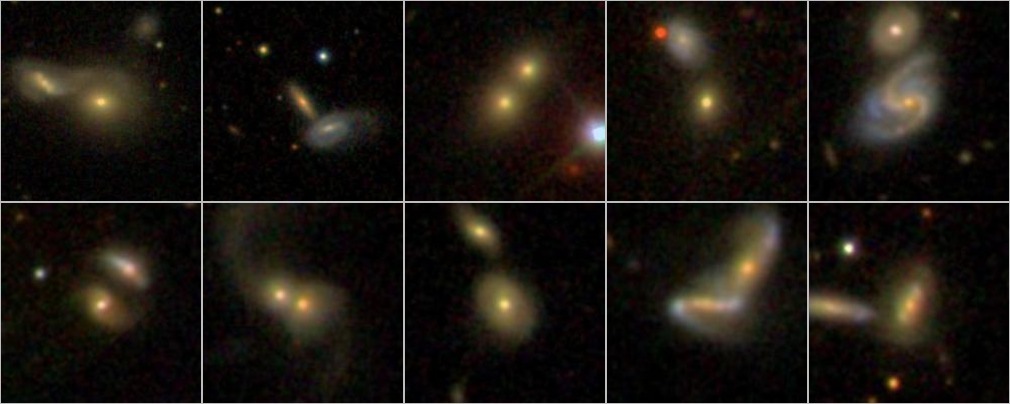}\repos{24.5ex}{2ex}{\Large\textcolor{white}{Q}}\repos{80.5ex}{2ex}{\Large\textcolor{white}{Q}}\repos{52.5ex}{30ex}{\Large\textcolor{white}{Q}}\repos{80.5ex}{30ex}{\Large\textcolor{white}{Q}}\repos{136.5ex}{30ex}{\Large\textcolor{white}{Q}}
\caption{SDSS gri colour images of the ten closest pairs in our sample.  Each image is 40$\times$40 kpc across and is centred on the more massive (host) galaxy.  A white ``Q'' denotes a quenched host galaxy (SSFR $< 10^{-11}$ yr$^{-1}$).  The separations shown here range from 5-15 kpc (median 11 kpc).}
\label{f:cp_sdss}
\end{figure*}

We use redshifts from the SDSS (Sloan Digital Sky Survey) Data Release 10 \citep{Ahn14}, which are $>$90 percent complete for galaxies brighter than $r = 17.77$.  In addition, we use median total stellar masses and total star formation rates (SFRs) from the MPA-JHU value-added catalog \citep{Kauffmann03,Brinchmann04}, updated for the imaging and spectroscopy in the SDSS Data Release 7 \citep{Abazajian09}. These stellar masses and SFRs were calculated assuming a \cite{Kroupa02} initial mass function (IMF), which we convert to a \cite{Chabrier03} IMF by dividing both by a factor 1.07.  In Appendix \ref{a:variation}, we also consider fibre star formation rates, D$_n$(4000) indices, the effects of BPT \citep{BPT} class, and using alternate stellar masses and redshifts from the NYU Value-Added Galaxy Catalog (NYU-VAGC; \citealt{Blanton05}).

The MPA-JHU catalog covers a spectroscopic area of 8032 deg$^2$ and 600,480 galaxy targets with nonzero stellar masses and redshifts $z>0.006$.  We find and remove duplicate targets separated by a projected distance less than 2 kpc \citep[see also][]{Ellison08,Tollerud11}, leaving 592,054 galaxies.  For our tests, host galaxies must be further than 2 Mpc from a survey boundary or region of significant incompleteness; we also exclude host galaxies with redshifts $z<0.01$ to avoid Hubble flow corrections \citep[e.g.,][]{Baldry12}.  We exclude such boundary regions (23 percent) from the sample selections, although we retain them for the purposes of counting neighbours or testing for larger nearby galaxies; this leaves 479,439 galaxies over 6201.8 deg$^2$ of sky.

Since the SDSS is magnitude-limited, cuts are needed to convert to a stellar mass--complete sample.  Galaxies at fixed stellar mass have a range of mass--to--light ratios (Fig.\ \ref{f:completeness}, top panel).  We find that more than $96$ percent\footnote{After weighting by inverse observable volume according to galaxy $r$-band absolute luminosity; this prevents bias against fainter galaxies at fixed stellar mass.} of galaxies in the SDSS between $10^{9.5}$ and $10^{10}\,\Msun$ (i.e., eligible paired galaxies) satisfy:
\begin{equation}
M_r < -0.25 - 1.9 \log_{10}\left(\frac{M_\ast}{\Msun}\right)
\end{equation}
where $M_r$ is the galaxy's Petrosian $r$-band absolute magnitude and $M_\ast$ is its stellar mass.

Similarly, for galaxies at redshift $z$, the corresponding apparent magnitude limit is
\begin{equation}
r < -0.25 -1.9 \log_{10}\left(\frac{M_\ast}{\Msun}\right) + 5 \log_{10}\left(\frac{D_L(z)}{10 \mathrm{pc}}\right)
\label{e:mag_limit}
\end{equation}
where $D_L(z)$ is the luminosity distance for our adopted cosmology.  We note that a given host galaxy can be included in our sample only if a paired galaxy (\S \ref{s:method}) is detectable---i.e., only if SDSS is complete at the same redshift to 0.5 dex less than the host galaxy's stellar mass.  As making a completely volume-limited catalog would then unacceptably reduce sample sizes, we instead weight selected host galaxies inversely by the observable volume for potential paired galaxies, obtained by inverting Eq.\ \ref{e:mag_limit} with $r=17.77$.  Galaxies included in this cut---i.e., galaxies for which $r<17.77$ according to Eq.\ \ref{e:mag_limit}---are shown in Fig.\ \ref{f:completeness}, bottom panel.

Fibre collisions can significantly reduce spectroscopic completeness \citep{Strauss02} when multiple targetings are not available ($\sim$ two-thirds of the SDSS footprint).  Following \cite{Patton13}, we reduce the observable volume by a factor 3.08 for galaxy pairs closer than 55'' on the sky; this increases the relative influence of such pairs in all calculations by an equal factor.  We have also checked pairs in the NYU-VAGC, which resolves fibre collisions by using the closest galaxy's spectroscopic redshift.  In that catalog, all close pairs within 55'' are 3.23 times more numerous than spectroscopic close pairs within 55'', corresponding to a very consistent weighting factor of 3.10 after correcting for the 4\% chance projections within 55" expected from our mock catalogs.

  The SDSS spectroscopic target selection algorithm  also introduces an important bias near clusters.  As shown in Fig.\ \ref{f:bcg_bias}, targeting a brightest cluster galaxy (BCG) can result in many fibre collisions with surrounding satellites, whereas targeting the satellites instead would result in many more available spectra in the same region.  Since the SDSS algorithm maximises the number of non-collided sources, BCGs are frequently not targeted.  This affects the galaxy isolation criteria in \S \ref{s:method}, so we have visually examined the 1,339 host galaxies with paired galaxies within 500 kpc and removed 89 with BCGs also found within this distance.

After applying the satellite exclusion criteria in \S \ref{s:method}, as well as the masking cuts and completeness corrections above, 7,303 host galaxies with stellar masses between $10^{10}\,\Msun$ and $10^{10.5}\,\Msun$ remain in the host sample; redshifts range from $0.01$ to $0.057$, with a median of  $z=0.03$.  Of these galaxies, 439 fall in the close pairs sample, and of these, 36 pairs are within 55'' of each other.  After observable volume and fibre-collision weightings, 6.7$\%$ of the host galaxies are in a close pair.  Fig.\ \ref{f:cp_sdss} shows the physically closest ten pairs in this sample.

\subsection{Mock Catalogs}

\label{s:mocks}

As a basis for mock catalogs, we use merger trees generated from the \textit{Bolshoi} simulation \citep{Bolshoi}.  Bolshoi follows 2048$^3$ particles ($\sim$ 8 billion) in a comoving volume 357 Mpc on a side from $z=80$ to $z=0$ using the \textsc{Art} code \citep{kravtsov_etal:97}.  The simulation's particle and force resolution are $1.94\times 10^{8}\,\Msun$ and 1.4 kpc, respectively, which correspond to a minimum resolvable halo mass of $10^{10}\,\Msun$.  Haloes were found using the \textsc{Rockstar} halo finder \citep{Rockstar}, an algorithm which determines particle-halo membership via a six-dimensional phase-space metric which is particularly suited to recovering haloes in major mergers and at close separations \citep{Knebe11,Knebe13b,Knebe13,Onions12,Pujol14}.  In Appendix \ref{a:mocks}, we also compare to halo catalogs generated using the \textsc{BDM} halo finder \citep{Klypin99,Riebe13}, which uses a three-dimensional density-based algorithm to assign particles.  For both halo finders, halo masses are calculated using the virial overdensity criterion of \cite{mvir_conv}.  Merger trees for both catalogs were generated using the \textsc{Consistent Trees} algorithm \citep{BehrooziTree}, which compares halo catalogs across multiple timesteps to repair halo finder inconsistencies; the algorithm yields significantly cleaner mass accretion histories as compared to many other methods \citep{Srisawat13}, especially when combined with the \textsc{Rockstar} halo finder \citep{Avila14}.

To generate mock catalogs, we use abundance matching, which assigns galaxy stellar masses to haloes with the same cumulative number density \citep{Nagai05,conroy:06}.  Several halo orderings have been explored (e.g., by mass or maximum circular velocity, $v_\mathrm{max}$, defined as the maximum of $\sqrt{\frac{GM(<R)}{R}}$ within the halo's virial radius) in \cite{Reddick12}.  Using present-day satellite halo mass dramatically underestimates galaxy clustering \citep{conroy:06,Reddick12}, as satellite haloes' dark matter is stripped much more rapidly than their galaxies' stars.  Instead, ordering haloes by decreasing peak historical $v_\mathrm{max}$ or peak historical mass give the best matches to galaxy autocorrelation functions and conditional stellar mass functions.  Specifically, \cite{Reddick12} finds that peak historical $v_\mathrm{max}$ gives the closest match; however, as noted in \cite{BehrooziMergers}, peak $v_\mathrm{max}$ is typically set during 1:5 mergers in halo mass, and models using time since peak $v_\mathrm{max}$ as a quenching proxy do not reproduce quenched galaxy distributions around clusters (Behroozi et al., in preparation).

Using peak historical mass avoids these latter problems, but then somewhat underpredicts observed galaxy autocorrelation functions \citep{Reddick12} as many satellites continue forming stars even after accretion onto a larger halo \citep{Wetzel11}.  We can approximate this physical process with a proxy that continues to grow for the typical galaxy's quenching time after a halo reaches its peak mass.  Combining the average satellite quenching time as a function of stellar mass inferred in \cite{Wetzel13b} with the stellar mass---halo mass relation in \cite{BWC13}, we obtain an average quenching time as a function of peak halo mass ($M_p$), which is well-approximated by a double power-law:
\begin{equation}
t_\mathrm{quench}(M_p) = \frac{1.584\times10^{10}\textrm{ yr}}{\left(\frac{M_p}{10^{10.63}\,\Msun}\right)^{-0.50} + \left(\frac{M_p}{10^{10.63}\,\Msun}\right)^{0.37}} 
\end{equation}
For each halo in \textit{Bolshoi}, we calculate its peak historical mass, as well as the time at which that mass was reached ($t_p$).  If the halo is not currently at its peak mass (i.e., $t_\mathrm{p} < t_\mathrm{now}$), we randomly select a continued mass accretion history from Bolshoi \citep[from tables calculated in][]{BehrooziND} starting at mass $M_p$ at time $t_p$ and ending at $t_p+t_\mathrm{quench}$ or $t_\mathrm{now}$, whichever is earlier.  We call the mass at the end of this appended history $M_q$; for haloes with $t_\mathrm{p} = t_\mathrm{now}$, we simply set $M_q = M_p$.

Abundance matching on $M_q$ therefore approximates the continued stellar mass growth expected to happen in satellite galaxies after their accretion, so we adopt this method for constructing our main mock catalogs.  For the haloes under consideration, 14.3\% have $\frac{M_q}{M_p}$ larger than 0.1 dex, and 3.4\% have $\frac{M_q}{M_p}$ larger than 0.3 dex.  For comparison, we have also constructed catalogs based on abundance matching with many other halo proxies (including peak mass and peak $v_\mathrm{max}$) to demonstrate that the catalog construction method does not affect our validity tests (Appendix \ref{a:mocks}).  For the source stellar mass function, we calculate observable volume-corrected galaxy number counts from the same region of the SDSS used in \S \ref{s:obs_data}; this process is detailed in Appendix \ref{a:smf}.  We incorporate a log-normal scatter of 0.2 dex in stellar mass at fixed halo mass or $v_\mathrm{max}$ using the iterative approach in \cite{Reddick12}.  Finally, we incorporate a Gaussian scatter of 30 km s$^{-1}$ in halo relative velocities to mimic SDSS spectroscopic redshift errors.

\subsection{Validations of the Mock Catalog and Observational Cuts}
\label{s:validation}

\begin{figure*}
\vspace{-10ex}
\plotssgrace{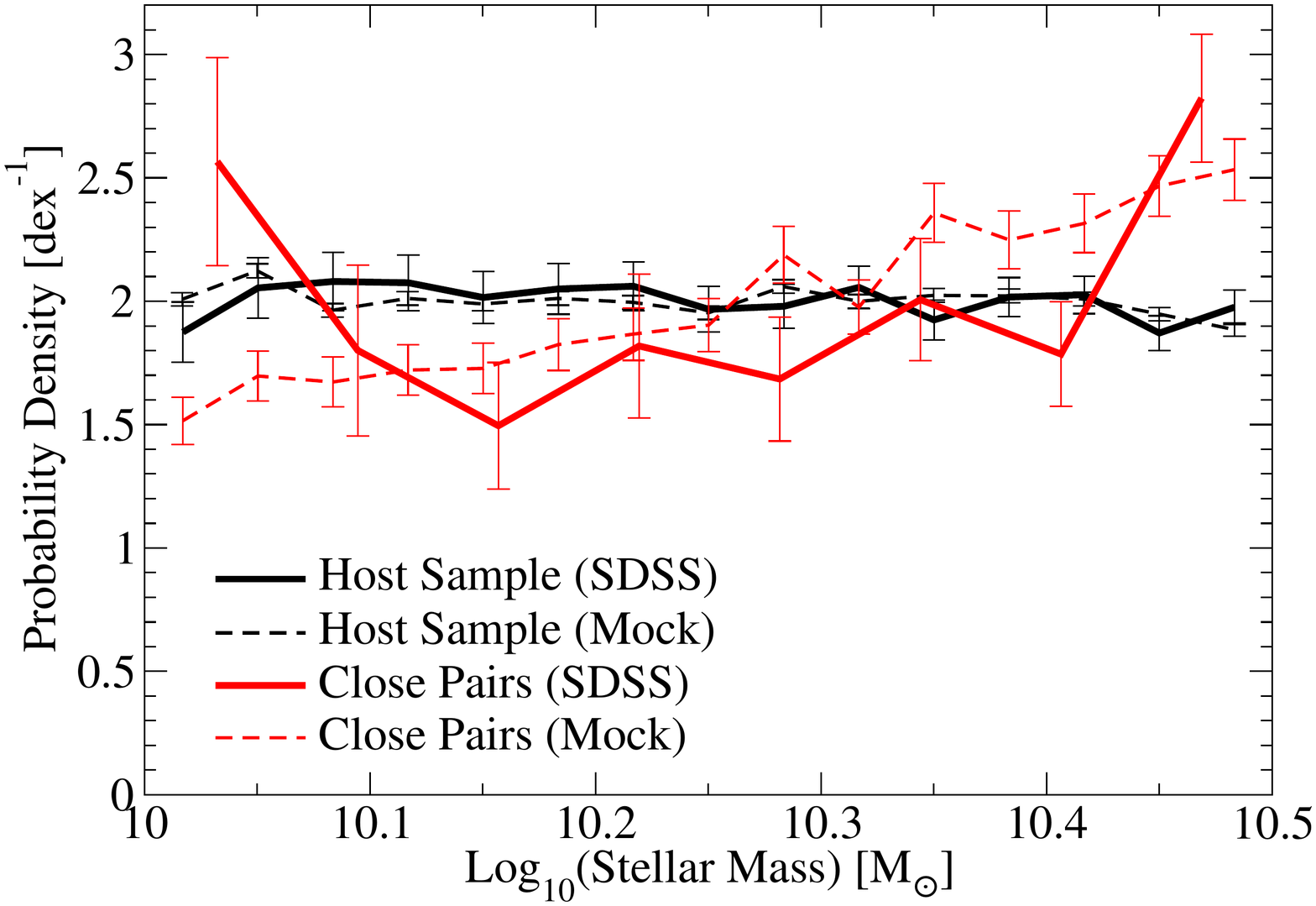}\plotssgrace{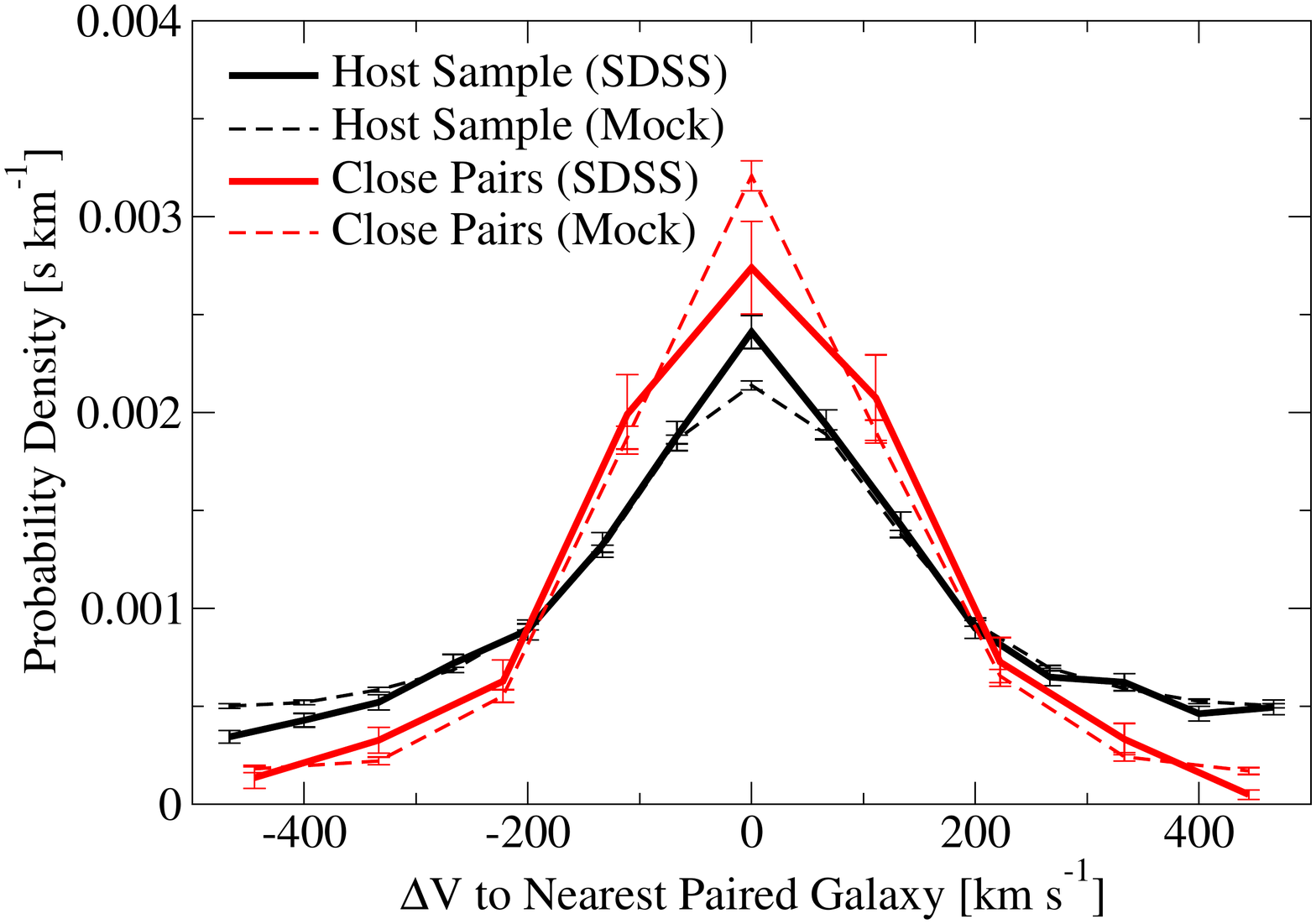}\\[-7ex]
\plotssgrace{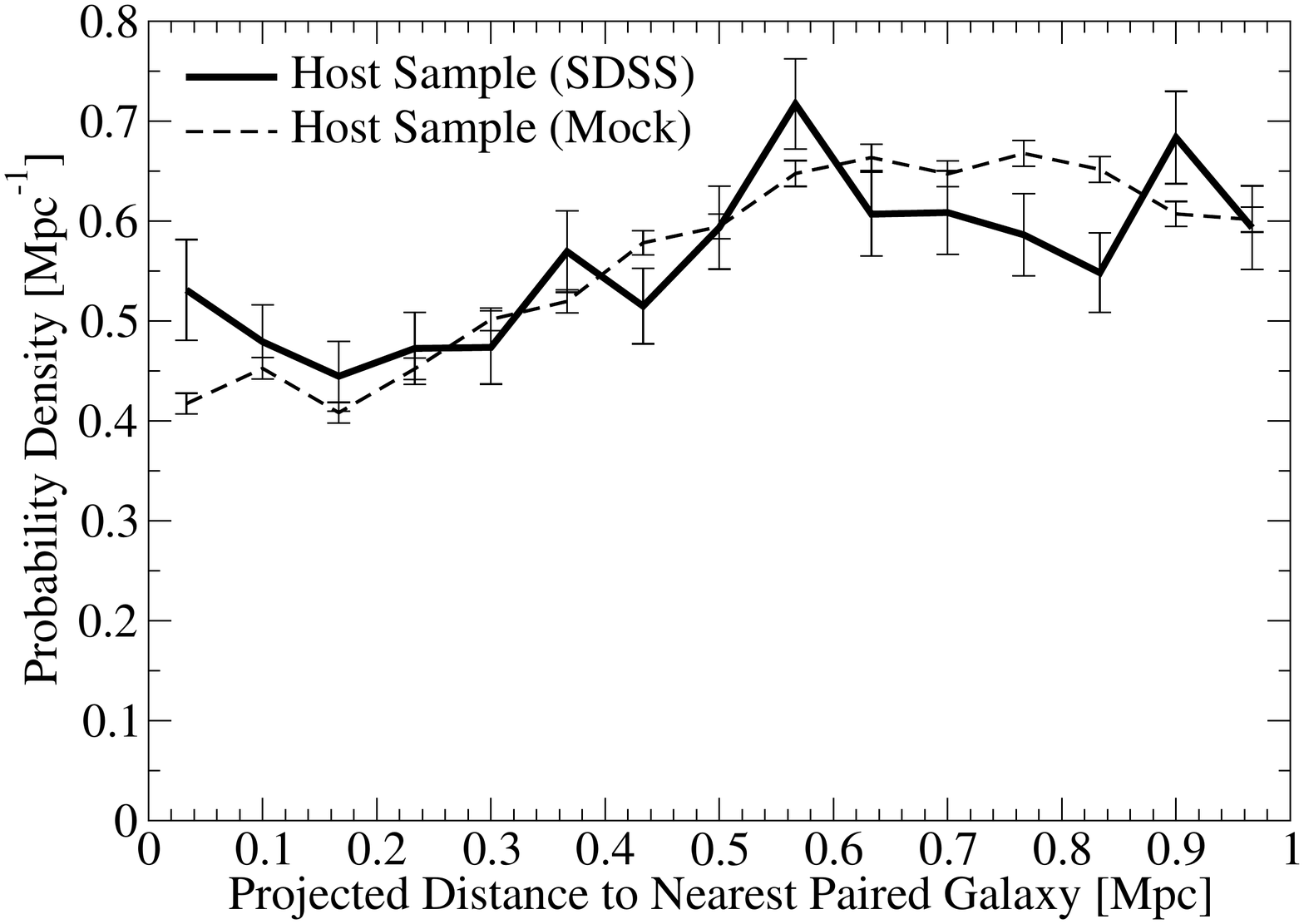}\plotssgrace{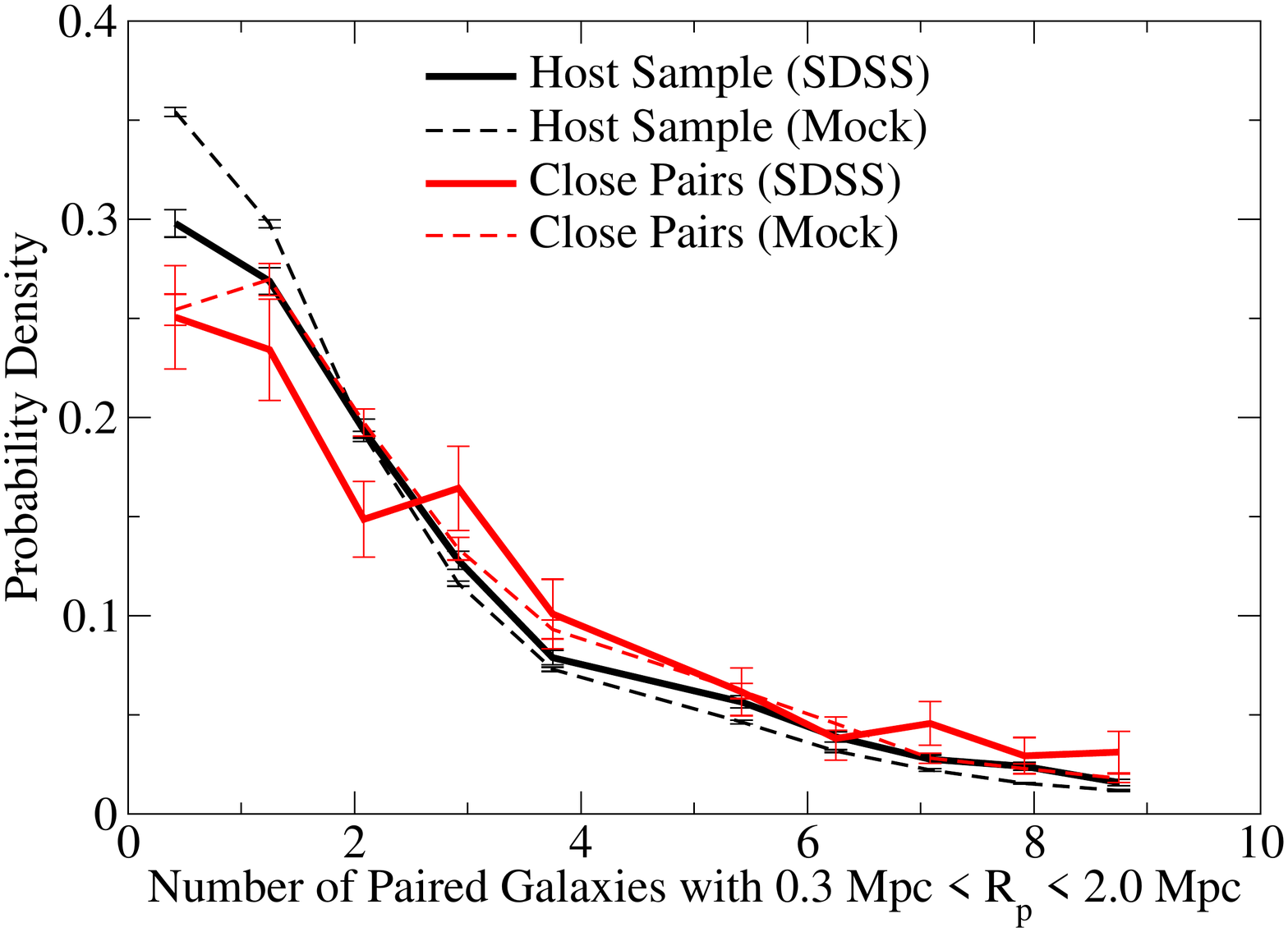}\\[-4ex]
\caption{Mock catalog comparisons with the SDSS.    \textbf{Top-left} panel: host galaxy stellar mass distributions in the host sample and in close pairs; \textbf{top-right} panel: distributions of the velocity separation between host galaxies and the nearest paired galaxies (i.e., smaller galaxies with $\Delta V < 500$ km s$^{-1}$ and $\Delta M_* <0.5$ dex from the host); \textbf{bottom-left} panel: distributions of the projected distance ($R_p$) between host galaxies and the nearest paired galaxy; \textbf{bottom-right} panel: distributions for the number of paired galaxies with projected distances $0.3 < R_p < 2.0$ Mpc, i.e., the larger-scale environment.  SDSS galaxies are weighted by observable volume and fibre collision rate.  Errors in all cases are jackknifed; mock catalog errors are smaller due to $\sim10$ times larger sampled volume.}
\label{f:mock_obs}
\vspace{-5ex}
\plotssgrace{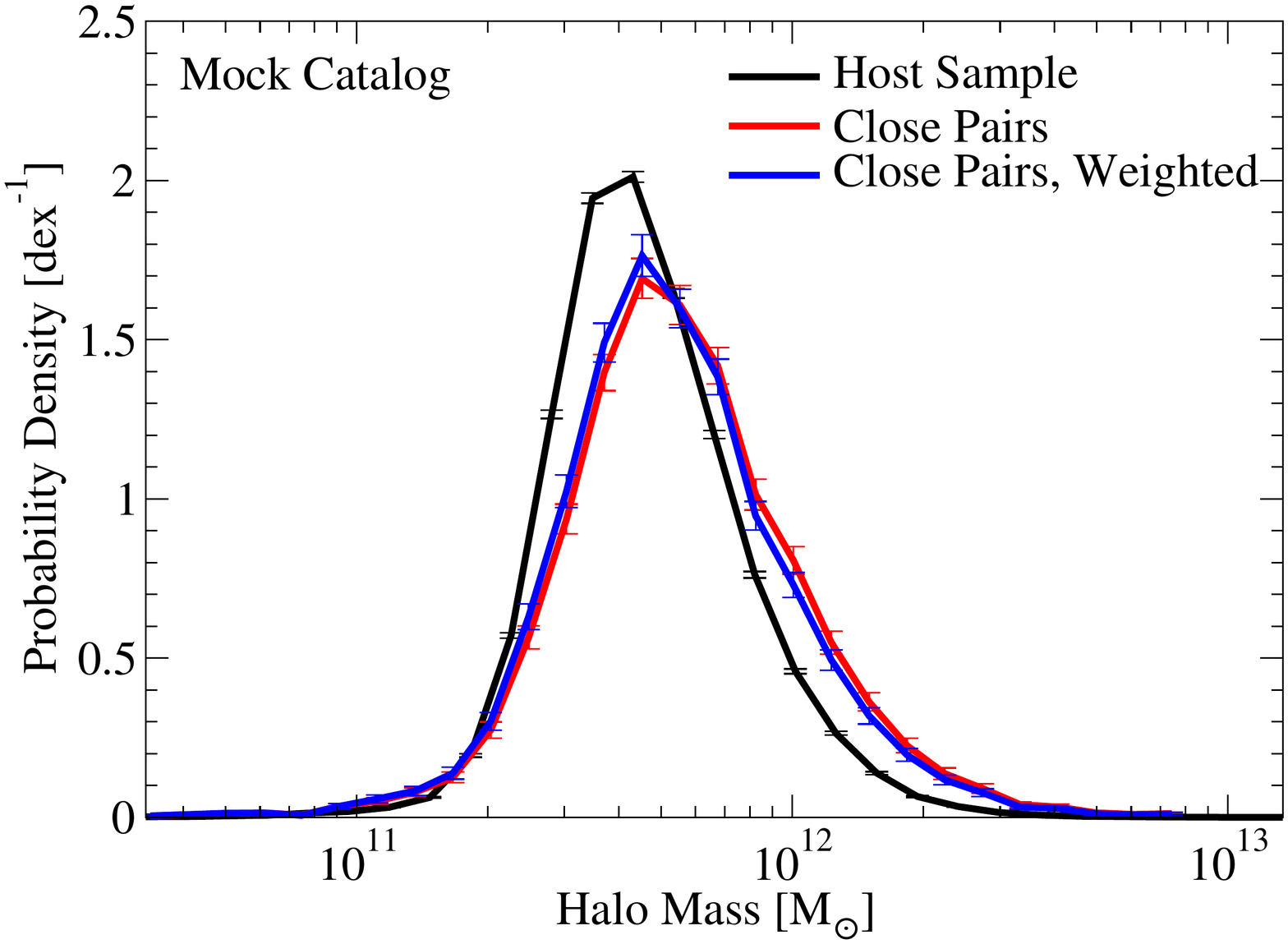}\plotssgrace{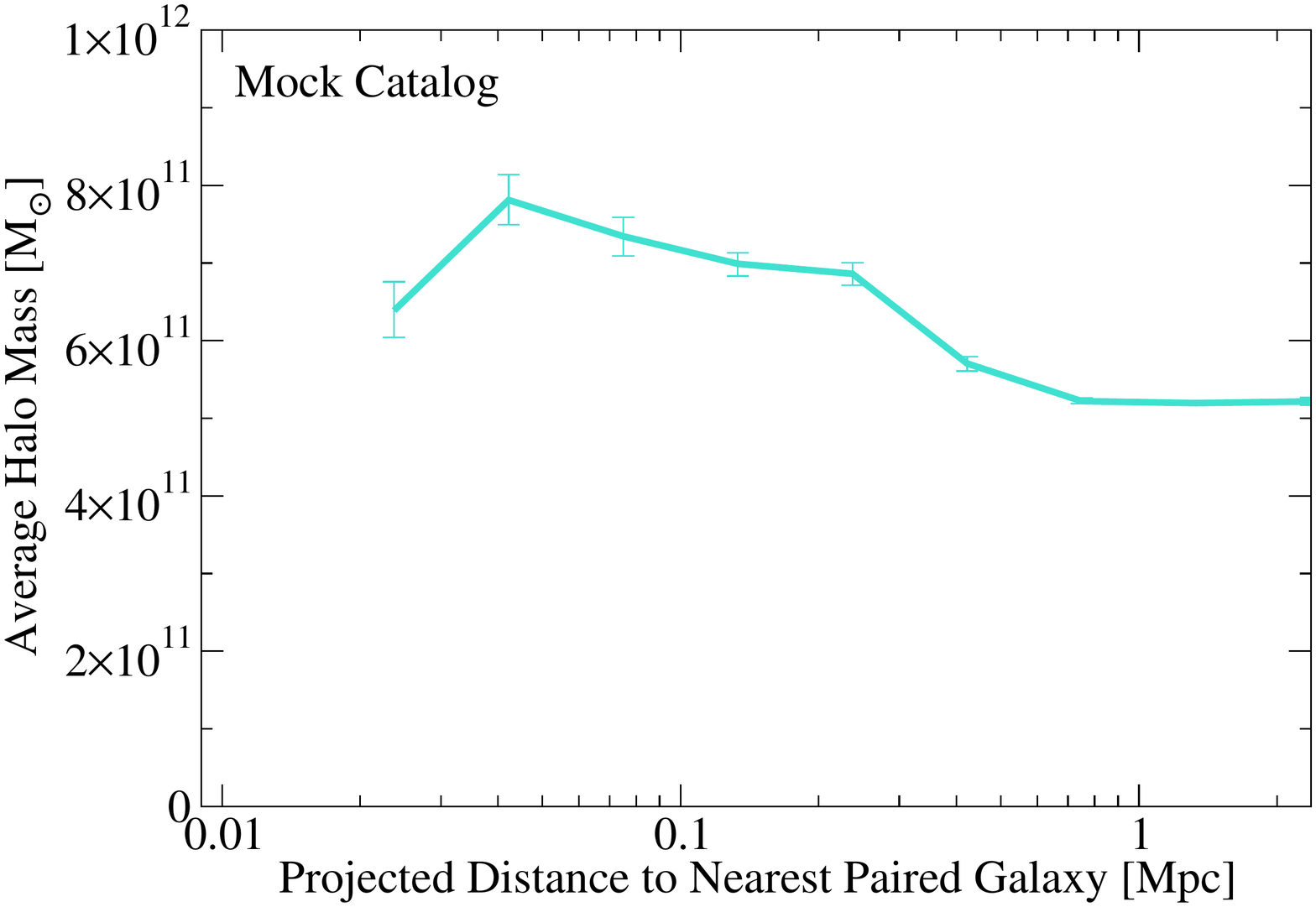}\\[-7ex]
\plotssgrace{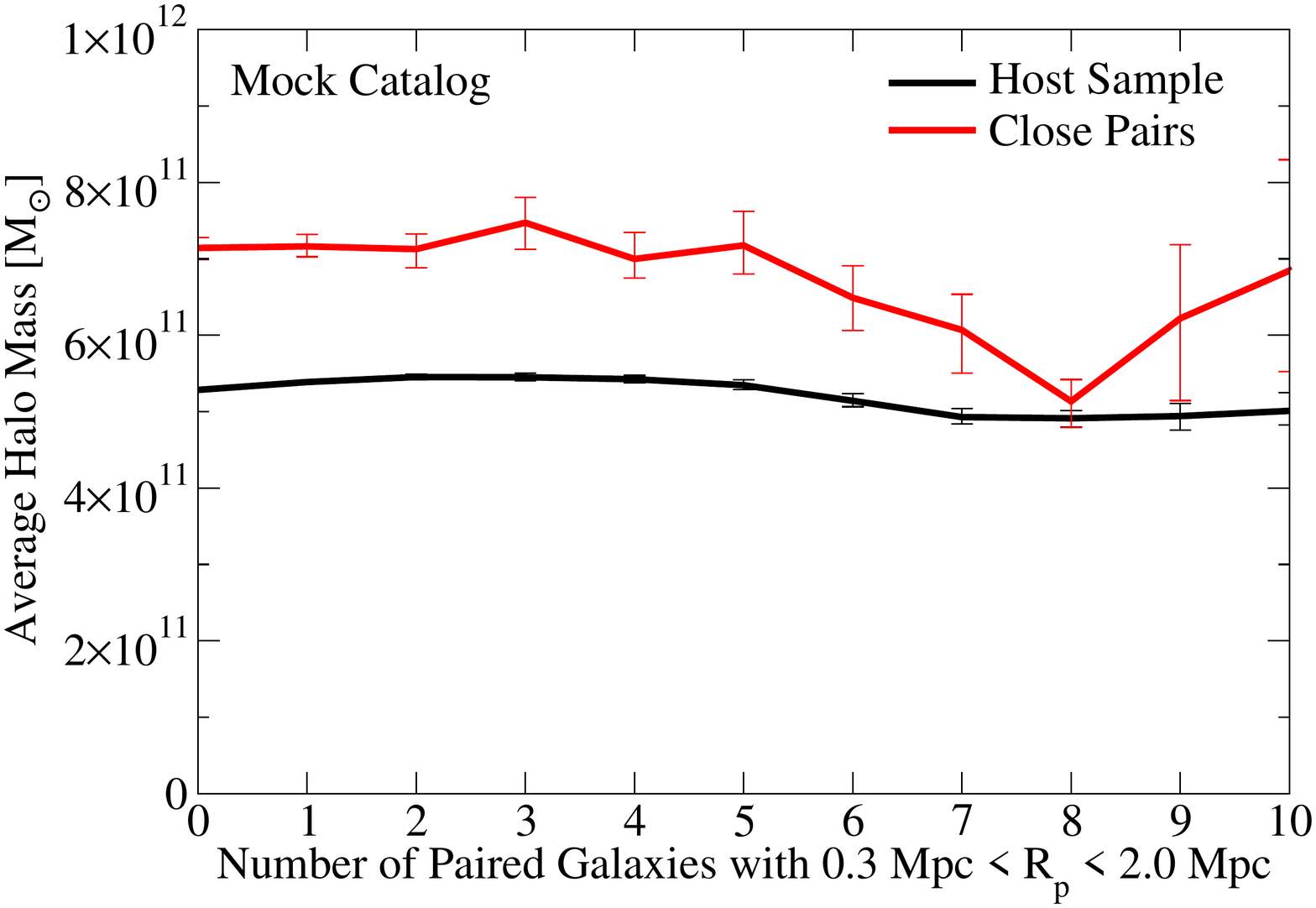}\plotssgrace{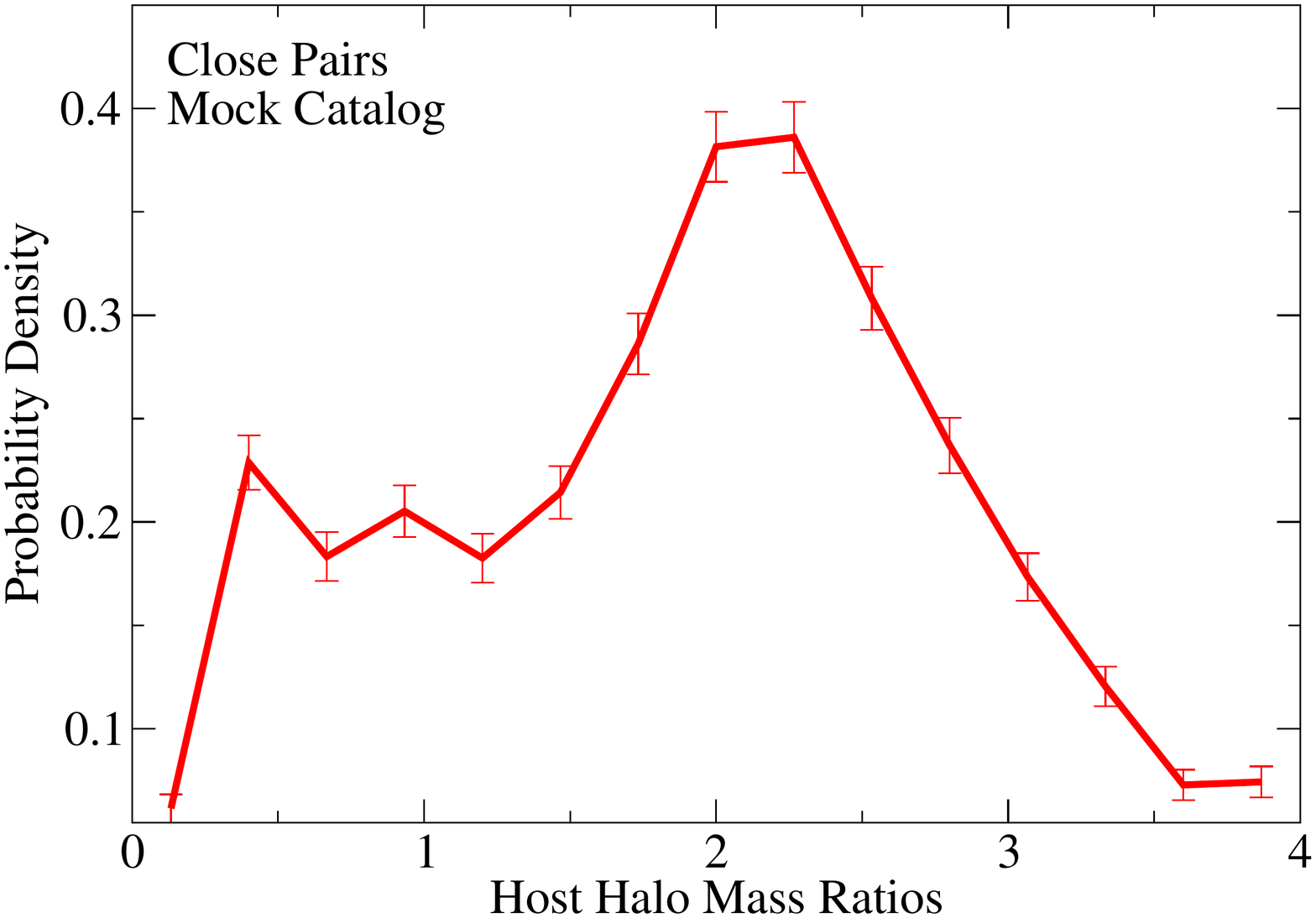}\\[-4ex]
\caption{\textbf{Top-left} panel: host halo mass distributions of close pairs compared to all galaxies in our sample, from the mock catalog.  Close pairs' host haloes are 20\% more massive, on average, than for the host sample (see explanation in \S \ref{s:dm_hosts}).  \textbf{Top-right} panel: host halo mass dependence on the distance to the nearest paired galaxy; \textbf{bottom-left} panel: host halo mass dependence on environment (number of nearby paired galaxies).  \textbf{Bottom-right} panel: halo mass ratios between the host galaxy and paired galaxy.  Errors in the top-left and bottom-right panels are jackknifed; errors on the other two panels are bootstrapped.}
\label{f:halo_masses}
\end{figure*}

\begin{figure*}
\vspace{-7ex}
\plotgrace{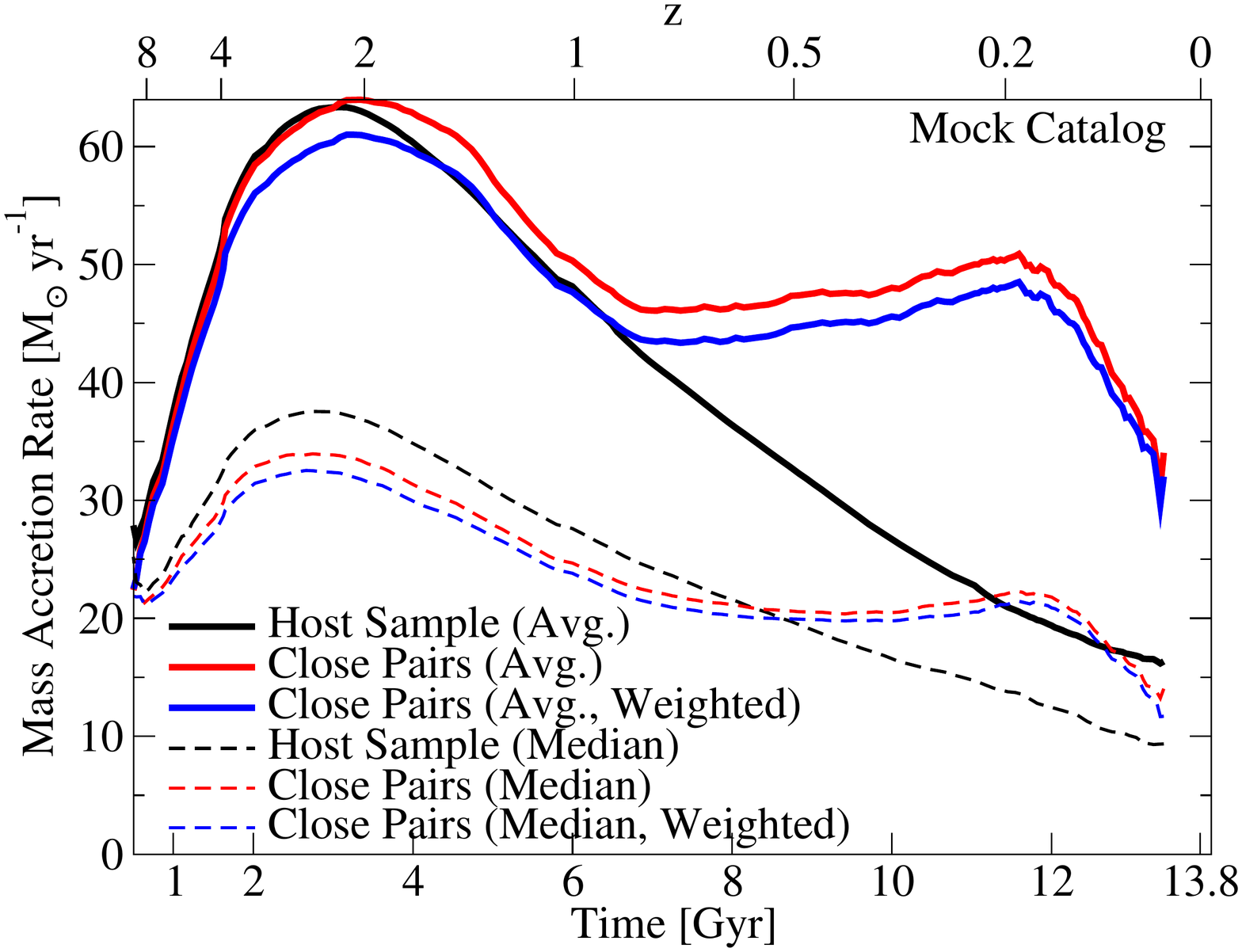}\plotgrace{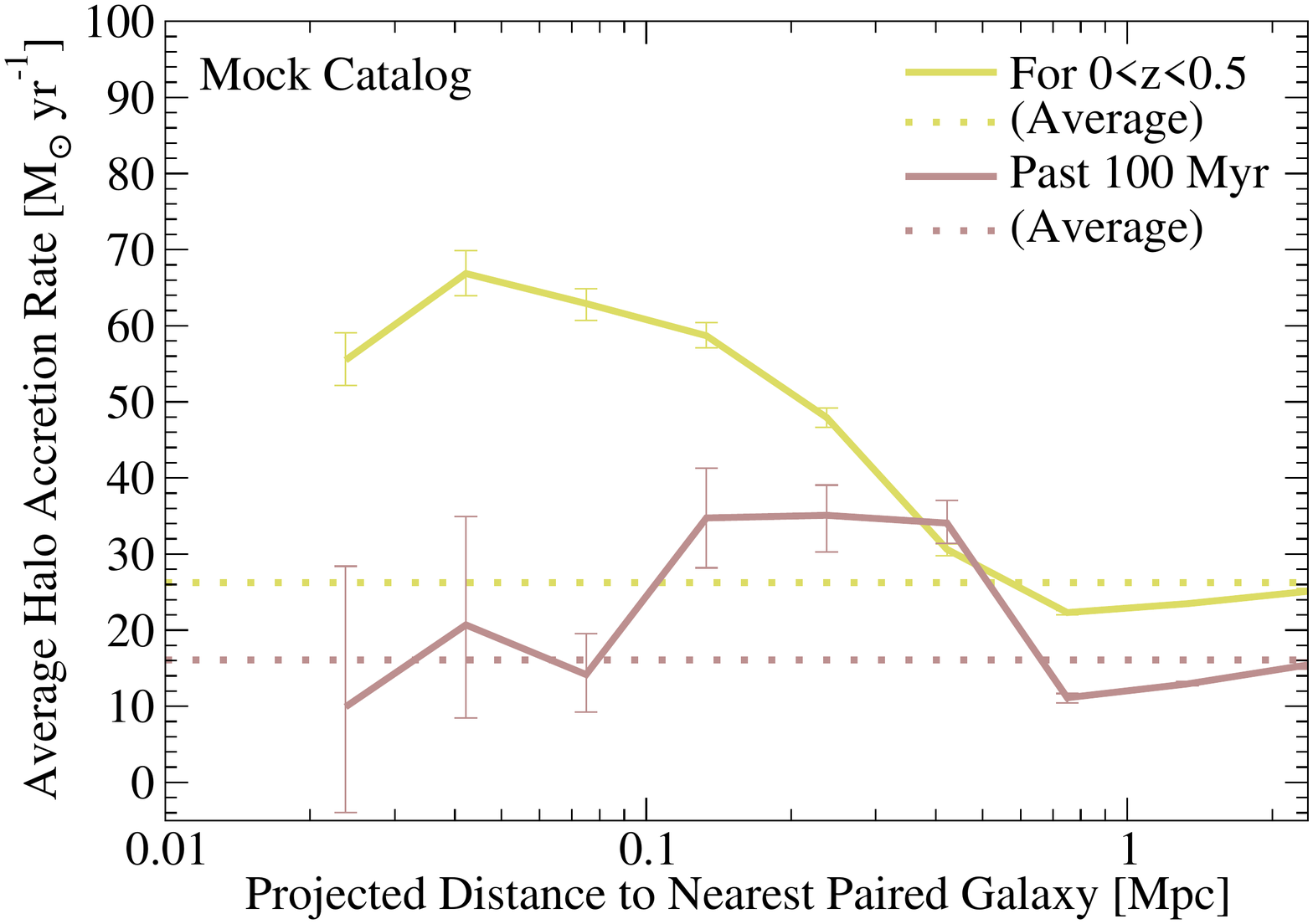}\\[-5ex]
\plotgrace{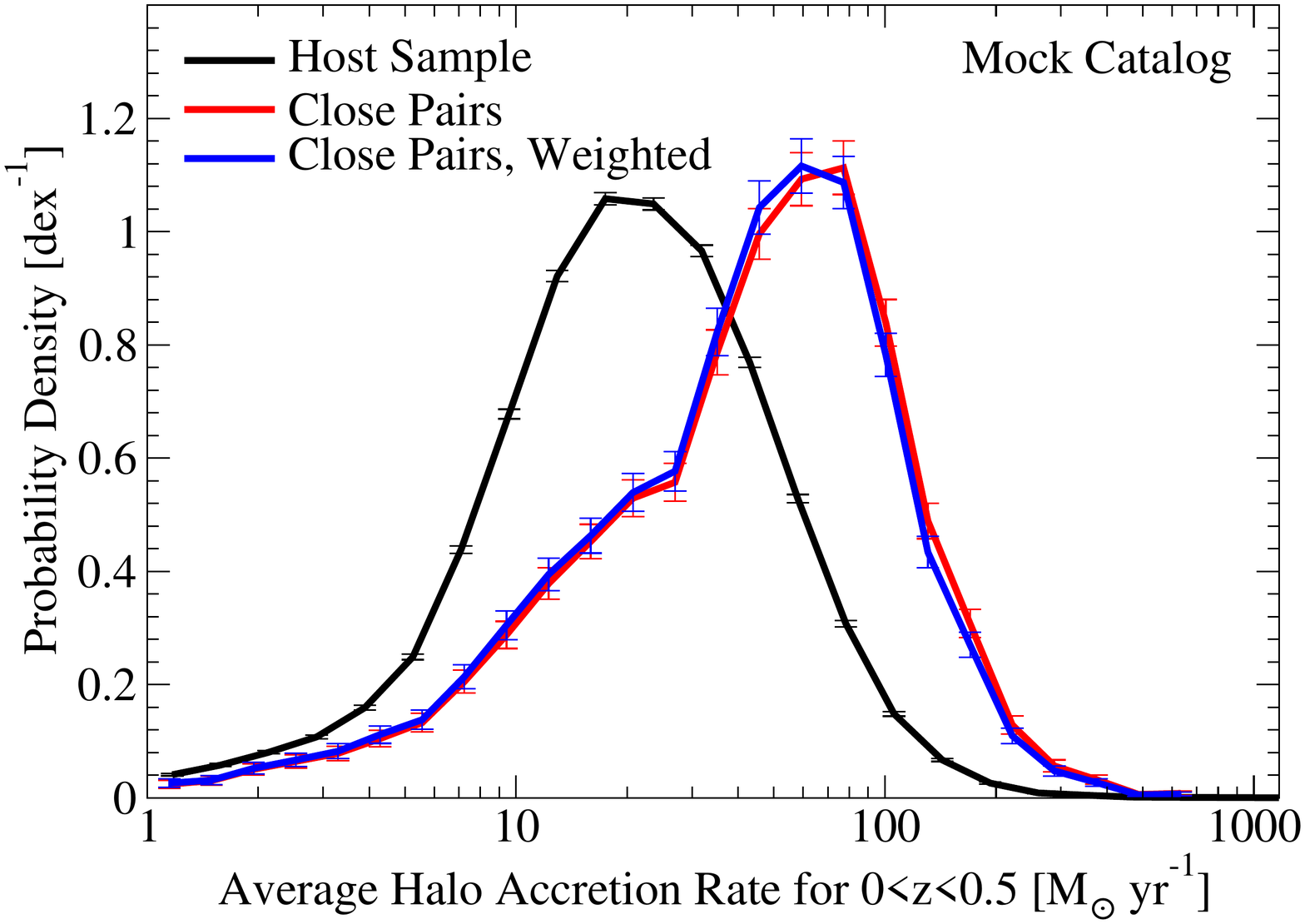}\plotgrace{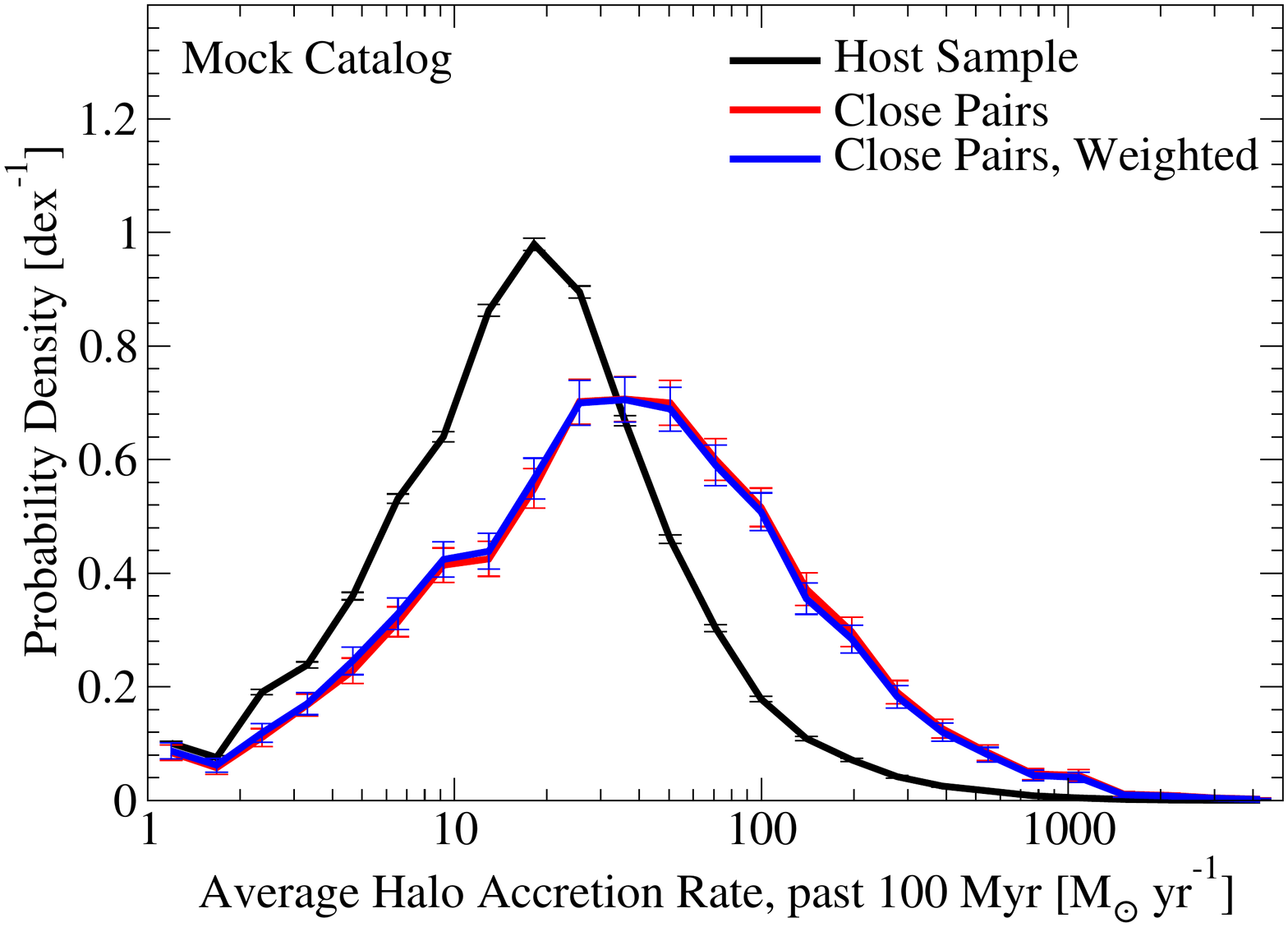}\\[-5ex]
\plotgrace{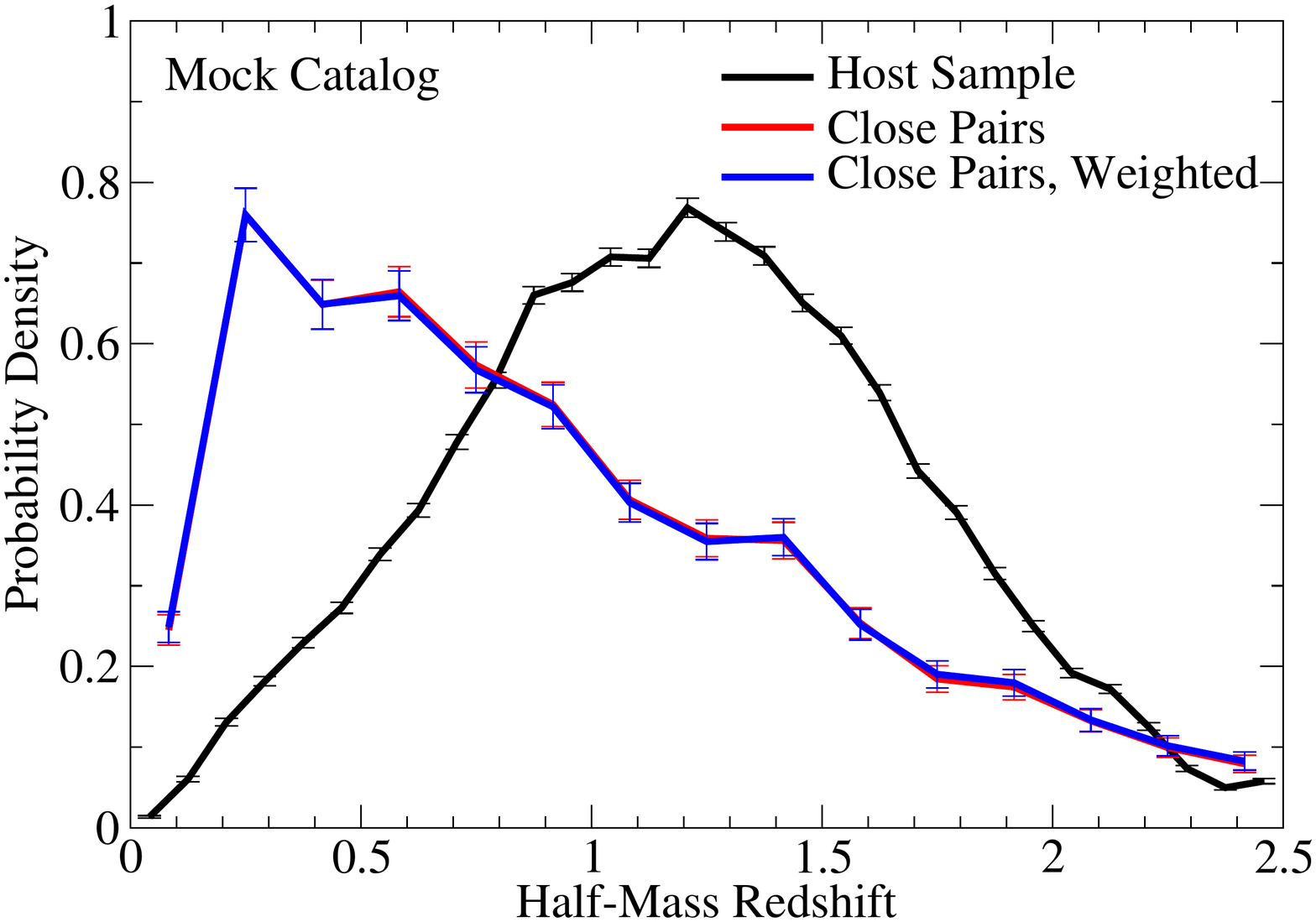}\plotgrace{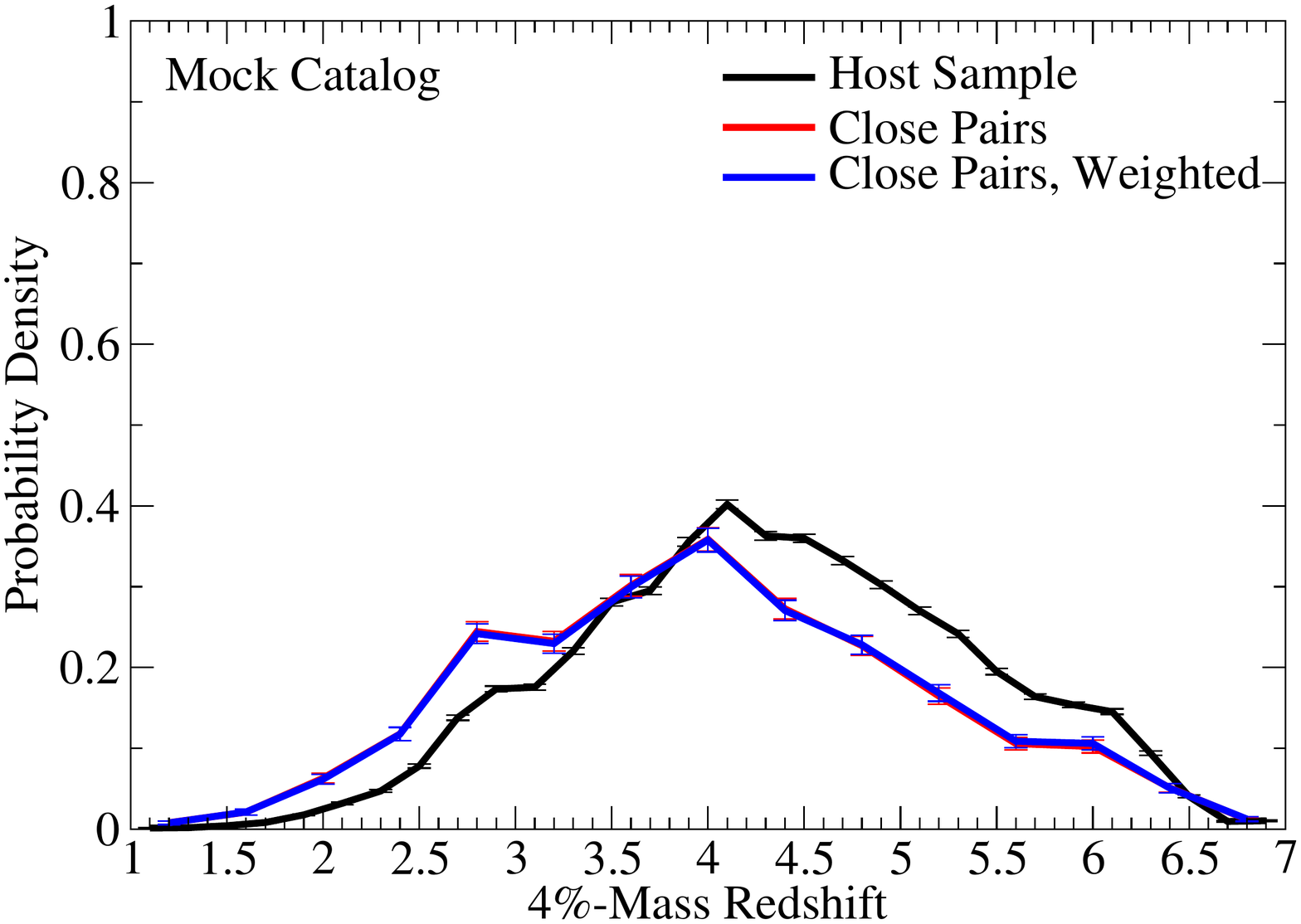}\\[-3ex]
\caption{Differences between total host halo mass accretion rates for close pairs and the host sample.  These accretion rates encompass mass growth from all sources (i.e., both clumpy and unresolved ``smooth'' accretion). \textbf{Top-left} panel: median and average (stacked) mass accretion histories for the host haloes of galaxies in the close pairs and host samples, from the mock catalog.  These show, for haloes selected at $z=0.03$, how their typical mass accretion histories change if a close pair is also present at $z=0.03$.   The large bump at $z\sim 0.2$ in the close pairs sample matches the typical redshift of first accretion for incoming major mergers.  \textbf{Top-right} panel: dependence of the averaged halo accretion rate on distance to the nearest paired galaxy.  Separate solid lines are shown for averaged accretion rates over the past 100 Myr and over the period $0<z<0.5$.  Dotted lines for each timescale show the average across all paired galaxy separations.   \textbf{Middle-left} panel: probability distribution of halo accretion rates averaged over $0<z<0.5$; \textbf{middle-right} panel: same, averaged over past 100 Myr.  \textbf{Bottom-left} panel: half-mass redshift (i.e., redshift at which haloes first obtained half their $z=0$ mass) distribution for close pairs and host sample; \textbf{bottom-right} panel: same, for the 4\%-mass redshift.  Errors on the averages in the top-right panel are bootstrapped; errors in the middle and bottom panels are jackknifed.}
\label{f:acc_rate}
\end{figure*}

We show several comparisons between the mock catalog (\S \ref{s:mocks}) and the observational galaxy sample (\S \ref{s:obs_data}) in Fig.\ \ref{f:mock_obs}.  We find excellent agreement in all cases between the mock catalog and observations, including the stellar masses of the host galaxies (nontrivial because of the satellite exclusion criterion), the velocity separation between host galaxies and the nearest paired galaxy, the distribution of projected distances between host galaxies and the nearest paired galaxy, and the large-scale environment of host galaxies, as measured by the number of paired galaxies within a projected distance of 0.3 -- 2.0 Mpc.

Encouragingly, the large-scale environment of hosts with close pairs is similar to that of the entire host sample.  We find that, while the overall satellite fraction in the host sample is small (2.6\%), galaxies with close pairs are slightly more likely to be satellites (13.3\%).  Most of these latter cases result from scatter in the stellar mass---halo mass relation, which implies that the smaller of the two merging haloes will occasionally contain the larger of the two galaxies.

The distribution of stellar masses is also similar between close pairs and the host sample, although our close pair selection criteria does have a slight bias towards larger galaxies.  It is possible to reverse this bias with an additional weighting function for close pairs:
\begin{equation}
W(M_\ast) = 1 - \log_{10}\left(\frac{M_\ast}{10^{10.25}\,\Msun}\right)
\label{e:sm_weight}
\end{equation}
This weighting is treated as a multiplicative change to galaxies' observable volumes; we present both weighted and unweighted results in \S \ref{s:results}.  

We note that the mocks slightly underpredict the number of close pairs within 200 kpc (Fig.\ \ref{f:mock_obs}, lower-left panel); host galaxies in the mock catalog have close pairs 6.2\% of the time as compared to 6.7\% of the time in observations, a 9\% difference.  Tests with a higher-resolution simulation (125 Mpc $h^{-1}$ on a side, 2048$^3$ particles, each with mass $2.58\times10^7\,\Msun$) show the same slight discrepancy, as do tests with the Millennium-II catalog from \cite{Tollerud11}.  Several other factors could be responsible, including errors partitioning light in merging galaxies (see Fig.\ \ref{f:cp_sdss}), slightly increased binding energies for galaxies as compared to their host dark matter haloes, scatters larger than 0.2 dex between stellar mass and halo mass for the host galaxies \citep{Busha11}, and sample variance; however, most of these factors are beyond the scope of this paper to address.

\subsection{Host Dark Matter Halo Masses and Formation Redshifts}
\label{s:dm_hosts}

\begin{figure*}
\vspace{-7ex}
\plotgrace{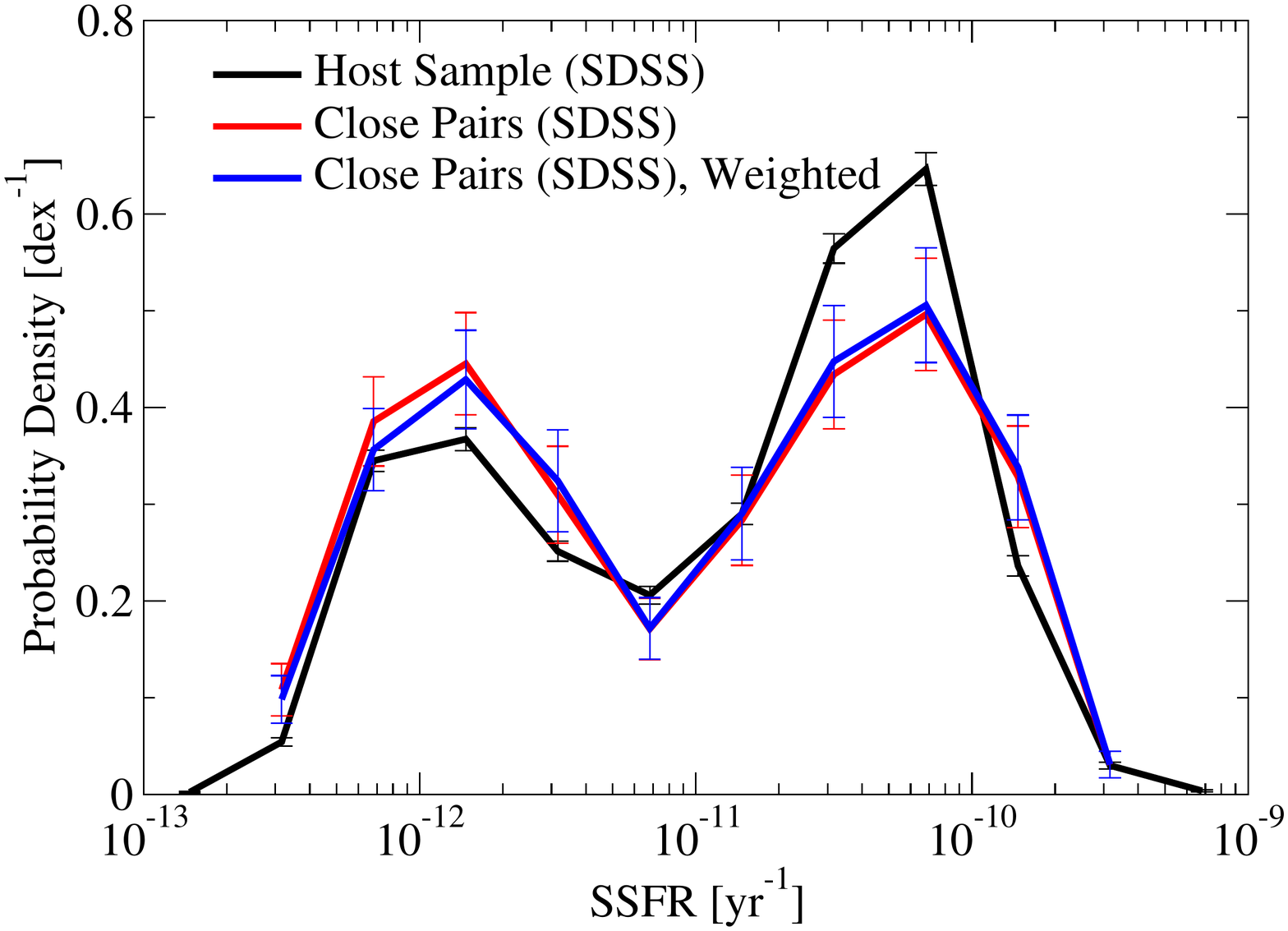}\plotgrace{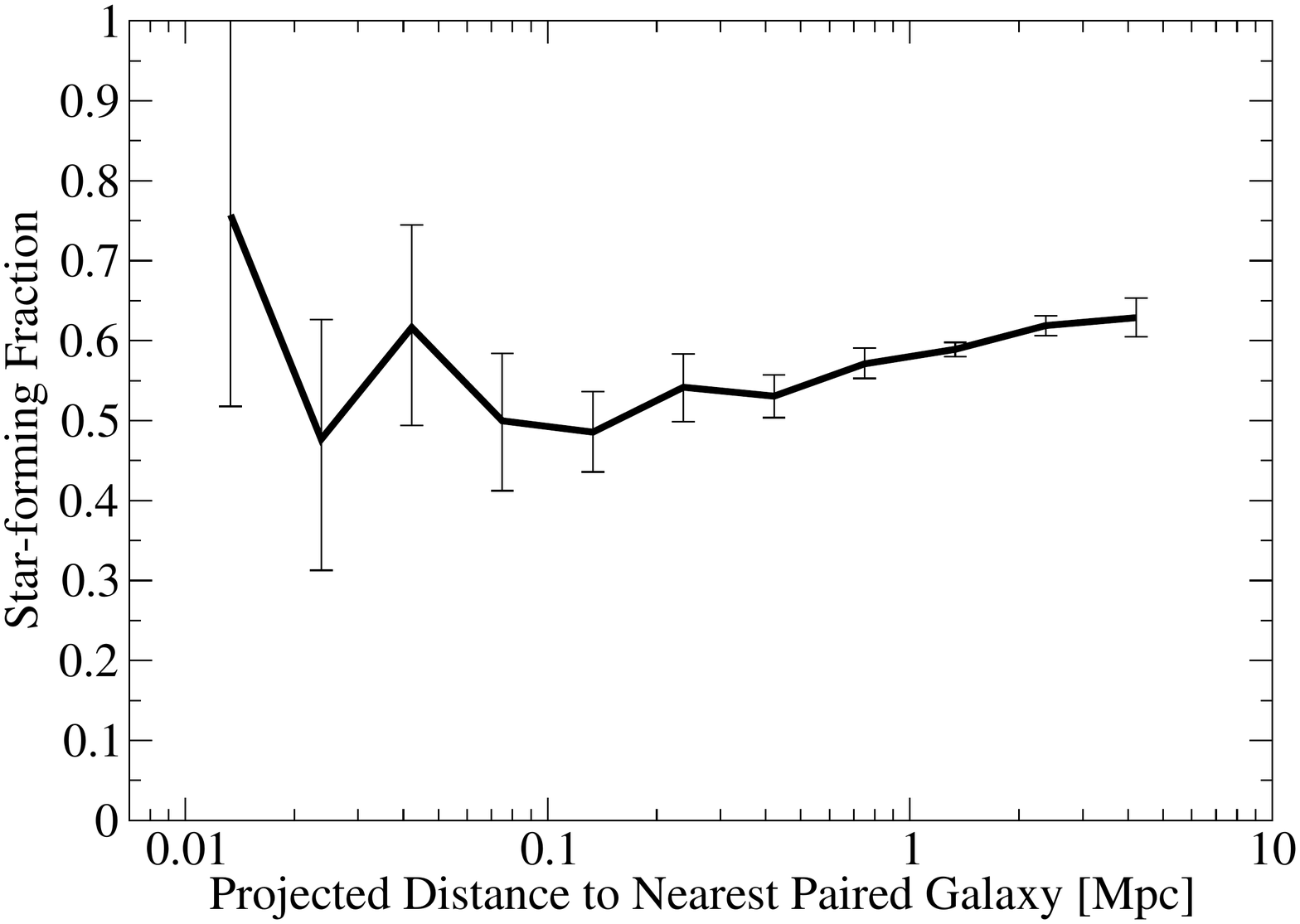}\\[-5ex]
\plotgrace{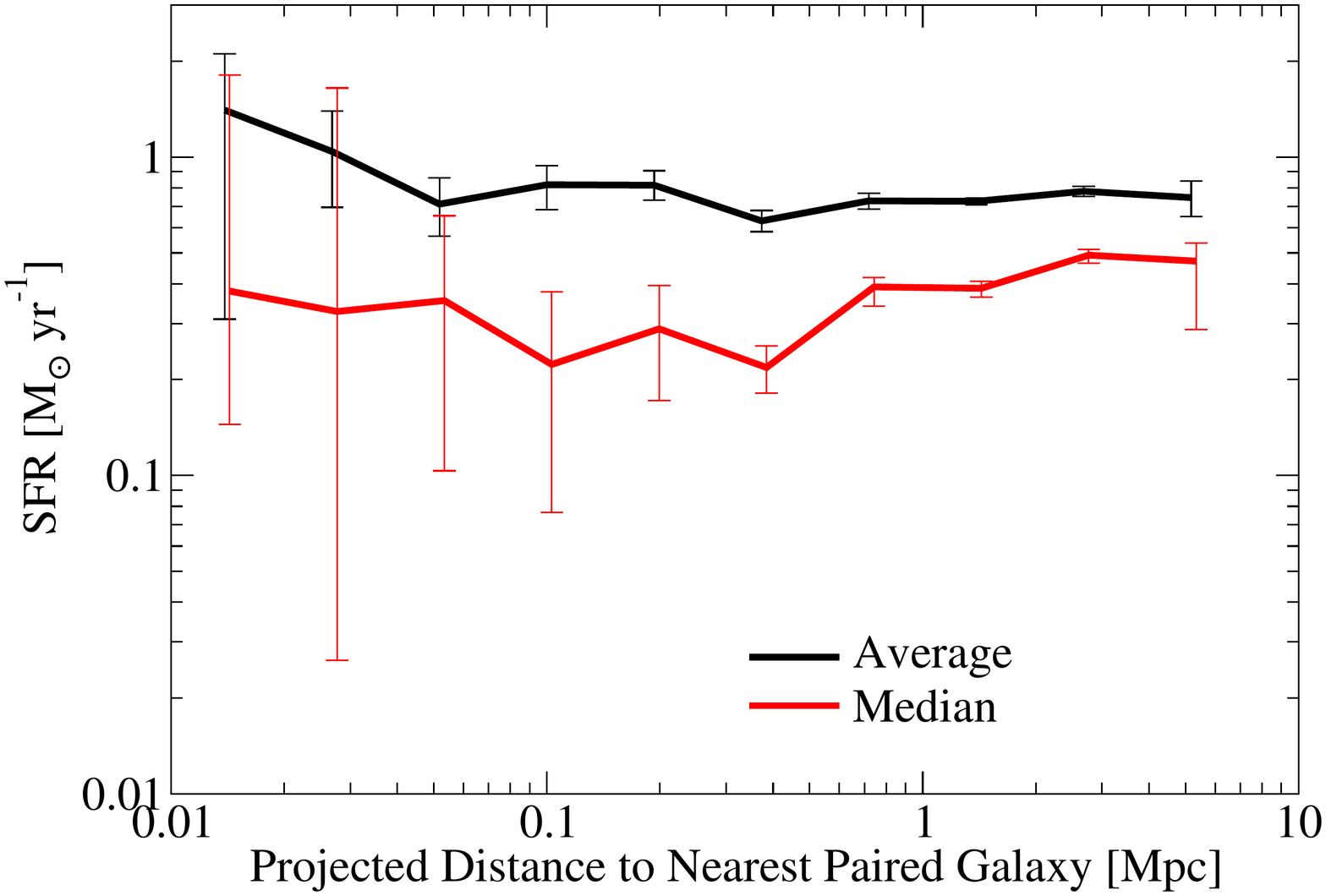}\plotgrace{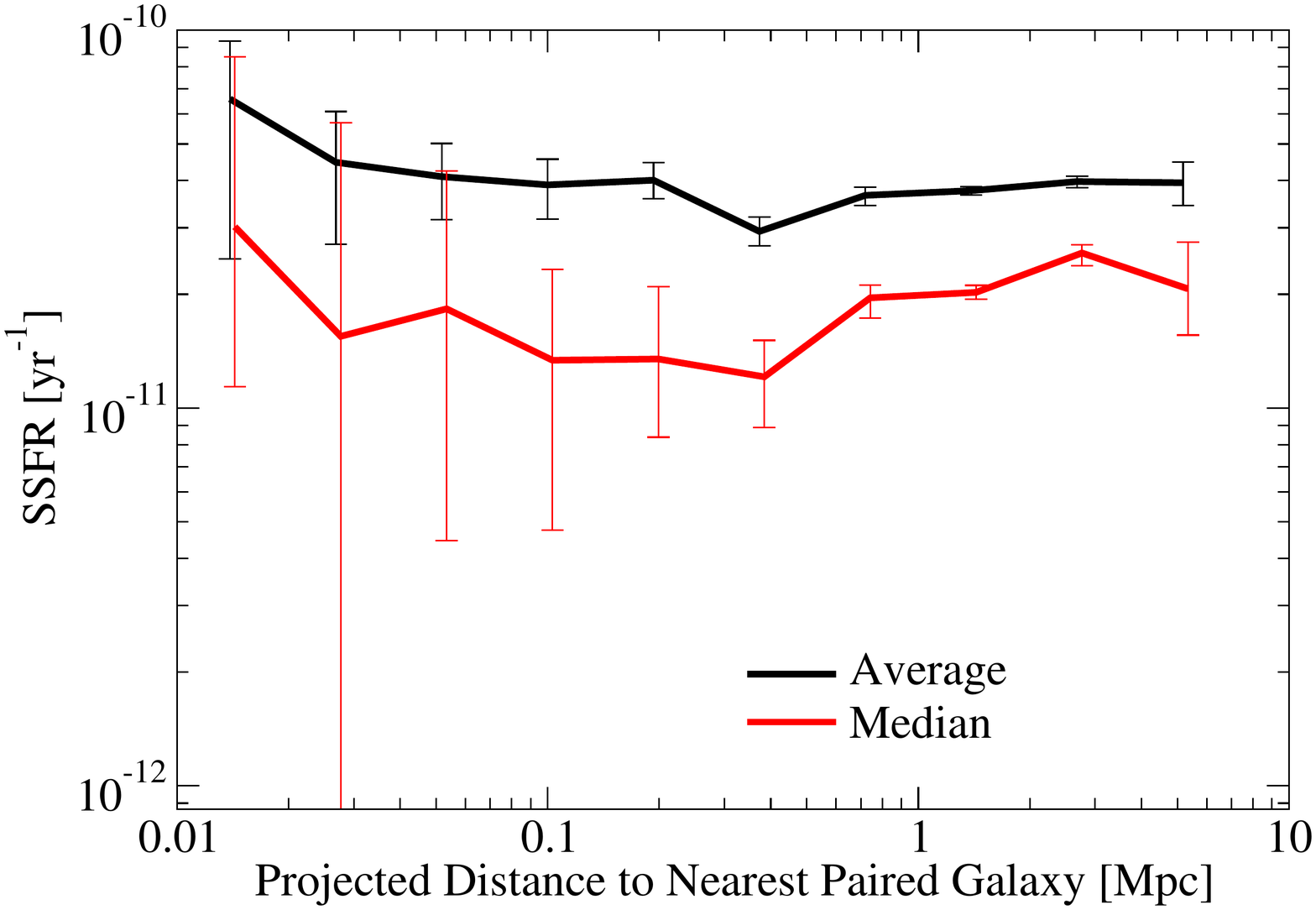}\\[-3ex]
\caption{\textbf{Top-left} panel: Specific star formation rate distribution for the host galaxies in close pairs and in the host sample, from SDSS; the weighting removes a small stellar mass bias for close pairs (\S \ref{s:mocks}).  \textbf{Top-right} panel: star-forming (SSFR $> 10^{-11}$ yr$^{-1}$) fraction for host galaxies as a function of distance from the host galaxy to the nearest paired galaxy (i.e., a galaxy with $\Delta V < 500$ km s$^{-1}$ and stellar mass between 0 to 0.5 dex less than the host galaxy).  \textbf{Bottom-left} and \textbf{bottom-right} panels: host galaxy star formation rates and specific star formation rates, respectively, as a function of distance between the host galaxy and the nearest paired galaxy.  Median SFRs and SSFRs have been slightly offset horizontally for clarity.  Errors are jackknifed for top-left panel, bootstrapped for all others.}
\label{f:sdss_sfr}
\end{figure*}

We show the host halo masses corresponding to galaxies in our host and close pairs samples in Fig.\ \ref{f:halo_masses}, as taken from our mock catalog.  Close pairs have very similar host halo masses to those for the host sample, with only $\sim$20\% larger masses (both average and median).  This mass bias is not affected by reweighting close pairs according to stellar masses (Eq.\ \ref{e:sm_weight}), since there is significant scatter in stellar mass at fixed halo mass.  Indeed, this scatter is fundamentally responsible for the close pairs mass bias.  Since host halo mass correlates with the number of satellites, a larger halo mass at fixed galaxy mass will result in a higher likelihood of the galaxy having a close pair.  Selecting galaxies with close pairs will therefore automatically select slightly larger haloes (Fig.\ \ref{f:halo_masses}, top-right panel).

Unfortunately, both stellar mass and environment (Fig.\ \ref{f:halo_masses}, bottom-left panel) correlate weakly with halo mass within our sample because of the highly restricted range of halo masses being considered.  Weighting galaxies by stellar mass or environment to further align the halo mass distributions of close pairs and the host sample would ruin the existing agreement between the stellar mass or environment distributions (Fig.\ \ref{f:mock_obs}).  Weighting galaxies in close pairs by their velocity separations may seem attractive, but satellite velocities at fixed halo mass are correlated with the assembly time of the halo (i.e., halo age), which would introduce an additional nontrivial correlation with star formation rate (Hearin et al., in preparation).  As the cure seems worse than the problem in this case, we do not add any weighting to target halo masses.

For galaxies with close pairs, the vast majority of their haloes' peak mass ratios are within $1:3$ (Fig.\ \ref{f:halo_masses}, bottom-right panel).  We find that for close pairs, the paired galaxy is actually within the host halo's virial radius 50.5\% of the time, with the remaining cases due to chance projections.  While no selection can be perfect, this represents a very strong preference for major halo mergers; in the host sample as a whole, the major merger fraction is only 3.2\%.

We show the corresponding halo mass accretion rates (i.e., net matter flux into the virial radius from both clumpy and unresolved ``smooth'' sources) in Fig.\ \ref{f:acc_rate}.  The host haloes of galaxies with close pairs experience over 100\% higher average accretion rates from $z=0.4$ to $z=0$ as compared to host haloes for the entire host galaxy sample (Fig. \ref{f:acc_rate}, top-left panel).  As noted above, close pairs reside in somewhat larger haloes, but this would result in an expected increase of only 20\% in their average accretion rates \citep{BehrooziHighZ}.  The enhanced accretion rates extend over two dynamical times before $z=0$; this is not only due to a range of merger timescales, but also due to correlated structure arriving along the same filament as the merging halo.

These enhanced accretion rates are only evident when a paired galaxy appears within the host halo's virial radius (Fig.\ \ref{f:acc_rate}, top-right panel).  The position of the paired galaxy within the virial radius does not strongly constrain infall time \citep{Oman13}, except for paired galaxies very close to their host's centres, which cannot have fallen in recently (Fig.\ \ref{f:acc_rate}, top-right panel).  The distributions of host halo accretion rates are shown in the middle panels of Fig.\ \ref{f:acc_rate}.  Since halo accretion rates on 100 Myr timescales have a standard deviation of $0.45$ dex (Fig.\ \ref{f:acc_rate}, middle-right panel), the distributions for the close pairs and host sample overlap.  However, the differences in the medians, means, and modes of the distributions for close pairs versus the host sample are still large: 0.3-0.4 dex, depending on the statistic.  This large log-normal scatter in instantaneous accretion rates also explains why average accretion rates are typically a factor of two higher than median accretion rates in the top-left panel of Fig.\ \ref{f:acc_rate}.

Finally, we show halo assembly times in the bottom panels of Fig.\ \ref{f:acc_rate}.  The half-mass assembly times for the haloes in the close pairs sample are (by construction) extremely skewed towards low redshifts, reflecting recent major mergers.  The median half-mass assembly redshift for close pairs is $z=0.83$, whereas it is $z=1.23$ for the host sample; 30\% of close pairs have formation redshifts $z<0.5$, whereas only 8\% of the host sample does.  However, the longer-term halo mass accretion histories (as probed by 4\%-mass assembly times) overlap significantly more between the close pairs and the host sample.

\section{Results}
\label{s:results}

\subsection{All Close Pairs}

\label{s:all_cp}

\begin{figure*}
\vspace{-7ex}
\plotgrace{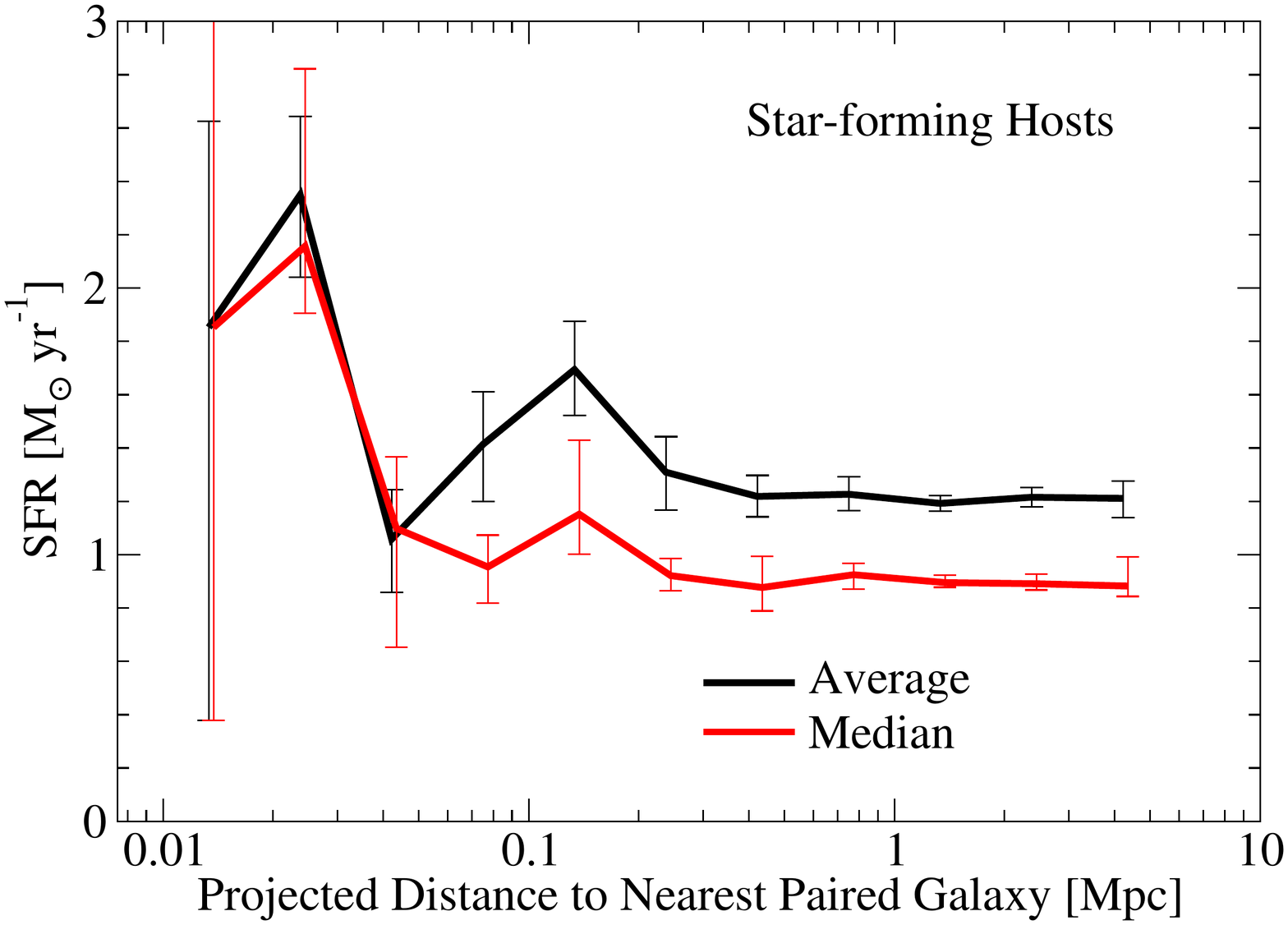}\plotgrace{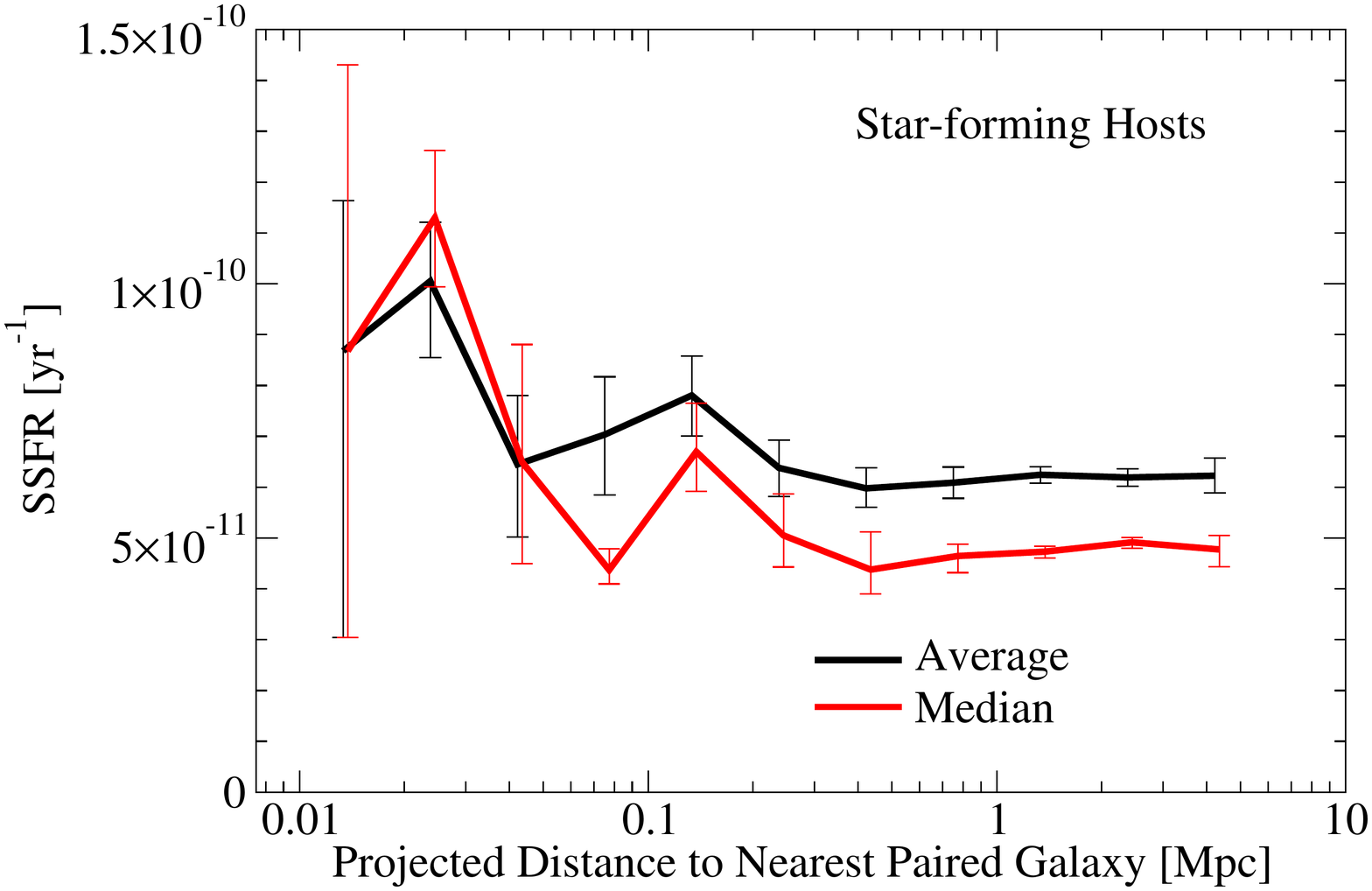}\\[-5ex]
\plotgrace{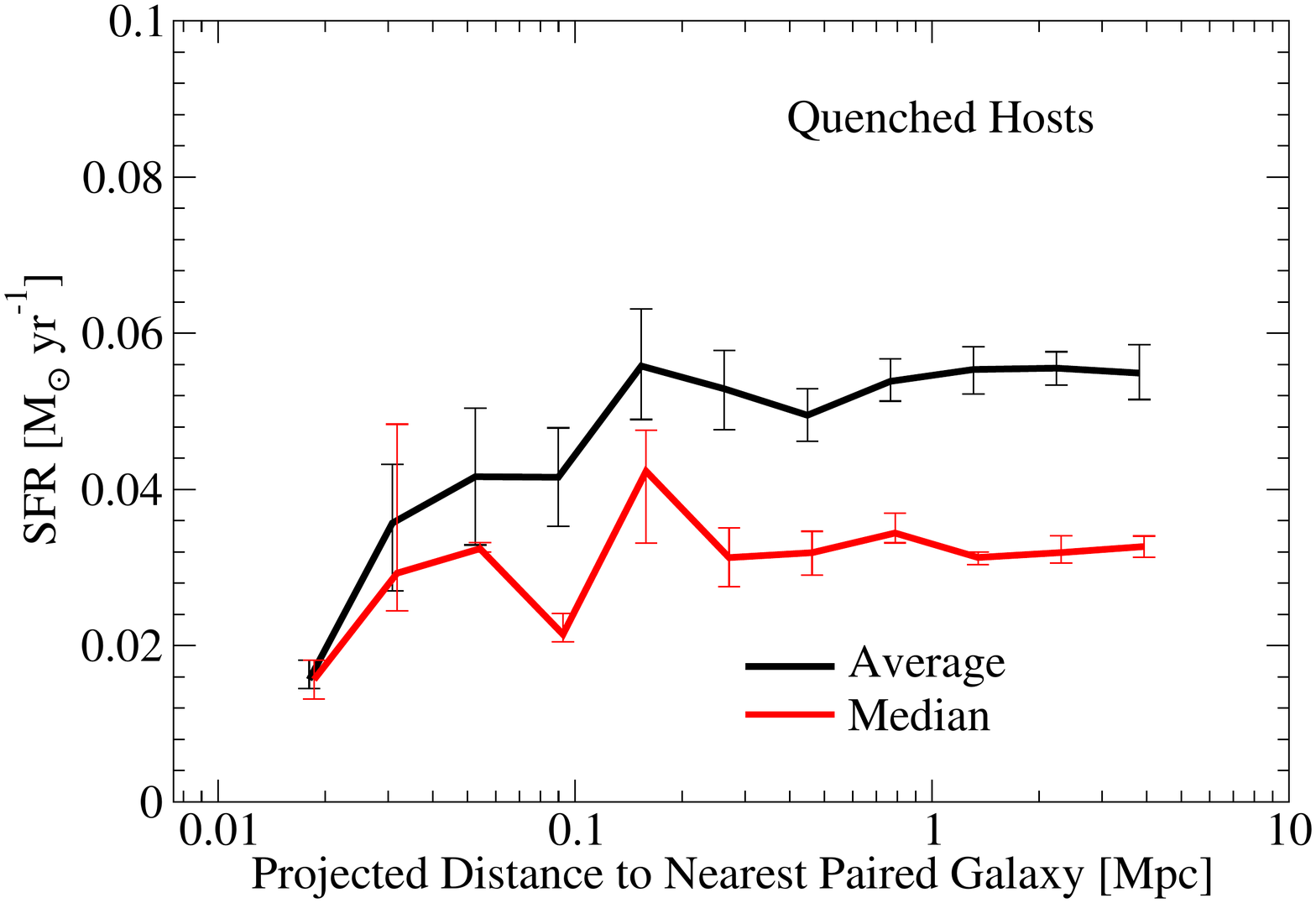}\plotgrace{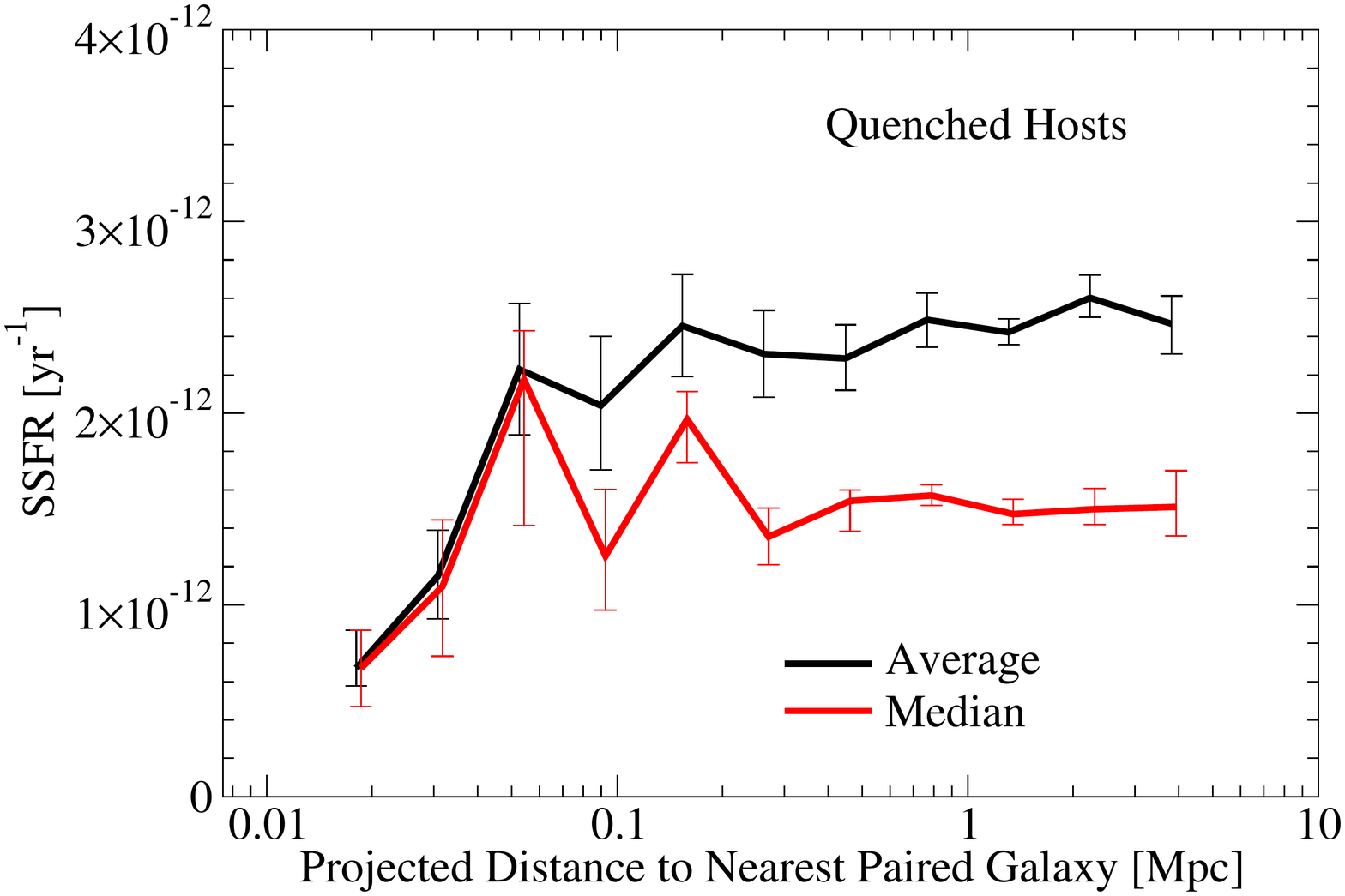}\\[-3ex]
\caption{\textbf{Top-left} panel: host galaxy star formation rates as a function of distance between \textbf{star-forming} host galaxies (SSFR $> 10^{-11}$ yr$^{-1}$) and the nearest paired galaxy.  \textbf{Top-right} panel: same, for host galaxy specific star formation rates.  \textbf{Bottom} panels: same as top panels, for \textbf{quenched} host galaxies.  The low SFRs and SSFRs for extremely close pairs ($<30$ kpc) near quenched hosts are not seen in larger samples (Fig.\ \ref{f:blanton} in Appendix \ref{a:variation}).  Errors are bootstrapped for all panels.}
\label{f:sdss_sf_q}
\end{figure*}

We show the distribution of host galaxies' specific star formation rates in the SDSS for those in close pairs and those in our full host sample in Fig.\ \ref{f:sdss_sfr}.  Galaxies with close pairs show no strong differences in SSFRs, even after reweighting to eliminate a small stellar mass bias (Eq.\ \ref{e:sm_weight}; Fig.\ \ref{f:sdss_sfr}, top-left panel).  The close pairs show a star-forming (SSFR$ > 10^{-11}$ yr$^{-1}$) fraction of 53\%, compared to 59\% for the host sample; while this is statistically very significant (99.6\% confidence for a lower star-forming fraction), the effect size is modest at best.  As 50\% of galaxies in close pairs are undergoing a major halo merger (\S \ref{s:dm_hosts}), major halo mergers can account for at most a 12\% reduction (to 47\%) in the star-forming fraction of central $L^*$ galaxies.  The true effect is likely smaller, as the galaxies with close pairs also have slightly increased host halo masses and satellite fractions (\S \ref{s:dm_hosts}), which would also lead to larger quenched fractions \citep{Wetzel11,Woo13}.

Average star formation rates may be enhanced during extremely close passages (i.e., projected distances $<$ 30 kpc, or 10-15\% of the virial radius), but no strong evidence for it is seen here (see, however, \S \ref{s:sf_v_q}).  Also, no evidence exists for any changes in the star-forming fraction of these extremely close pairs just prior to merging (see also Fig.\ \ref{f:cp_sdss}).  We have tested resolving fibre collisions with the redshift of the nearest galaxy to increase sample sizes, which further restricts the possibility of an enhanced star-forming fraction (Fig.\ \ref{f:blanton} in Appendix \ref{a:variation}).  We have also tested expanding the range of host stellar masses to range from $10^{10}\,\Msun$ to $10^{11}\,\Msun$, which expands the sample size to 25,364 host galaxies, but we do not find any stronger evidence for enhancement of the star-forming fraction at close radii.

\subsection{Star-forming versus Quenched Host Galaxies}

\label{s:sf_v_q}

Several previous studies have considered star-forming host galaxies only.  For completeness, we show the radial dependence of average and median SFRs and SSFRs for star-forming and quenched host galaxies in Fig.\ \ref{f:sdss_sf_q}. 

First, we note strong enhancements (60--100\%) in SFRs and SSFRs for star-forming hosts with a close pair separated by $<30$ kpc.  There is also a very modest average enhancement of $9.4^{+6.4}_{-6.0}$\% when a close pair appears within the host halo's virial radius (i.e., at about 200 kpc).  This enhancement is also visible for the unsplit sample (Fig.\ \ref{f:sdss_sfr}), although at lower significance.  The quenched sample shows no such enhancement, perhaps by definition; however, we note that the fraction of star-forming galaxies also does not show any enhancement (Fig.\ \ref{f:sdss_sfr}).  This excludes chance projections as a complete explanation for quenched galaxies' behaviour.  If quenched galaxies with close pairs were all due to chance projections, and all true (3D) close pairs rejuvenated star formation in their hosts, then the star-forming fraction would be gradually increasing with decreasing pair separation---i.e., as the fraction of chance projections decreases---when in fact the opposite occurs (Fig.\ \ref{f:sdss_sfr}).  This may imply that star-forming galaxies (but not quiescent ones) experience a weak accretion-associated increase in SSFRs during major mergers (see \S \ref{s:discussion}).  We note that while quenched hosts appear to show a decrease in SFRs and SSFRs at close separations ($<30$ kpc), this is due to sample variance, and it is not seen in the larger NYU-VAGC sample (Fig.\ \ref{f:blanton} in Appendix \ref{a:variation}). 

\section{Discussion}
\label{s:discussion}

In \S \ref{s:results}, we found that galaxies with close pairs have only modestly increased quenched fractions, that close pairs do not appear to rejuvenate star formation even at extremely close separations, and that star-forming host galaxies with close pairs show increased specific star formation rates (SSFRs), especially at extremely close separations.  We discuss how major mergers impact galaxy quenching in \S \ref{s:mm_q}, how halo growth correlates with galaxy growth in \S \ref{s:hma}, the permanence of quenching in \S \ref{s:perm_q}, comparisons with semi-analytic models in \S\ref{s:semi_a}, and comparisons with previous literature results in \S \ref{s:prev}.

\subsection{Impact of Major Mergers on Galaxy Quenching}

\label{s:mm_q}

From \S \ref{s:all_cp}, central $L^{*}$ galaxies with a close pair within 200 kpc have a slightly (6\%) reduced star-forming fraction, compared to central $L^*$ galaxies as a whole.  Based on the fraction of close pairs which are true (3D) major mergers, major mergers could reduce the star-forming fraction by at most 12\% (\S \ref{s:all_cp}).  The true effect is likely smaller, as the slightly larger host halo masses for close pairs (Fig.\ \ref{f:halo_masses}) and slightly larger satellite fractions (\S \ref{s:validation}) could also result in very modestly increased quenching.  To prevent confusion, we note again that a major halo merger means only that the smaller halo has arrived within the virial radius of the larger halo, \textit{not} that the two haloes have become indistinguishable.

If clump-induced gravitational heating were a major trigger for quenching, it is surprising that the quenched fraction changes so little.  Close pairs are associated both with major mergers and enhanced accretion rates (Fig.\ \ref{f:acc_rate}, top-left panel); the peak of the merging activity occurs at $z\sim 0.2$, about one dynamical time ago.  This is exactly the same timescale on which quenching would be expected to occur after suddenly shutting off cooling---for good reason: as both processes are driven by gravity, the characteristic timescales are expected to be very similar.  Without a mechanism to delay the cooling shutoff for several dynamical timescales, gravitational heating is likely not an initial trigger for quenching, although it may help sustain pre-existing quenching.

It also seems unlikely that major deficits in the host halo's hot gas reservoir are the initial cause for quenching.  Mass loss from infalling haloes accelerates rapidly once they pass within the virial radius of the host halo \citep[e.g.,][]{Tormen98b,Kravtsov04b,Knebe06}; hot gas can also be stripped by ram pressure \citep{Kimm10,Bahe13}.  70\% of the paired galaxies are star-forming (independent of projected distance), meaning that ample supplies of hot (and cold) gas should be available during major halo mergers \citep{Popping14,Ellison15}.  Quenched $L^*$ galaxies are also surrounded by significant amounts of cold gas \citep{Thom12,Tumlinson13}.  These results suggest that quenching in central galaxies may be the result of cold gas collapse stalling outside the galaxy \citep[e.g.,][]{Thom12}, rather than a lack of gas within the halo.

If this is the case, then tidal torques from passing galaxies may accelerate (and perhaps rejuvenate) star formation as the forces help funnel cold gas to galaxy centres \citep{Mihos94,Barnes96,Hopkins08,Patton13}.  This hypothesis is partially consistent with the results in this study.  The presence of an extremely close pair appears to strongly enhance the star formation \textit{rate} in star-forming galaxies (Fig.\ \ref{f:sdss_sf_q}), 
but not to change the star-forming \textit{fraction} (Figs.\ \ref{f:cp_sdss}, \ref{f:sdss_sfr}, and \ref{f:blanton}).  This may imply that the cold gas around quenched galaxies is too far removed from the disk to be brought in by tidal torques, or that the mechanism which quenches $L^*$ galaxies is strong enough to delay cold gas accretion until at least the physical merger of the two galaxies.  That said, future studies using photometric stellar masses for extremely close pairs may be able to test for a weaker effect than can be constrained with our analysis.  Unfortunately, using merger features to study this process is difficult, because morphological disturbances last for different amounts of time depending on the gas contents, morphologies, and trajectories of the progenitors \citep{Lotz11}.

\subsection{The Relation between Halo Mass Assembly and Galaxy Star Formation}

\begin{figure*}
\vspace{-7ex}
\plotgrace{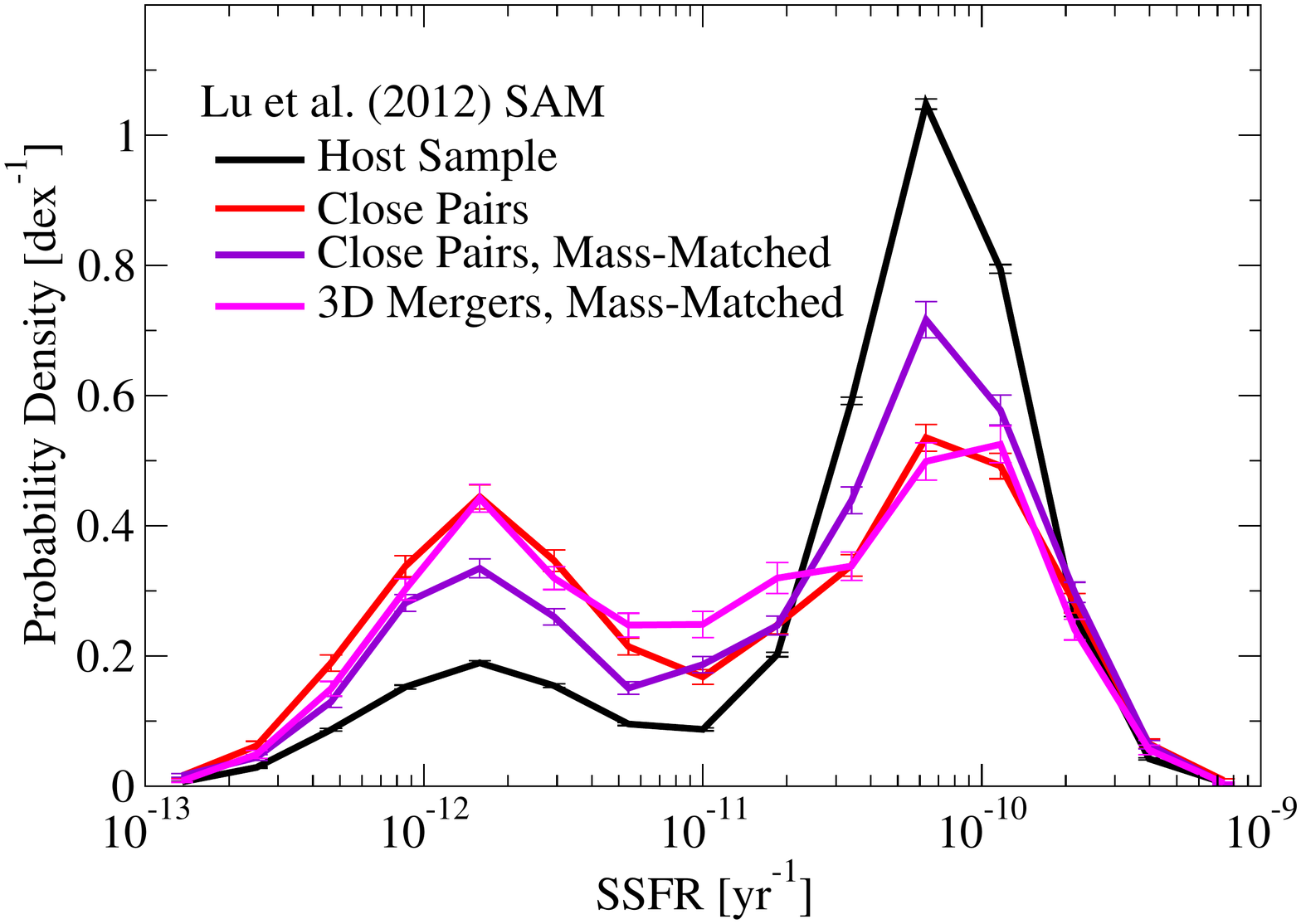}\plotgrace{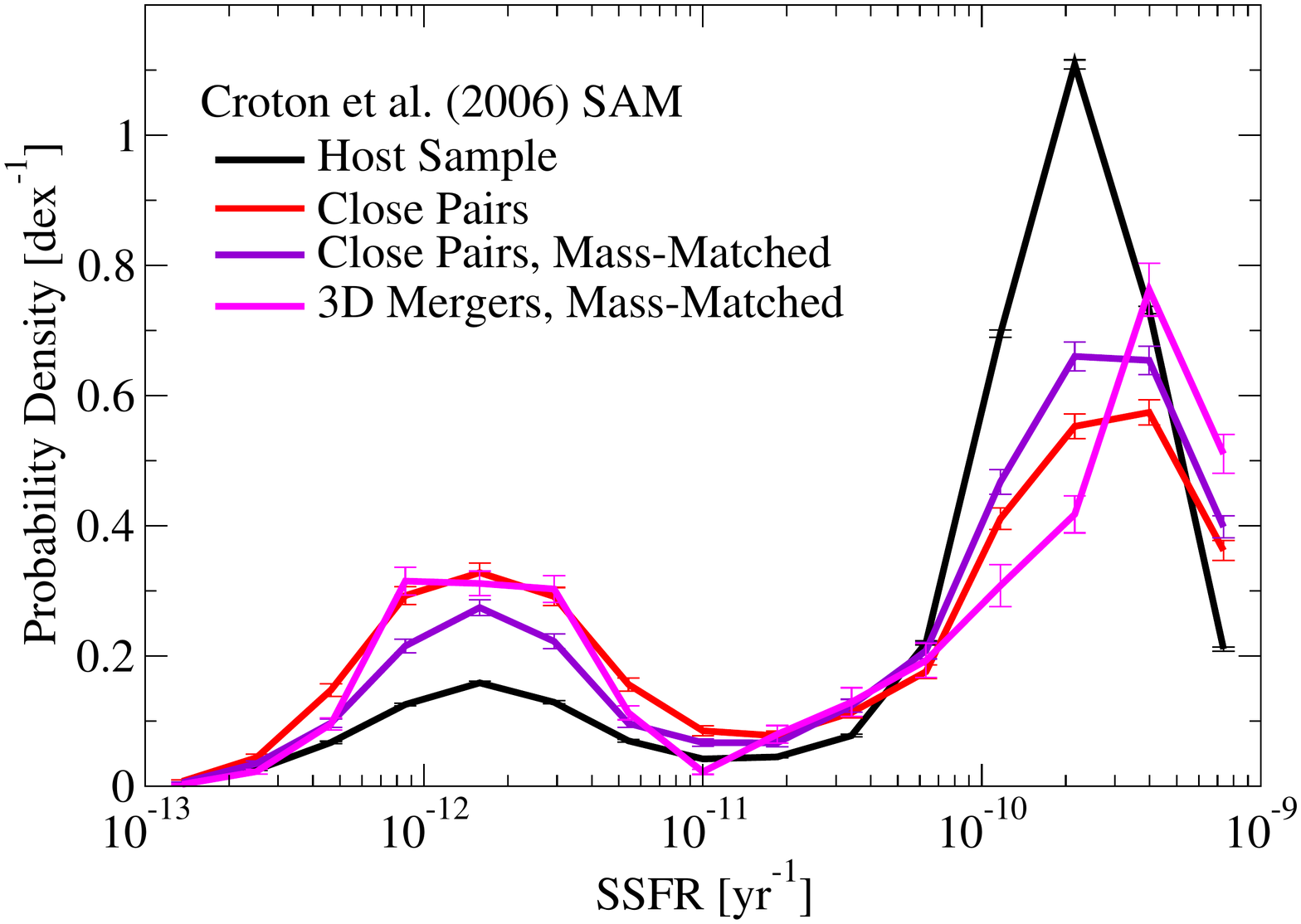}\\[-3ex]
\caption{Distributions of specific star formation rates in all host galaxies and galaxies with close pairs, using identical selection as in Fig.\ \ref{f:sdss_sfr} on mock catalogs from two semi-analytical models \citep{Lu12,Croton06}.  Also shown are SSFR distributions for host galaxies with close pairs drawn so as to have identical joint probability distributions for host halo mass and stellar mass as the host sample (``Close Pairs, Mass Matched''), as well as haloes undergoing true (3D) major mergers (``3D Mergers, Mass Matched'').  These latter two distributions allow disentangling the effects of halo mass biases in the close pairs sample from the effects of major mergers.}
\label{f:ssfr_sam}
\end{figure*}

\label{s:hma}
We have found that, for $L^*$ galaxies, a significant recent halo merger event does not imply significant recent galaxy star formation.  In addition, a lack of halo accretion does not quench star formation in $L^*$ and smaller satellite galaxies for several Gyr \citep{Wetzel13b,Wheeler14,Phillips14b}.  The galaxy's halo mass and the recent assembly history (e.g., half-mass formation scale) therefore are not sufficient information to predict galaxy star formation rates.  This suggests that, at low redshifts, $L^*$ \textit{galaxy evolution may have become decoupled from halo evolution}.  In this scenario, the depth of the host halo's potential well would limit the galaxy's average star formation rate, but further growth of the halo would not correlate with growth of the galaxy.\footnote{This is independent of pseudo-evolution \citep{Diemer13}, as major mergers contribute mass at all radial scales.}  If true, this would imply that the similarity between the recent time evolutions of galaxy star formation rates and average halo mass accretion rates \citep{Behroozi13} is a coincidence.  At the same time, it is clear that stellar mass alone cannot predict quenching in central galaxies: the weighted close pairs sample has an identical stellar mass distribution as the host sample, but a statistically significant difference in its quenched fraction (Fig.\ \ref{f:sdss_sfr}, top-left panel).\footnote{While it may be concerning that the close pairs sample has a $\sim$10\% higher satellite fraction than the host sample (\S \ref{s:validation}), making the isolation criteria stricter (for example) does not reduce the difference in quenched fractions (Fig.\ \ref{f:ssfr_var}, Appendix \ref{a:variation}).}

In the age-matching framework of \cite{Watson14}, star formation rates are correlated with halo assembly time (as measured either through concentration or mass accretion history); however, no previous simple measure of assembly time accounts for all the effects we observe.  Halo concentrations (used in \citealt{Watson14}) have the issue that major mergers have significantly \textit{increased} concentrations during close passages \citep{BehrooziMergers}.  As \cite{Watson14} place quenched galaxies in more concentrated haloes, this model cannot reproduce star formation enhancements in very close pairs ($<30$ kpc, Figs.\ \ref{f:sdss_sfr} and \ref{f:sdss_sf_q}).  Using halo ages (e.g., Fig.\ \ref{f:acc_rate}, bottom panels) would also not be able to reproduce strong star formation enhancements in very close pairs.

However, we note that SSFR enhancements for star-forming hosts with close pairs are hard to interpret because of uncertainty in how quenched and star-forming galaxies populate haloes.  While star-forming hosts undergoing major mergers could have up to a 20\% boost to their specific star formation rates, it is not clear whether this is directly attributable to the major merger, or whether the presence of a major merger correlates with differences in the longer-term mass accretion history (Fig.\ \ref{f:acc_rate}, bottom-right panel).  

\subsection{The Permanence of Quenching}

\label{s:perm_q}

The lack of increase in the star-forming fraction for close pairs within 30 kpc (Figs.\ \ref{f:sdss_sfr} and \ref{f:blanton}) suggests that close pairs cannot rejuvenate star formation prior to merging.  Nonetheless, Fig.\ \ref{f:cp_sdss} suggests that quenching may not be permanent.  That is, star formation in an incoming smaller galaxy is not quenched even at extremely close separations; when the two galaxies merge, the star-forming regions in the smaller galaxy would be transferred to the larger galaxy, contributing to a pseudo-rejuvenation of star formation in the merger remnant.

Recent stellar population analyses have shown that many early-type galaxies consist of a small fraction of young stars (e.g., \citealt{Trager00b,Schiavon07,Zhu10}, suggesting some low-level star-formation has occurred recently. Imaging observations in the UV/IR confirmed that star formation indeed occurs in many early-type galaxies (e.g., \citealt{Yi05,Kaviraj07,Salim10}. The details,  such as the extent and level of star formation, are not yet well determined, and the responsible mechanisms, e.g., whether or not major mergers are important, are not fully understood. The SAURON IFU (Integral Field Unit) observations \citep{Shapiro10} of 48 early-type galaxies suggest that this low-level star formation could be either due to a large molecular gas content brought in by (mostly minor) mergers \citep{Kaviraj14}, or rejuvenation within the previously quenched systems. Ongoing IFU surveys, such as the MANGA \citep{Bundy15} and SAMI \citep{Croom12} surveys, will help constrain the frequency of these scenarios with coverage of over 10,000 nearby galaxies.

\begin{table*}
\caption{Recent findings for star formation enhancement in close pairs.}
\begin{tabular}{lccccccc}
\hline
Reference & Host Selection & Pair Cut &  Redshifts & Boost at $<30$ kpc & at 30-200 kpc & Potential Confounds\\
\hline
\cite{Lambas03} & All 2dF 100K & $\Delta m_b < 0.75$ & $z \le 0.1$ & 1.2-1.4$\times$ & N/A & CHES$_b$\\ %http://arxiv.org/pdf/astro-ph/0212222v2.pdf
\cite{Nikolic04} & All DR1 & $\Delta z < 2$  & $0.03<z<0.1$ & 2-3$\times$ &Yes & CHES$_b$\\ %http://lanl.arxiv.org/pdf/astro-ph/0407289v2.pdf
\cite{Woods06} & CfA2, Zw $< 15.5$ & $\Delta R < 2$ & $0.008 < z < 0.055$ & Yes & N/A & CHS$_b$\\ %http://lanl.arxiv.org/pdf/astro-ph/0603175v2.pdf
\cite{Barton07} & 2dF, B$_\mathrm{j}<-19$ & Any Host & $0.010 < z < 0.088$ & Yes & N/A & CHS$_b$\\ %http://lanl.arxiv.org/pdf/0708.2912v1.pdf
\cite{Lin07}$^a$ & $10^{10} < M_*/\Msun < 10^{11}$ & Any Host  & $0.1 < z < 1.1$ & 2-4$\times$ & Yes & H\\ %http://lanl.arxiv.org/pdf/astro-ph/0607272v2.pdf
\cite{Li08} & SF, DR4 & $\Delta r < 1.4$ & $0.01 < z < 0.3$ & 1.7-2.4$\times$ & No &CS$_o$\\ % http://lanl.arxiv.org/pdf/0711.3792v2.pdf
\cite{Ellison08} & SF, DR4 & $\Delta M_* < $ 0.3 dex & $0.01 < z < 0.16$ & 1.1-1.7$\times$ & N/A & CHS$_o$\\ % http://lanl.arxiv.org/pdf/0803.0161v1.pdf
\cite{Rogers09} & E, DR6 & Any Host & $0 < z < 0.15$ & Yes & N/A & CHES$_a$\\ %http://lanl.arxiv.org/pdf/0905.3386v2.pdf
\cite{Robaina09} & $M^* > 10^{10}\Msun$ & $\Delta M_* < 0.6$ dex & $0.4<z<0.8$ & 2.0-2.5$\times$ & No & CHES$_b$ \\ %http://lanl.arxiv.org/pdf/0907.3728v1.pdf
\cite{Woods10} & All SHELS & $\Delta R < 1.75$  & $0.08 < z < 0.38$ & 1.4-2.1$\times$ & N/A & CHS$_b$\\ %http://lanl.arxiv.org/pdf/1002.0386v1.pdf
\cite{Wong11} & SF, PRIMUS, $i<22.5$ & Any Host & $0.25 < z < 0.75$ & 1.3$\times$ & N/A & CHS$_o$ \\ %http://lanl.arxiv.org/pdf/1012.1324v1.pdf
\cite{Scudder12} & SF, DR7 & $\Delta M_* < $ 0.5 dex & $0.02 < z < 0.15$ & 1.4-2$\times$ & N/A & CHS$_o$\\ %http://lanl.arxiv.org/pdf/1207.4791v1.pdf
\cite{Xu12b} & SF, $10^{10.4} < M_*/\Msun < 10^{11}$ & $\Delta M_* < 0.4$ dex & $0.2 < z < 1$ & 1-1.2$\times$ & N/A & S$_o$\\ %http://lanl.arxiv.org/pdf/1210.2362v1.pdf
\cite{Ellison13} & All DR4 & $\Delta M_* < $ 0.6 dex & $0.01 \le z \le 0.2$ & 2.2-2.8$\times$ (SF) & N/A & CS$_b$\\ %http://arxiv.org/pdf/1308.3707v1.pdf
\cite{Patton13} & SF, DR7 & $\Delta M_* < $ 1 dex & $0.02 < z < 0.2$ & 1.5-2.2$\times$ & Yes & CS$_o$\\ %http://lanl.arxiv.org/pdf/1305.1595v2.pdf
\cite{Robotham13} & GAMA, $M^* > 10^{10}\Msun$ & $\Delta M_* < 0.3$ dex  & $0.01 < z < 0.089$ & 1-5$\times$ & N/A & HE\\%http://lanl.arxiv.org/pdf/1301.7129v1.pdf 
\cite{Scott14} & All DR7+GALEX & Any Host & Median $z\sim 0.07$ & 2.4-2.5$\times$ & N/A & CHS$_b$ \\ %http://lanl.arxiv.org/pdf/1310.5148v1.pdf
This Work & DR7, $10^{10} < M_*/\Msun < 10^{10.5}$ & $\Delta M_* < 0.5$ dex & $0.01 < z < 0.057$ & 1.5-2$\times$ (SF) & 1.1$\times$ (SF)\\
\hline
\end{tabular}
\parbox{2.15\columnwidth}{\textbf{Notes.} When a reference considered multiple pair criteria, the result listed is for the most major mass ratio considered.  Abbreviations: DR\# = SDSS (Sloan Digital Sky Survey) Data Release \# Spectroscopic Galaxy Sample, CfA2 = Center for Astronomy Redshift Survey 2, 2dF = 2-degree Field Galaxy Redshift Survey, 2dF 100K = 2dF 100K public release, SHELS = Smithsonian Hectospec Lensing Survey, PRIMUS = PRIsm MUlti-object Survey, GAMA = Galaxy And Mass Assembly, GALEX = Galaxy Evolution Explorer, SF = star-forming galaxies only, E = elliptical galaxies only.  Confounds: C = galactic Conformity (due to luminosity delta in pair selection, luminosity threshold for sample, or star-forming paired galaxy bias), H = Halo mass (due to wide selection of host galaxy properties and/or to allowing any host galaxy to be a paired galaxy), E = Environmental effects (no isolation or environment matching for host galaxies; however, according to \citealt{Li08}, only minor environmental effects apply to SF-only selections), S$_b$ = bias towards Star-forming galaxies (via spectroscopic requirements, sample luminosity threshold, and/or incomplete volume corrections), S$_a$ = bias against Star-forming galaxies, S$_o$ = only Star-forming galaxies.  $^{a}$: \protect\cite{Lin07} uses redshift-dependent mass thresholds in the DEEP2 Galaxy Redshift Survey.}
\label{t:prev}
\end{table*}

\subsection{Comparison with Semi-Analytic Model Predictions}

\label{s:semi_a}

We have also tested mock catalogs from two semi-analytical models \citep{Croton06,Lu12,YuLu14b}, which have both been run on the \textit{Bolshoi} simulation (Fig.\ \ref{f:ssfr_sam}).  These models have been tuned to match $z<0.2$ stellar mass functions \citep{YuLu14}, and indeed we find that statistics such as the fraction of galaxies with close pairs (6.7\% and 7.2\% for the Croton and Lu semi-analytical models, respectively) match the SDSS result (6.7\%) extremely well.  These models have not been tuned to match star formation rates, which account for the different quenched fractions between Figs.\ \ref{f:sdss_sfr} and \ref{f:ssfr_sam}.  This remains true after simulating observational uncertainties in recovering SSFRs; we do so by adding 0.2 dex of log-normal scatter to star-forming galaxies' SSFRs and drawing quenched galaxies' SSFRs from a log-normal distribution with 0.35 dex scatter centered at 10$^{-11.8}$ yr$^{-1}$ (matching the quenched galaxy SSFR distribution in Fig.\ \ref{f:sdss_sfr}).

Despite these absolute differences, the \textit{relative} impact of a close pair on the quenched fraction in the two models remains interesting.  Both models exhibit 20--25\% larger quenched fractions for galaxies in close pairs as compared to the host sample (Fig.\ \ref{f:ssfr_sam}).  Encouragingly, this implies that our selection approach can recover differences (when they exist) in star formation activity during major mergers.  On the other hand, this implies that both models overestimate quenching efficiency compared to real galaxies.

Using semi-analytic models allows disentangling the contributing effects (halo mass biases, increased satellite fractions, and major mergers) for increased quenched fractions.  The effects of halo mass bias can be removed by creating mass-matched samples of close pair galaxies (i.e., all close pair galaxies with host halo and stellar masses within 0.1 dex of the target galaxy in the host sample).  Similarly, the effects of major mergers can be extracted by considering only 3D major mergers (i.e., where the close pair is within the 3D physical halo radius of the host galaxy); for clarity of interpretation, we also mass-match this sample to the host sample galaxies' halo and stellar mass distribution.  The resulting SSFR distributions for these mass-matched samples are shown in Fig.\ \ref{f:ssfr_sam} for both semi-analytic models.

In both semi-analytic models, the difference between mass-matched and non-mass-matched close pairs samples suggest that half of the increase in the quenched fraction for close pairs is due to larger host halo masses.  The remaining increase is due to larger fractions of ``host'' galaxies being satellites in the close pairs sample.  In the \cite{Lu12} model, quenching is implemented as a cooling shutoff which is a function of the host halo mass only, so the entire difference between the host sample and the mass-matched 3D major mergers sample arises from the much larger satellite fraction in the mass-matched major mergers sample (50\% versus 4\% in the overall host sample).\footnote{This satellite fraction is much larger than in the abundance-matched mock catalogs, which suggest that the satellite fraction in the close pairs sample should be 13.3\% (\S\ref{s:validation}), and that the satellite fraction in the mass-matched 3D major mergers sample should be 19.0\%.}  In the \cite{Croton06} model, higher satellite fractions also yield larger quenched fractions for the mass-matched major mergers sample.  However, in this model, galaxies are quenched more indirectly, using black hole feedback.  Since cooling is not shut off, the presence of a major merger increases the amount of gas available to cool onto the galaxy, increasing SSFRs in star-forming host galaxies with true (3D) major halo mergers by a factor of 2 compared to star-forming galaxies in the host sample.  In the original close pairs sample, the \cite{Croton06} model predicts SSFRs enhanced by 34\% for star-forming galaxies compared to the host sample, which is significantly larger than our findings in the SDSS (enhancements of $\sim10\%$; \S \ref{s:sf_v_q}).

\subsection{Comparison with Previous Work}

\label{s:prev}

Table \ref{t:prev} summarises several recent literature results on star formation enhancements in close pairs.  The majority of previous results preferentially include star-forming galaxies in their samples.  There are good reasons for doing so, especially for improved statistics (due to star-forming galaxies being brighter at fixed mass).  While conformity and halo mass biases may be present in these past works, they appear to have a modest impact on measured enhancements for extremely close pairs.  Compared to our results, \cite{Li08} found $\sim 20\%$ larger enhancements using a luminosity delta for extremely close pairs (resulting in a conformity bias; see \S \ref{s:method}) while using similar stellar masses and star formation rates from earlier MPA-JHU catalogs.  That said, biases on the $20\%$ level can be important when studying star formation enhancements at larger separations.

Among these past results, there is general agreement that extremely close pairs (separated by $<30$ kpc) show star formation enhanced by 50-150\%, although the reported values range from 20\% to 400\%.  Less agreement exists about whether enhancements persist at larger radii, but this may be due to methodological limitations.  Papers which report no enhancement \citep{Li08,Robaina09} use photometric identification of pairs; however, within 150 kpc, the ratio of true satellites to background contaminants is $\sim 15\%$ \citep{Liu10}.  Hence, the weak enhancements (5-20\%) reported in the papers that use spectroscopic pair selection \citep{Nikolic04,Lin07,Patton13} would be reduced to percent-level effects in the photometric selections.  While not considered in this paper, past results suggest that star formation enhancements are lower for minor mergers \citep{Woods06,Woods10,Scudder12} and for more massive galaxies \citep{Li08,Robotham13}.

\cite{Robotham13} is the only paper in Table \ref{t:prev} that also creates a stellar mass--limited local pair sample including both star-forming and non-star-forming host galaxies.  Unfortunately, due to limited statistics, their constraints on star formation enhancements in extremely close pairs are weak (0-400\%).  With modestly better statistics, our results constrain this to be between 0\% and 100\%; with the still better statistics in Fig.\ \ref{f:blanton}, this becomes 50-100\%.  \cite{Robotham13} report halo mass (FoF group-based) differences between their close pairs and control samples, and also do not isolate their host galaxies; however, both effects would be expected to reduce any enhancements, rather than increase them.

\section{Conclusions}
\label{s:conclusions}

We have presented a method for selecting central $L^*$ SDSS galaxies whose host haloes are preferentially undergoing major mergers.  Tests with mock catalogs (\S \ref{s:validation}) suggest that our selection can identify host galaxies with near-identical host halo masses, stellar masses, and environments, but with average halo mass accretion rates higher by 0.3 dex over the past 5 Gyr (\S \ref{s:dm_hosts}).  Additionally, 50\% of galaxies selected in this way are undergoing major halo mergers, as compared to 3\% of isolated $L^*$ galaxies (\S \ref{s:dm_hosts}).\\

Our findings include:
\begin{enumerate}
\item The subsample with 50\% major mergers has a 6\% lower star-forming fraction than the whole isolated host galaxy sample, implying at most a 12\% effect in a pure major mergers sample.  (\S \ref{s:all_cp}).
\item This latter finding limits how gravitational heating or gas reservoir transfers in mergers can affect central galaxy star formation rates (\S \ref{s:hma}).
\item Consistent with previous research, star-forming host galaxies show 70\% larger SSFRs when an extremely close pair is present ($<$30 kpc), but only $\sim$ 10\% larger SSFRs when the paired galaxy is between 30--200 kpc in projected separation (\S \ref{s:sf_v_q}).
\item Extremely close pairs ($<$30 kpc) do not appear to rejuvenate star-formation for quenched host galaxies (\S \ref{s:all_cp}).
\item Quenching for central $L^*$ galaxies does not depend on their stellar mass alone (\S \ref{s:hma}).
\item Previous halo age-based or concentration-based methods for matching galaxy star formation rates to haloes have a difficult time reproducing all SFR enhancements found for SDSS galaxies (\S \ref{s:hma}).
\item Current semi-analytical models over-predict the impact of major mergers on galaxy quenching (\S \ref{s:semi_a}).
\end{enumerate}

\section*{Acknowledgements}

We thank Eric Bell, Fabio Governato, Brice M\'enard, Joel Primack, and Alex Szalay for insightful discussions during the preparation of this paper.  Support for PSB was provided by a Giacconi Fellowship from the Space Telescope Science Institute, which is operated by the Association of Universities for Research in Astronomy, Incorporated, under NASA contract NAS5-26555.  GZ acknowledges support provided by NASA through Hubble Fellowship grant \#HST-HF2-51351.001-A awarded by the Space Telescope Science Institute.  DC acknowledges receipt of a QEII Fellowship by the Australian Research Council.  The \cite{Croton06} model used in this work was generated using Swinburne University's Theoretical Astrophysical Observatory (TAO; \citealt{TAO}). TAO is freely accessible at https://tao.asvo.org.au/.  The \textit{Bolshoi} simulation was carried out by A. Klypin, J. Primack and S. Gottloeber at the NASA Ames Research Centre. The simulation and data products can be found at http://astronomy.nmsu.edu/aklypin/Bolshoi/.  We are also grateful to Matt Becker for running the higher-resolution \textit{Lb125} simulation used for convergence tests.  PSB, APH, and DFW thank the Aspen Center for Physics and the NSF (Grant \#1066293) for hospitality during the editing of this paper.  PSB, APH, and DFW are also grateful for the Braunschweiger Institute Summer Science Programme, supported by the foundations of Dmitry Ledov, Sam Steem, Bennett Gettu, Daccia Paugh, Alby Bach, and Alan Schaefer.  This research has made use of NASA's Astrophysics Data System.  Funding for SDSS-III has been provided by the Alfred P. Sloan Foundation, the Participating Institutions, the National Science Foundation, and the U.S. Department of Energy Office of Science. The SDSS-III web site is http://www.sdss3.org/.  SDSS-III is managed by the Astrophysical Research Consortium for the Participating Institutions of the SDSS-III Collaboration including the University of Arizona, the Brazilian Participation Group, Brookhaven National Laboratory, Carnegie Mellon University, University of Florida, the French Participation Group, the German Participation Group, Harvard University, the Instituto de Astrofisica de Canarias, the Michigan State/Notre Dame/JINA Participation Group, Johns Hopkins University, Lawrence Berkeley National Laboratory, Max Planck Institute for Astrophysics, Max Planck Institute for Extraterrestrial Physics, New Mexico State University, New York University, Ohio State University, Pennsylvania State University, University of Portsmouth, Princeton University, the Spanish Participation Group, University of Tokyo, University of Utah, Vanderbilt University, University of Virginia, University of Washington, and Yale University.  

{\footnotesize
\bibliography{master_bib}
}

\appendix

\section{Variation in Selection Parameters}
\label{a:variation}

\begin{figure*}
\vspace{-3ex}
\plotsmallgrace{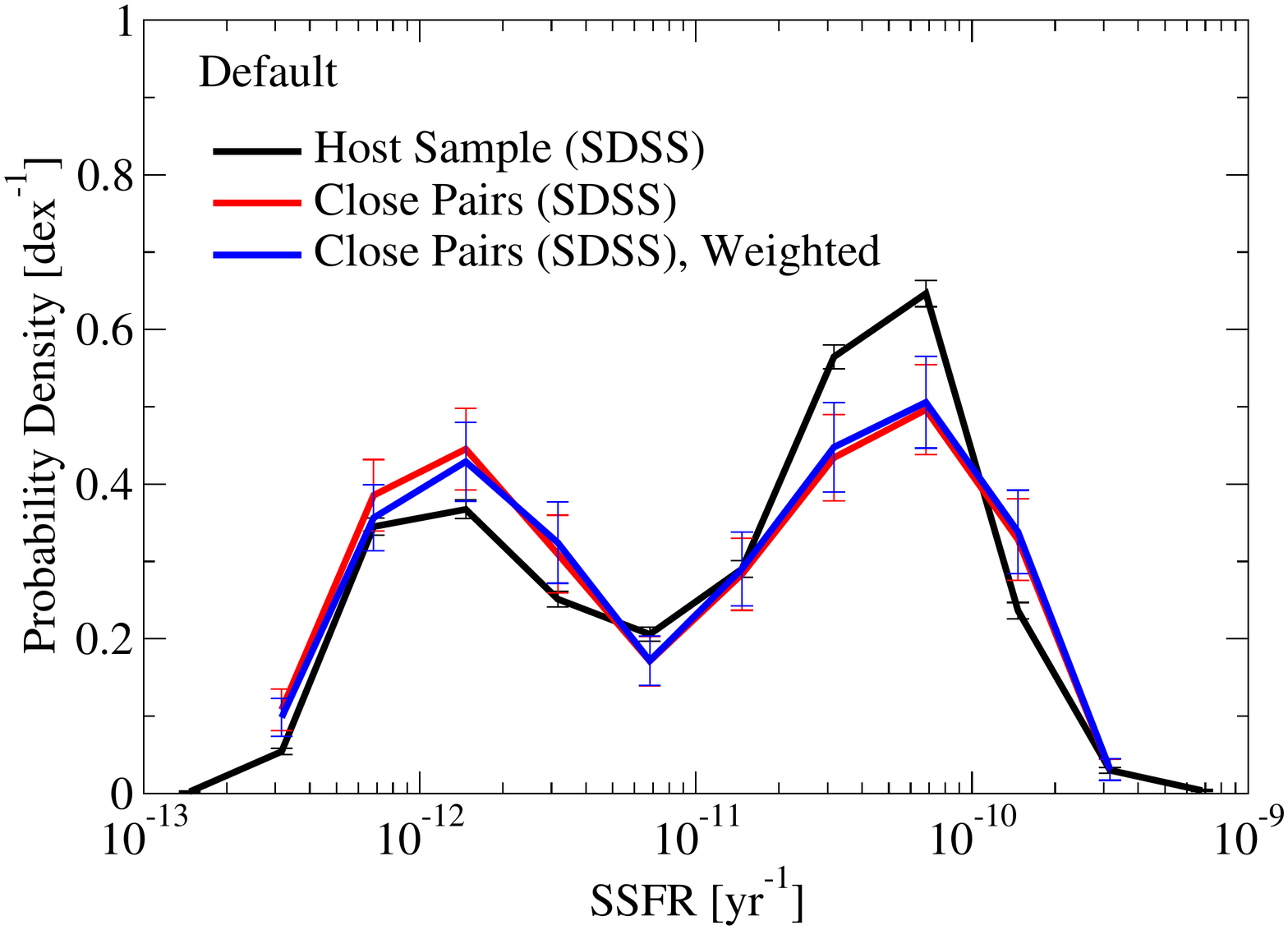}\plotsmallgrace{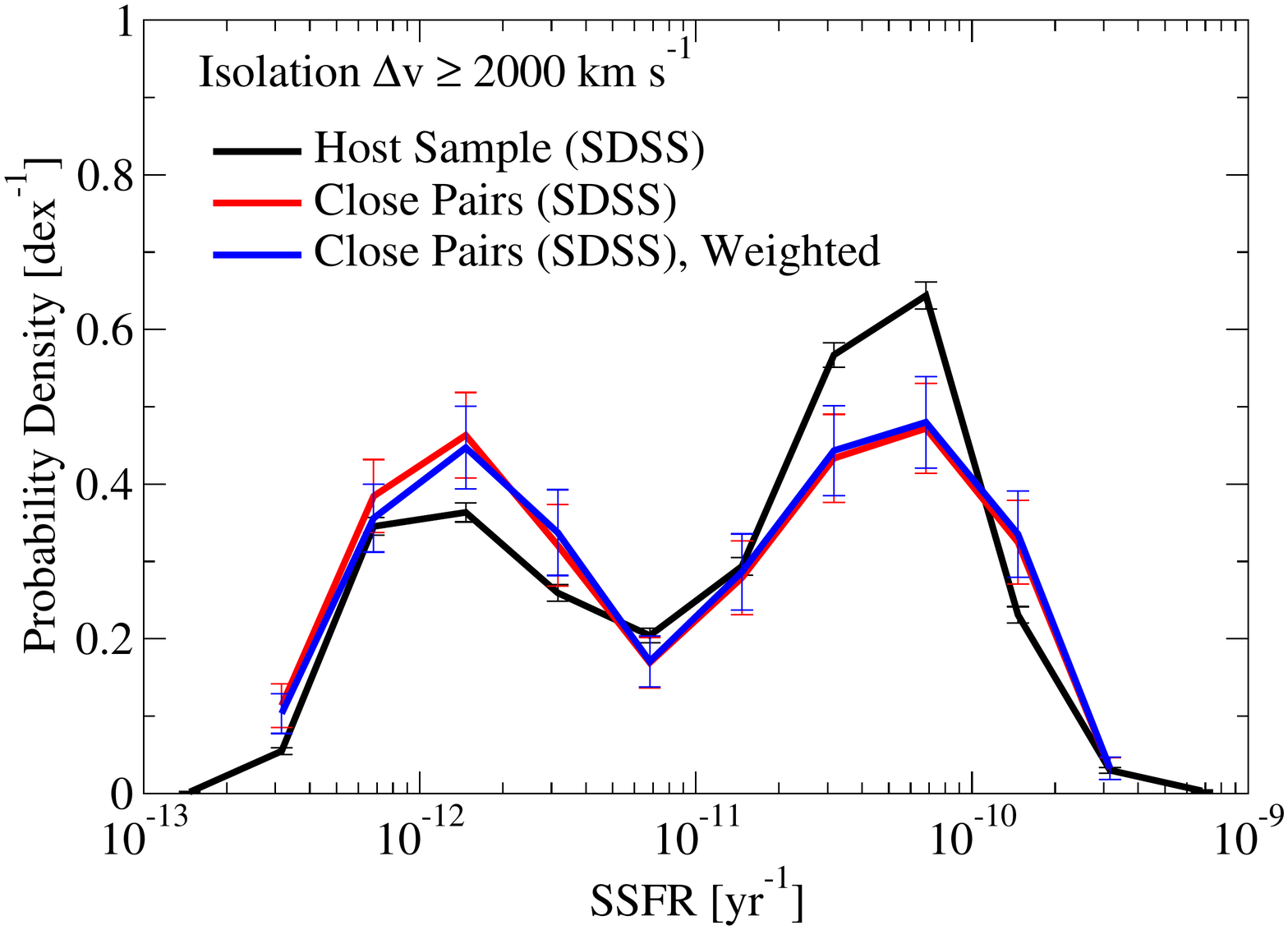}\plotsmallgrace{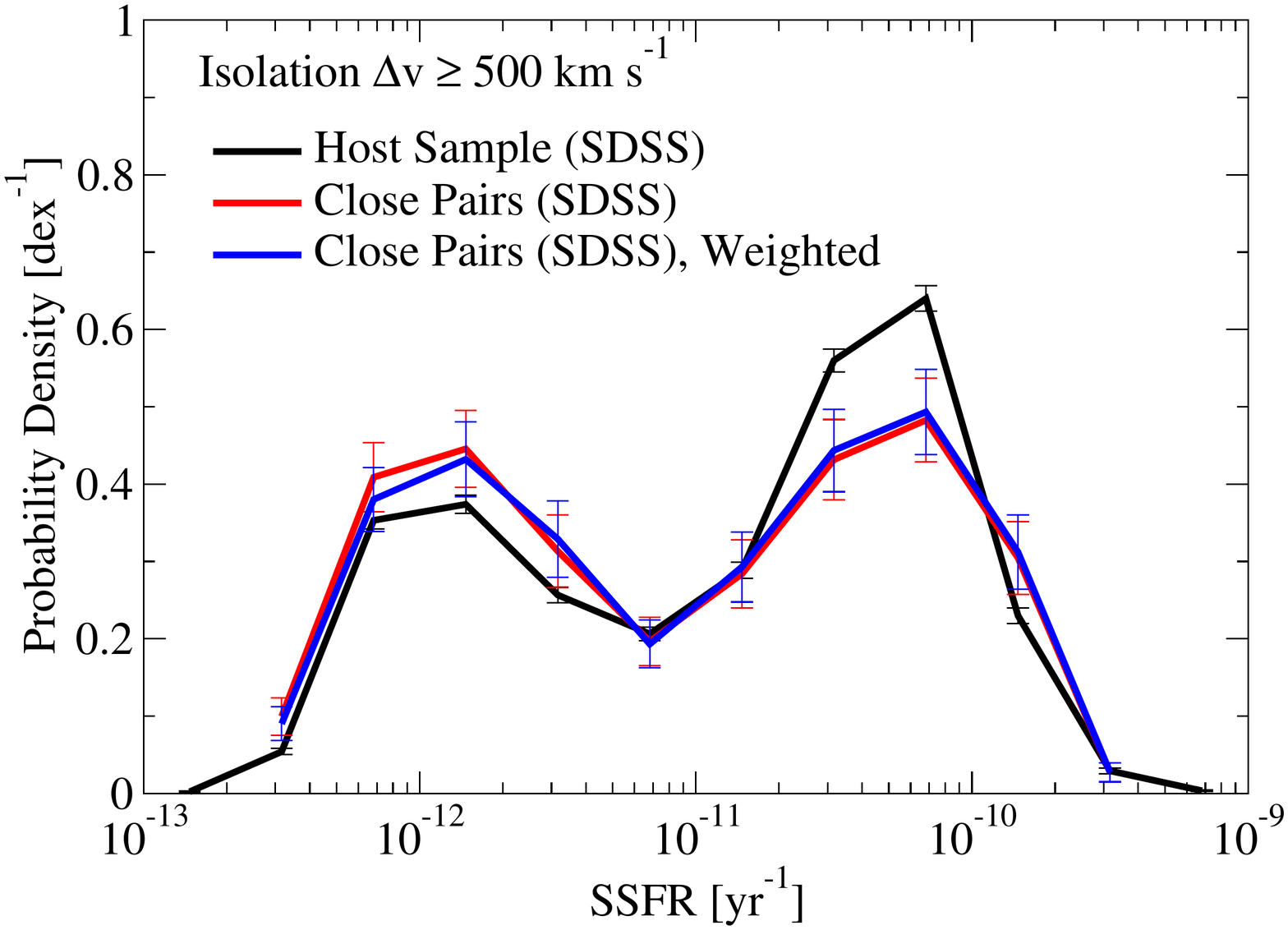}\\[-3ex]
\plotsmallgrace{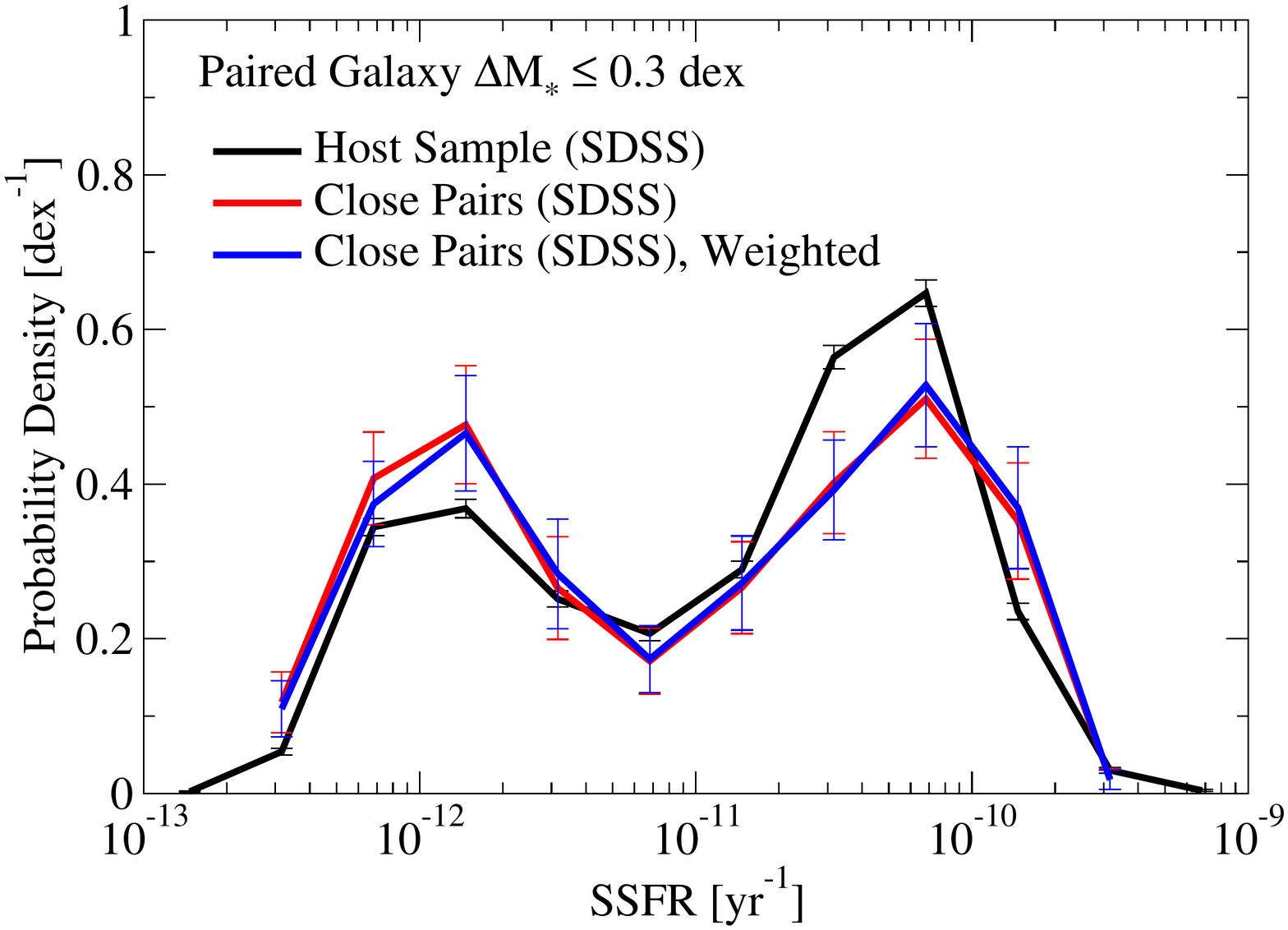}\plotsmallgrace{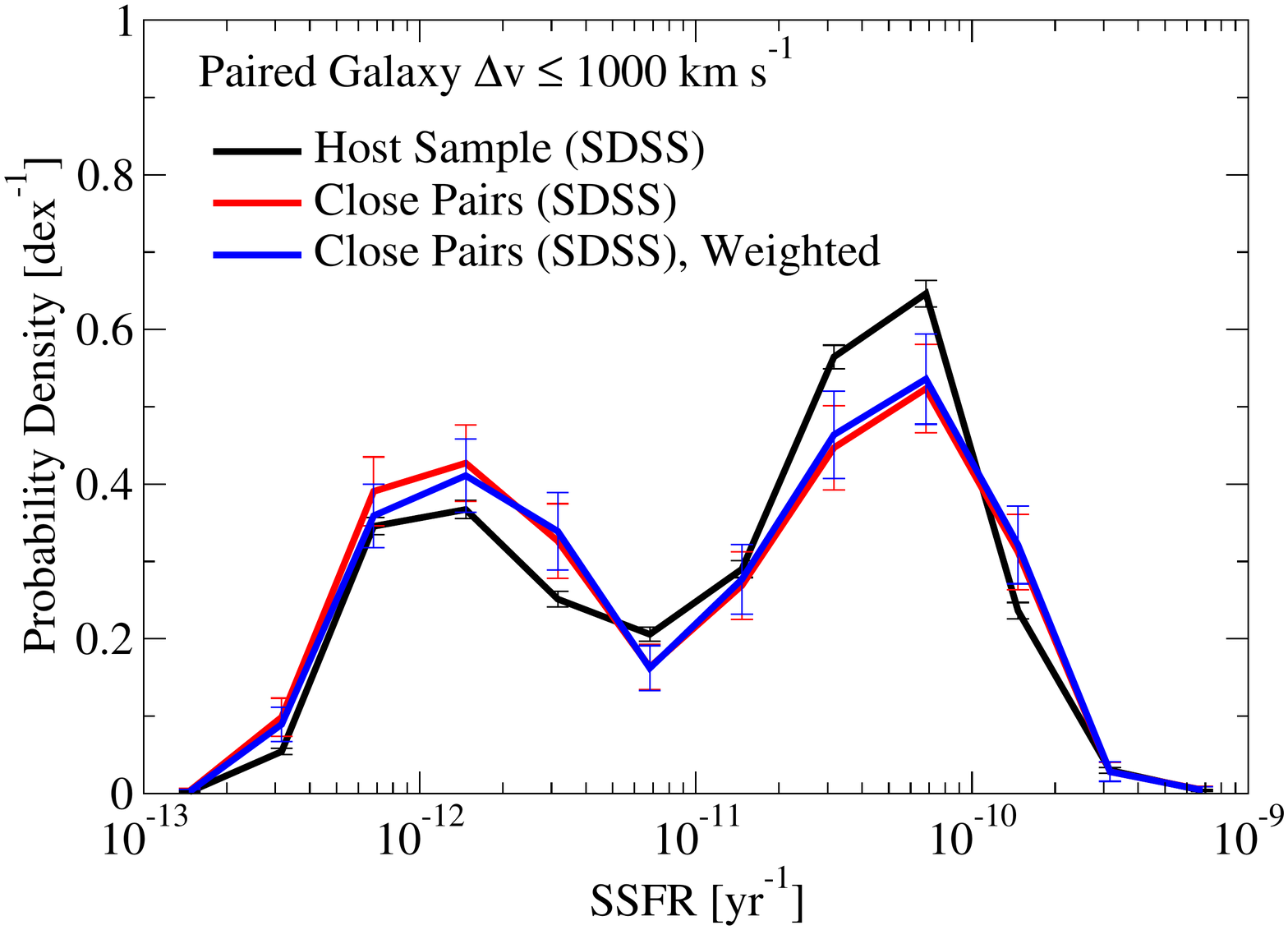}\plotsmallgrace{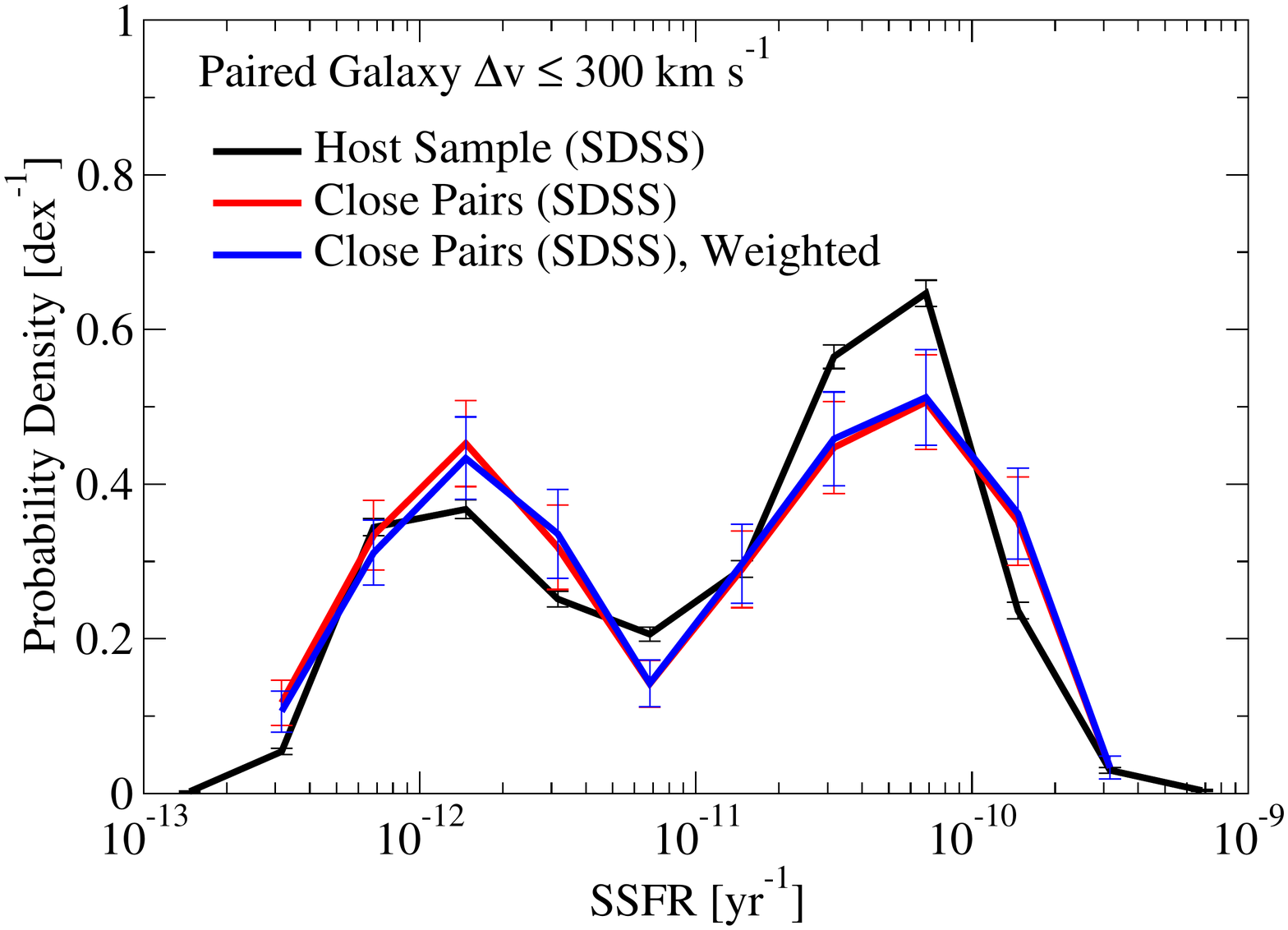}\\[-3ex]
\hspace{0.66\columnwidth}\plotsmallgrace{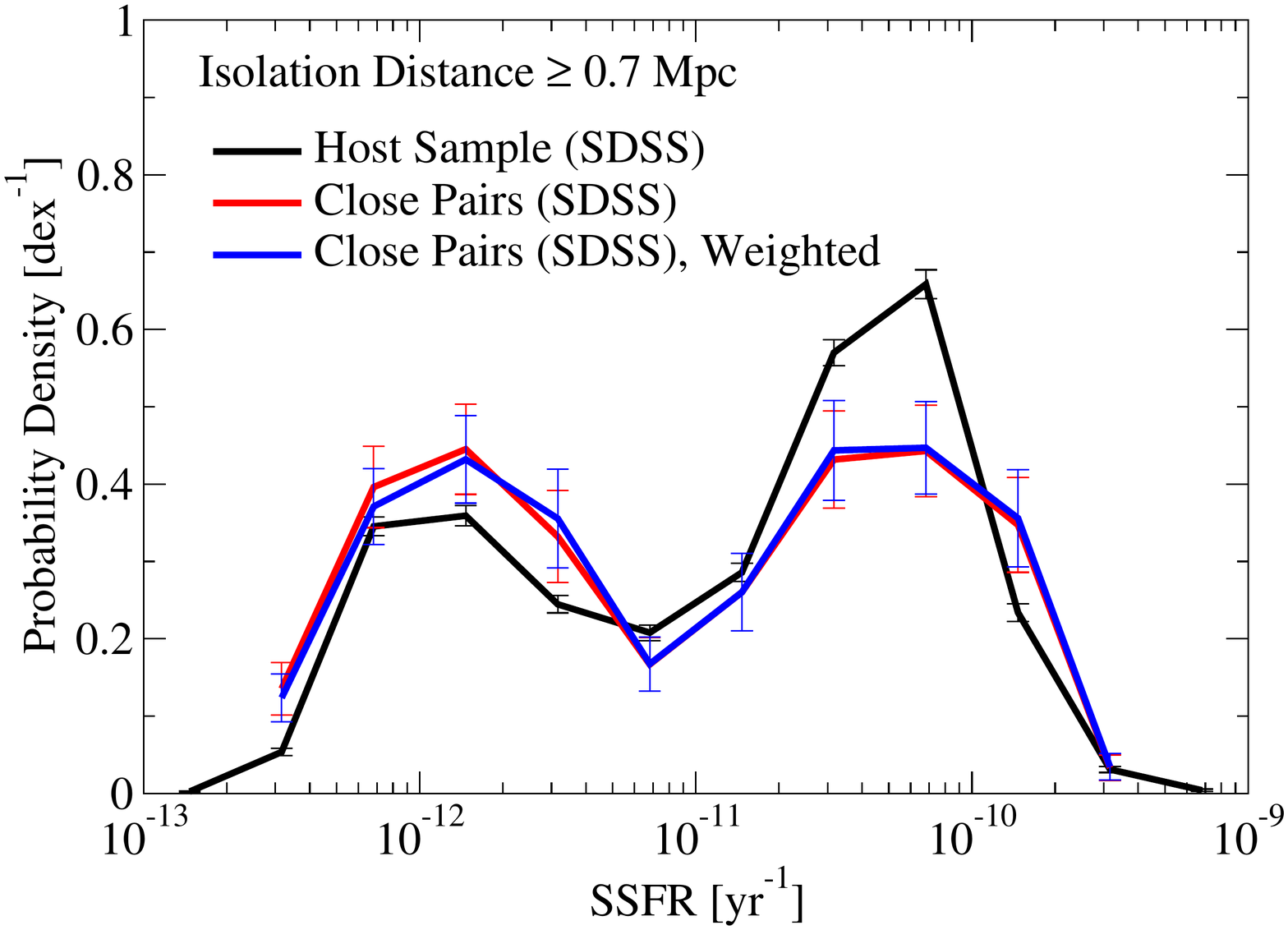}\plotsmallgrace{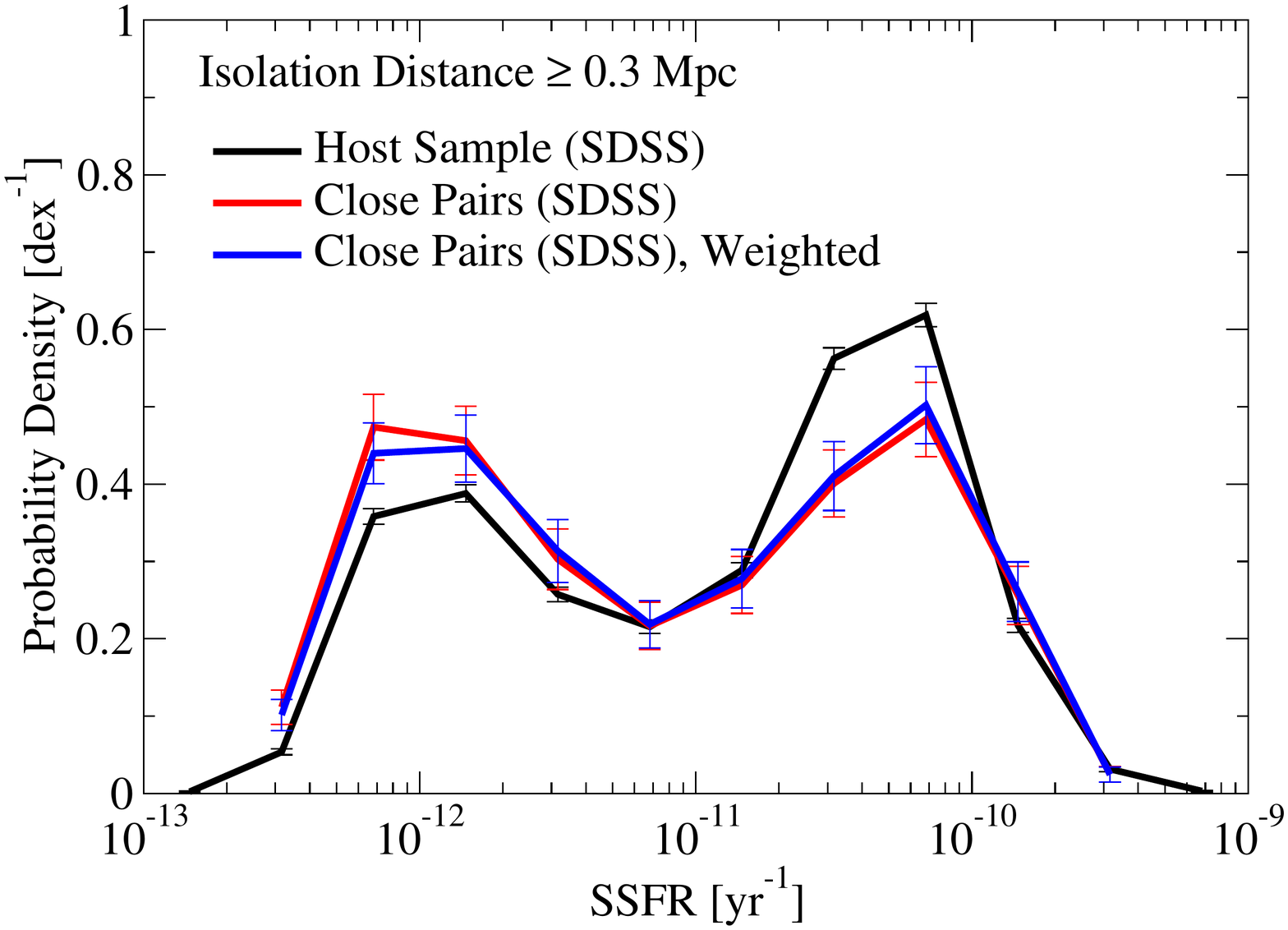}
\caption{Specific star formation rate distributions for all host galaxies and those with close pairs when selection criteria are varied.  For reference, the default selection criteria for host galaxies are: no larger galaxy within 1000 km s$^{-1}$ in redshift (``Isolation $\Delta$v'') or 0.5 Mpc in projected distance (``Isolation Distance''); for a smaller galaxy to be called a paired galaxy, it must be within 500 km s$^{-1}$ in redshift (``Paired Galaxy $\Delta$v'') and 0.5 dex in stellar mass (``Paired Galaxy $\Delta M_*$'').  In all panels, close pairs are defined as a paired galaxy within 200 kpc in projected distance from a host galaxy.  For alternate definitions of close pair distances, see Fig.\ \ref{f:sf_frac_var}.}
\label{f:ssfr_var}
\end{figure*}

\begin{figure*}
\vspace{-3ex}
\plotsmallgrace{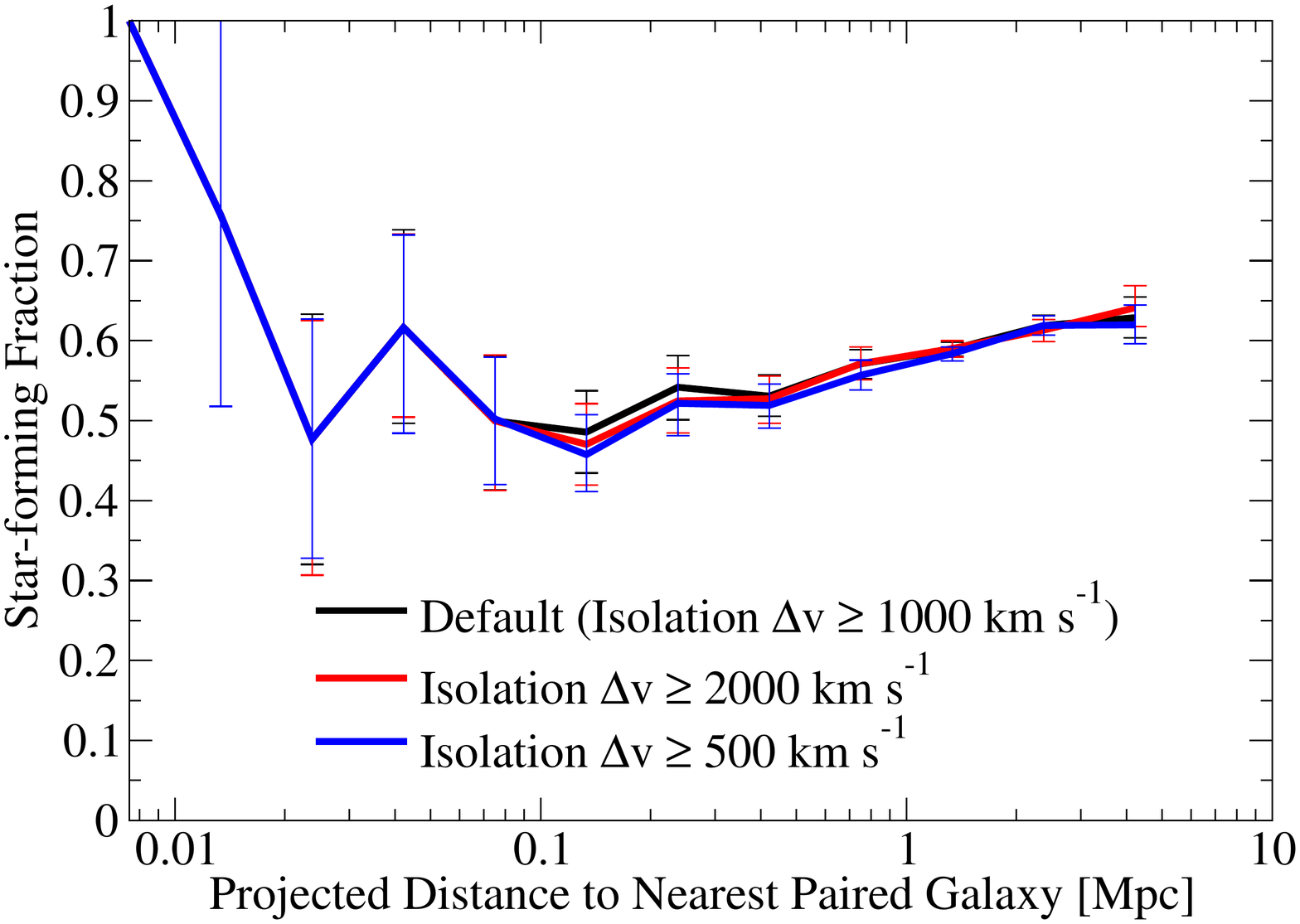}\plotsmallgrace{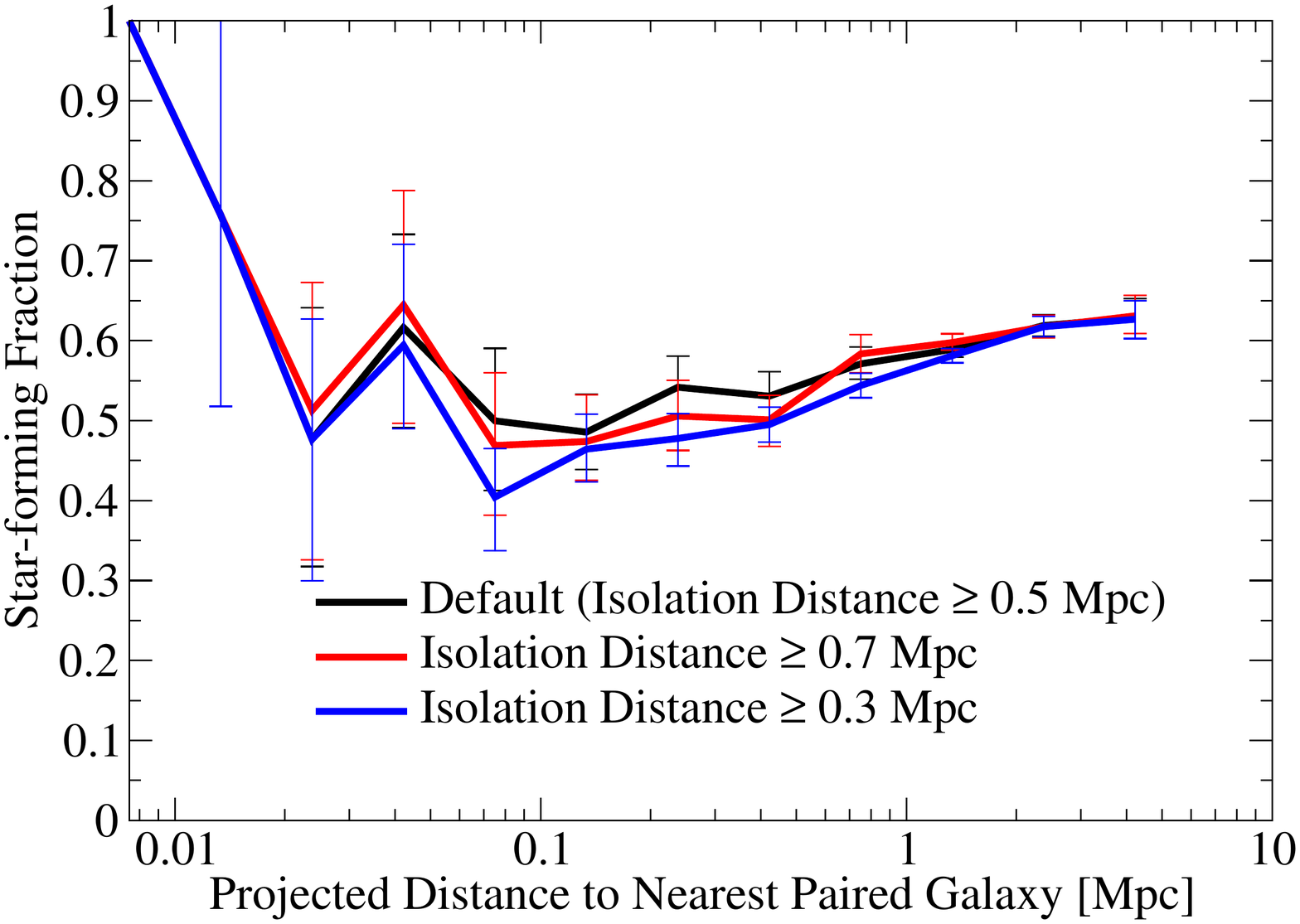}\\[-3ex]
\plotsmallgrace{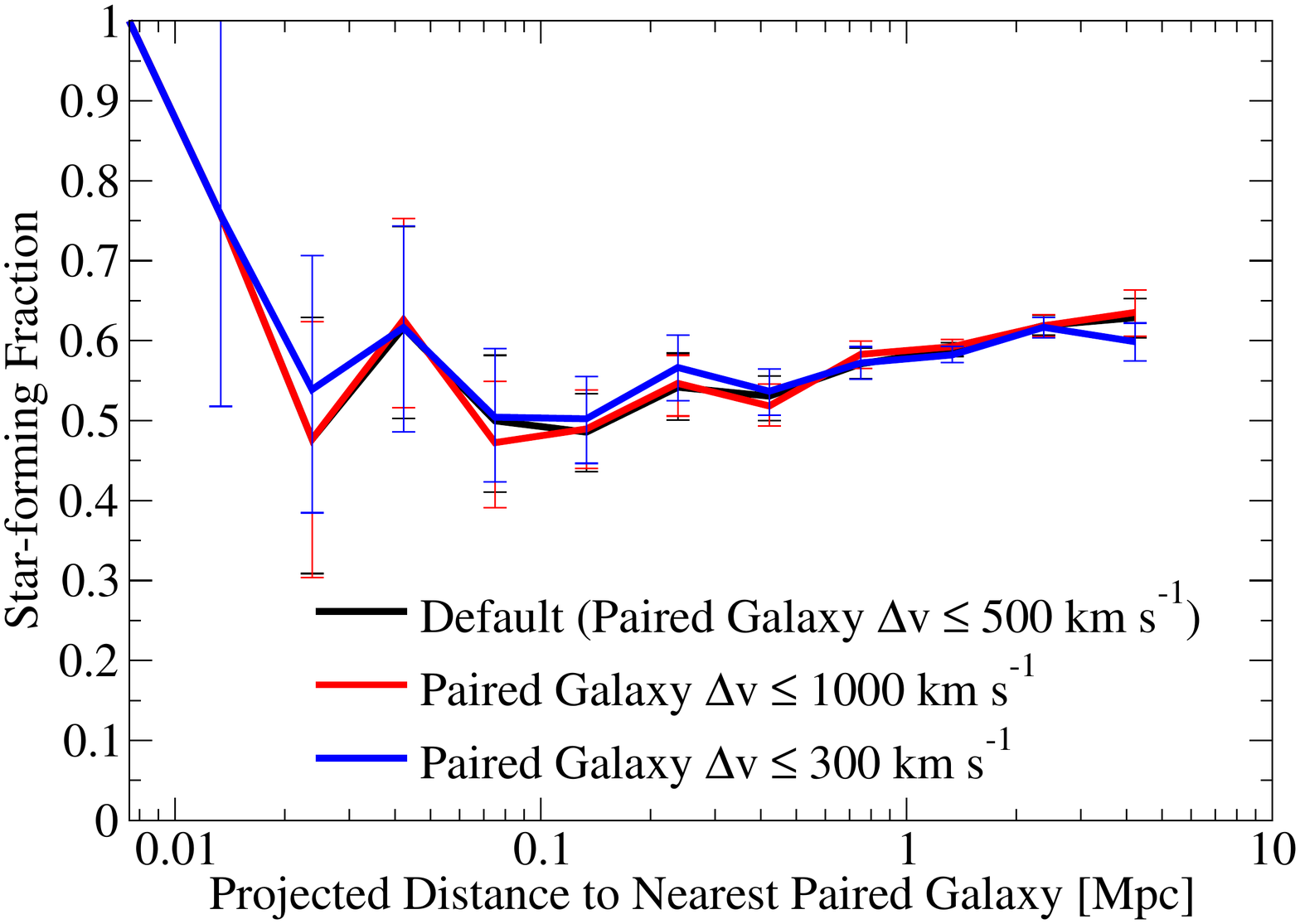}\plotsmallgrace{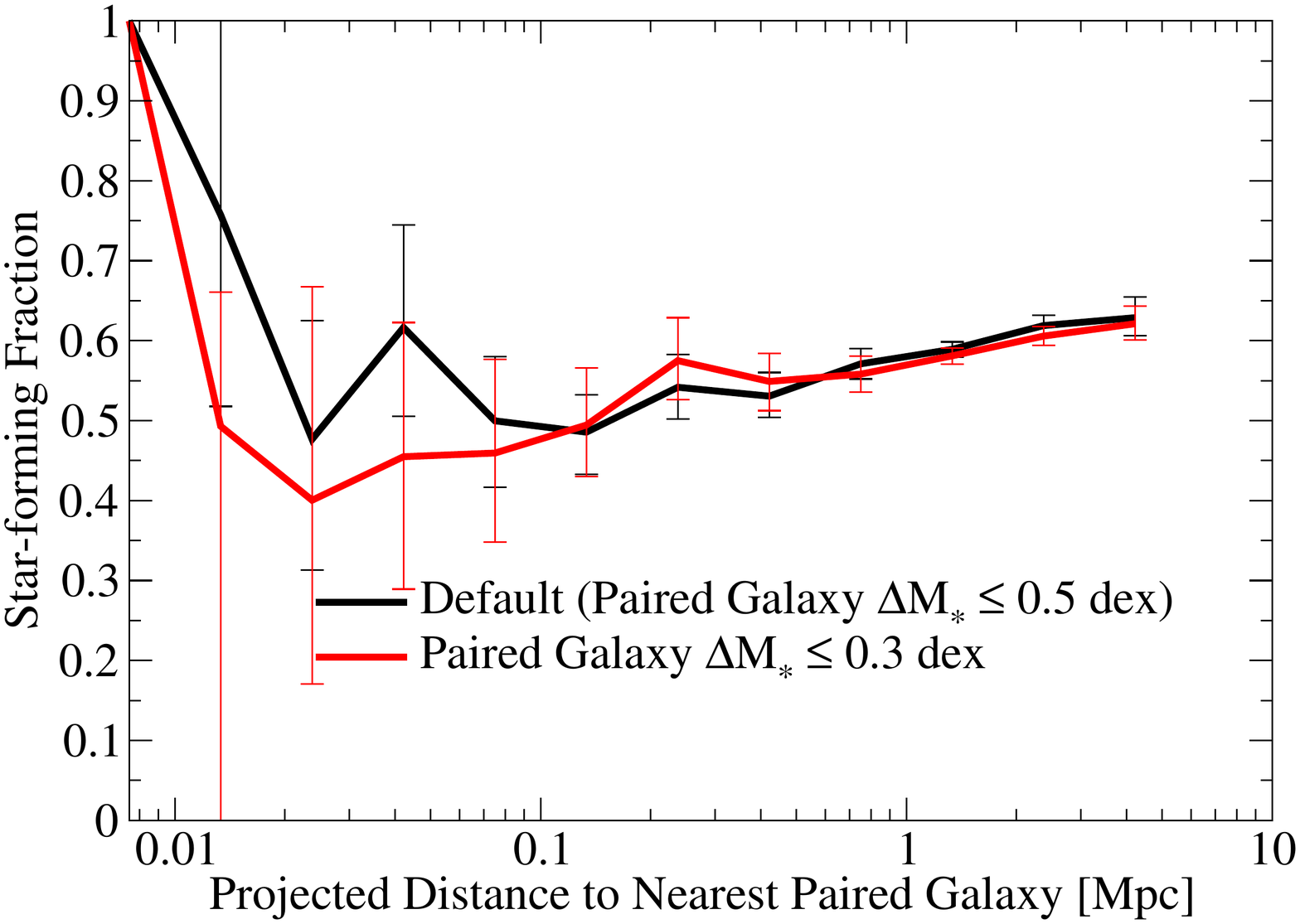}\\[-3ex]
\caption{Star-forming fraction for host galaxies as a function of distance to the nearest paired galaxy when selection criteria are varied, as in Fig.\ \ref{f:ssfr_var}.}
\label{f:sf_frac_var}
\end{figure*}

\begin{figure*}
\vspace{-3ex}
\plotgrace{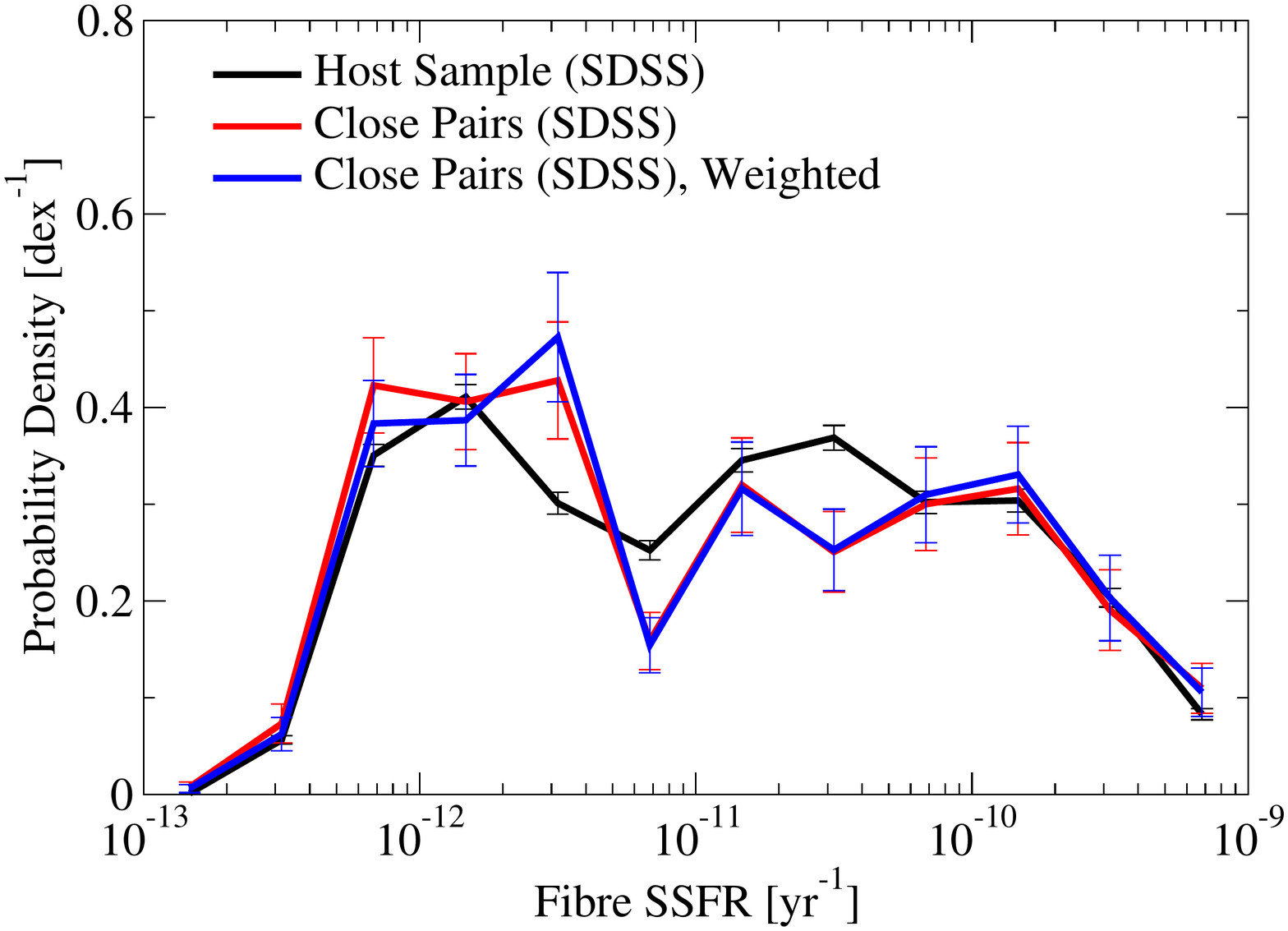}\plotgrace{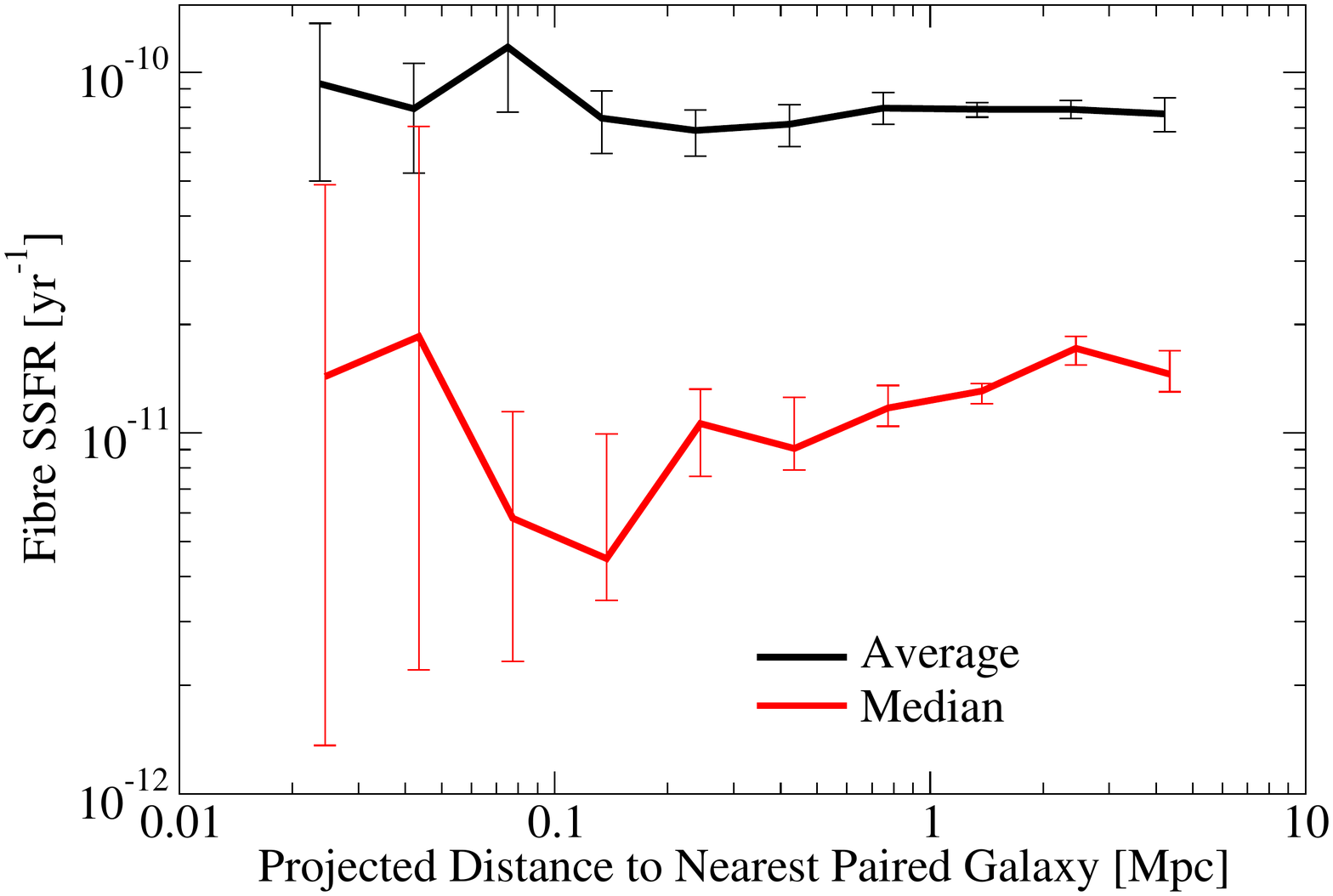}\\[-3ex]
\caption{Left panel: distribution of fibre SSFRs for all host galaxies and those in close pairs.  Right panel: fibre SSFRs as a function of distance to the nearest paired galaxy.  The diameter of an SDSS fibre is 3", corresponding to 0.43 kpc at $z=0.01$ and 2.4 kpc at $z=0.057$.  For our stellar mass range, typical host galaxy half-light radii are 1--2.5 kpc \citep{Kravtsov13}.}
\label{f:fibre_ssfrs}
\end{figure*}

\begin{figure*}
\vspace{-3ex}
\plotgrace{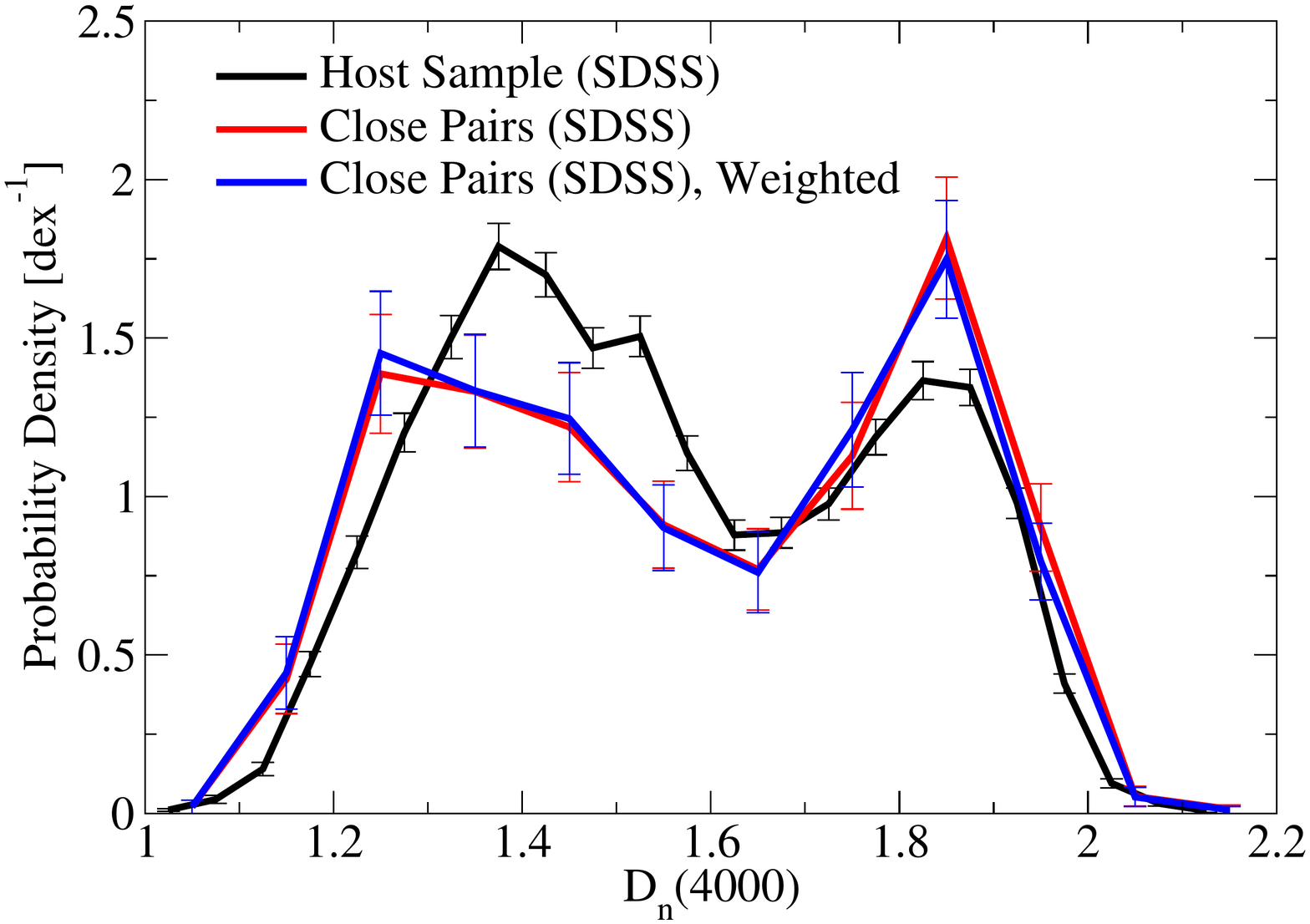}\plotgrace{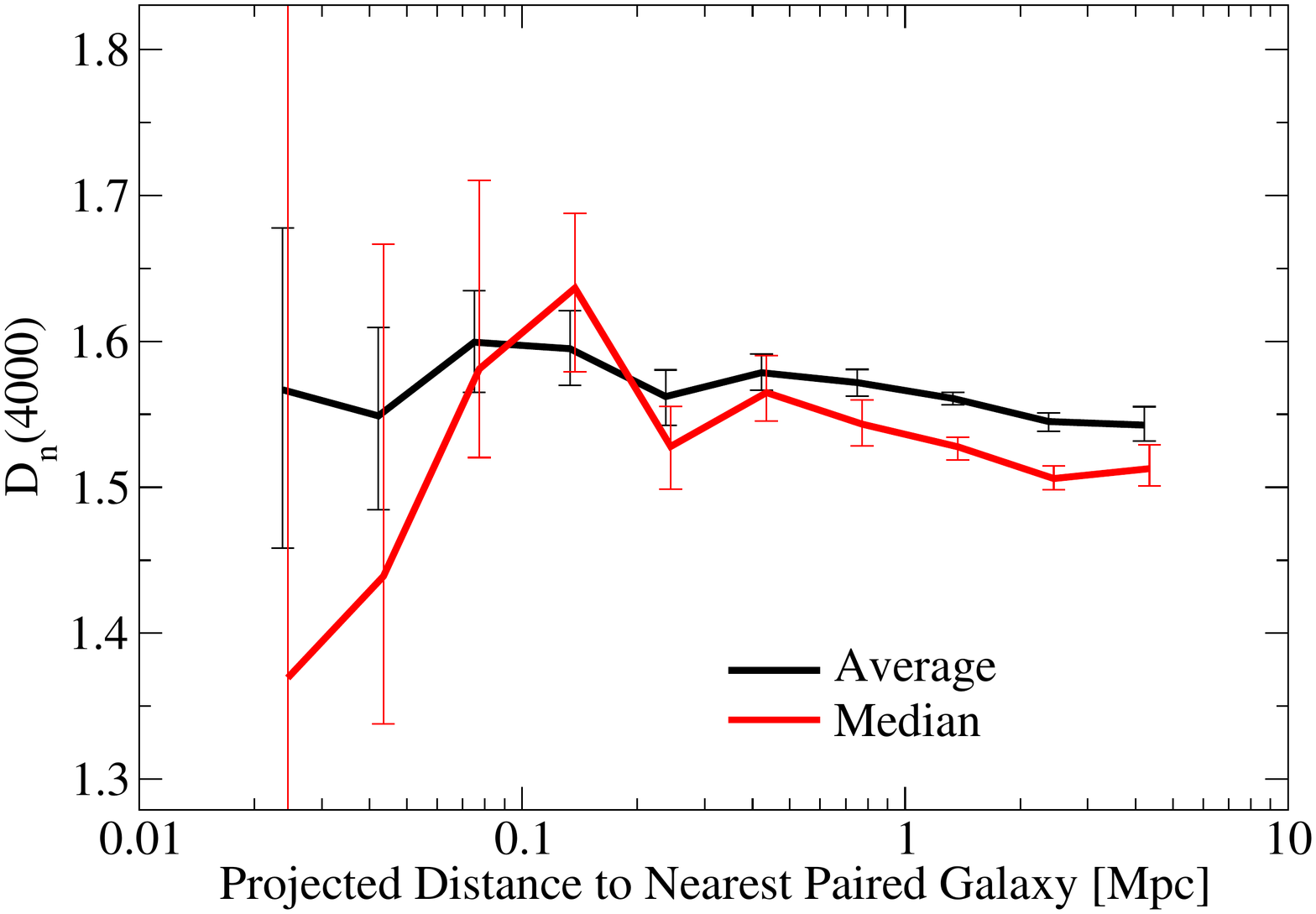}\\[-3ex]
\caption{Left panel: distribution of D$_n$(4000)---i.e., 4000\AA\ break strength---for all host galaxies and those in close pairs (\citealt{Balogh99} definition).  A weak 4000\AA\ break (D$_n$(4000)$< 1.6$) signifies a young stellar population, and a strong one (D$_n$(4000)$> 1.7$) signifies an older stellar population. Right panel: D$_n$(4000) as a function of distance to the nearest paired galaxy.  The diameter of an SDSS fibre is 3", corresponding to 0.43 kpc at $z=0.01$ and 2.4 kpc at $z=0.057$.  For our stellar mass range, typical host galaxy half-light radii are 1--2.5 kpc \citep{Kravtsov13}.}
\label{f:dn4000}
\end{figure*}

\begin{figure*}
\vspace{-3ex}
\plotgrace{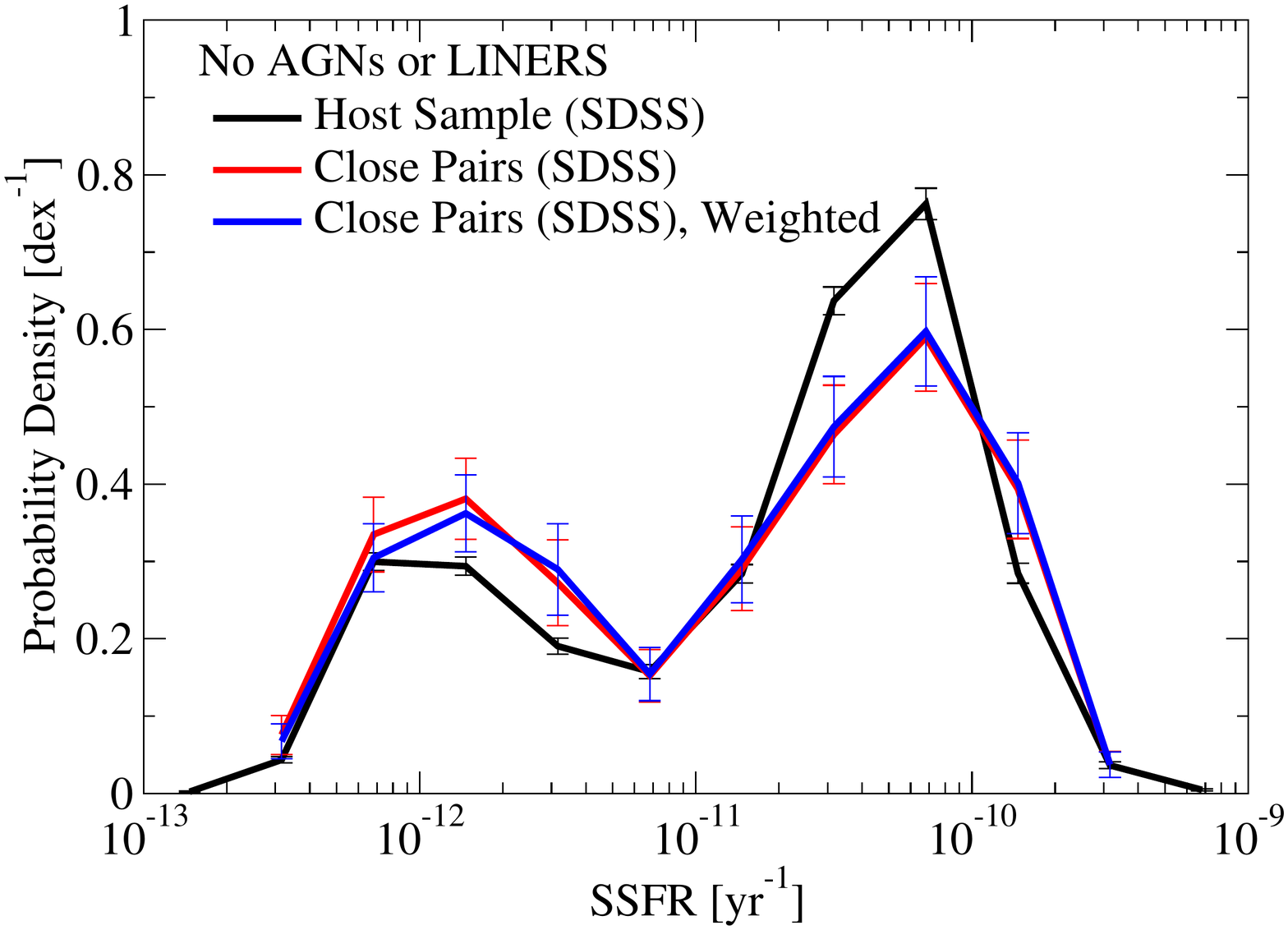}\plotgrace{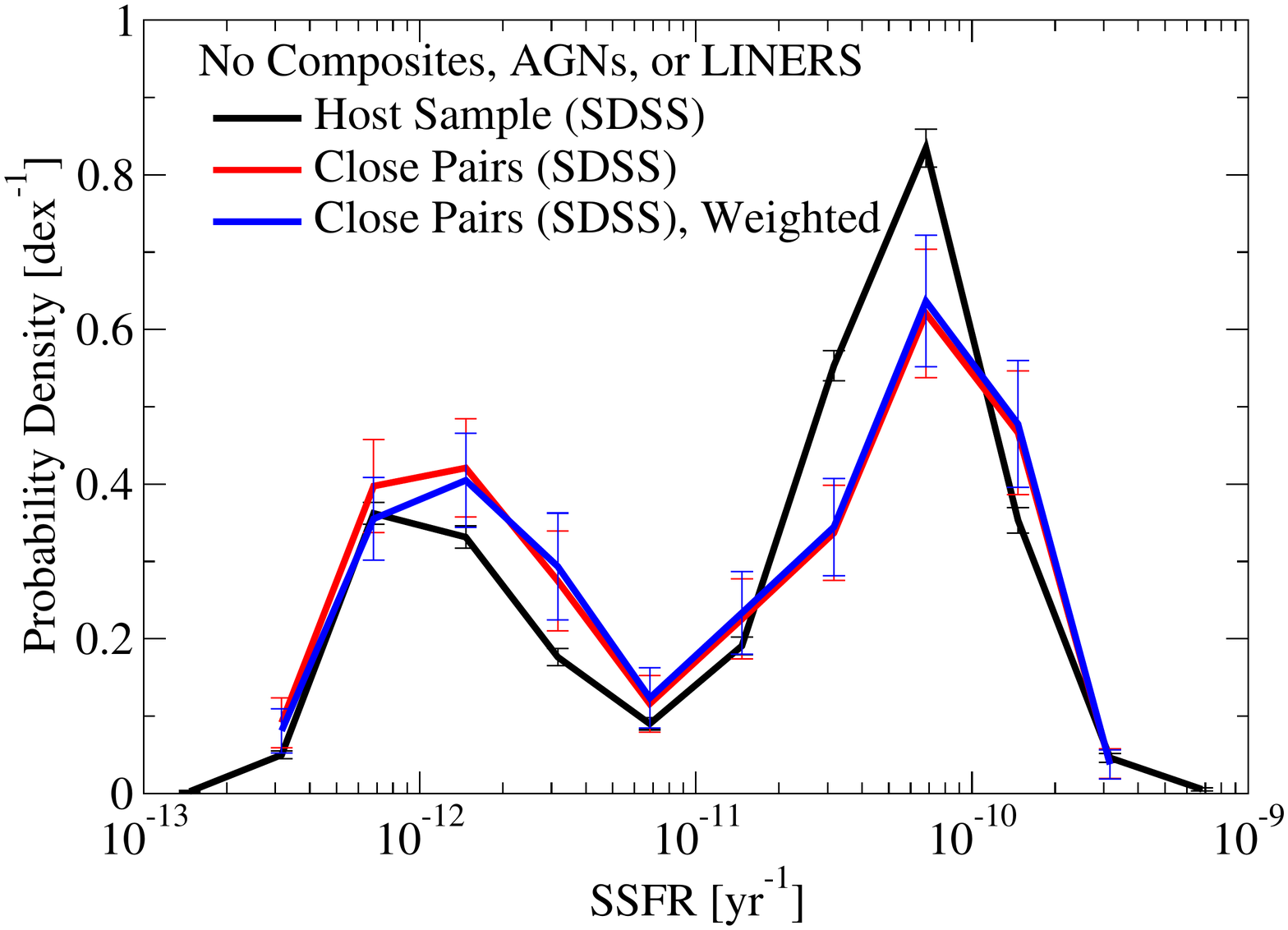}\\[-3ex]
\caption{Left panel: distribution of SSFRs for all host galaxies and those in close pairs, excluding hosts classified as AGN or LINERs according to the BPT diagram.  Right panel: same, except also excluding composite host galaxies.}
\label{f:agn}
\end{figure*}

\begin{figure*}
\vspace{-3ex}
\plotgrace{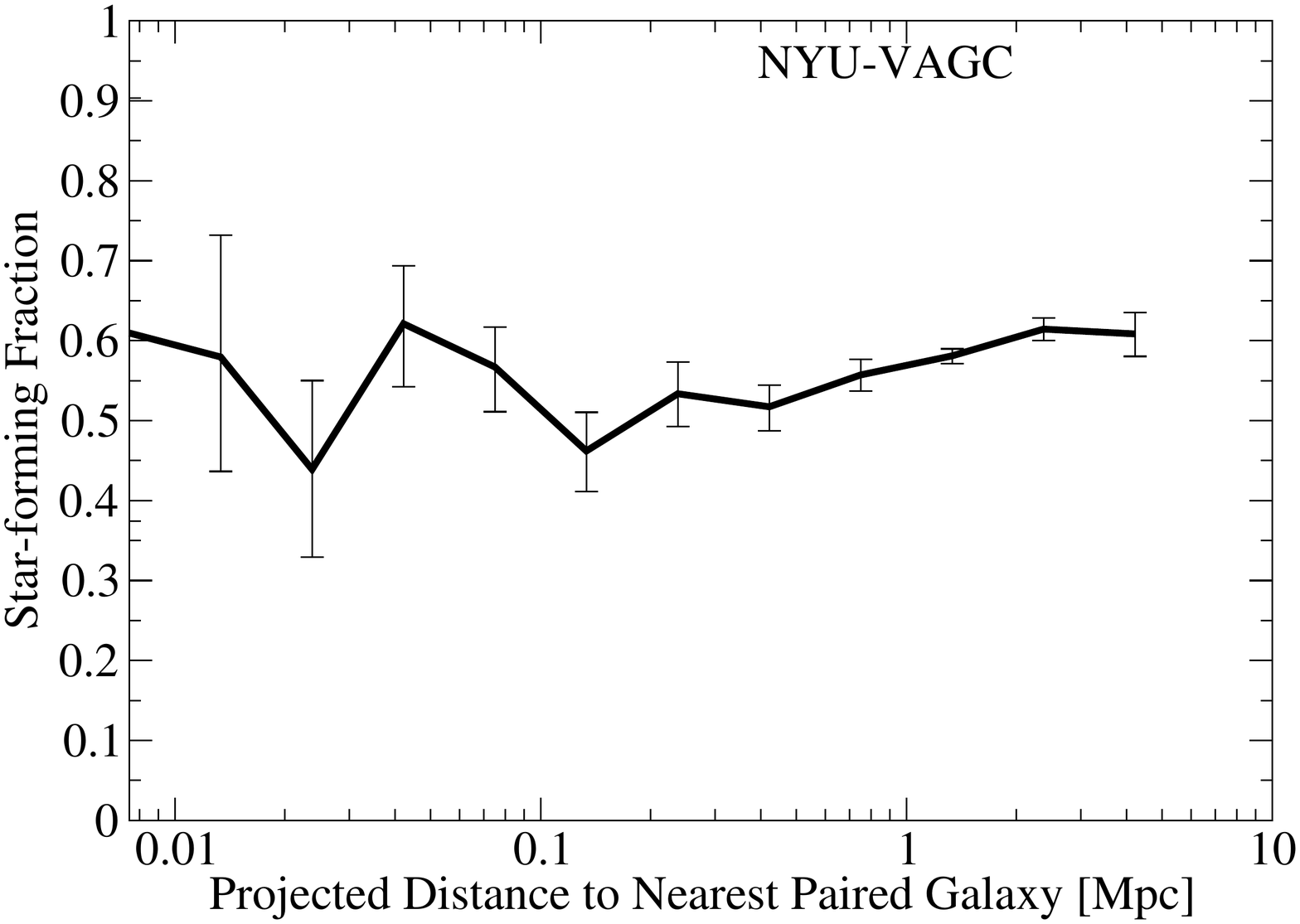}\plotgrace{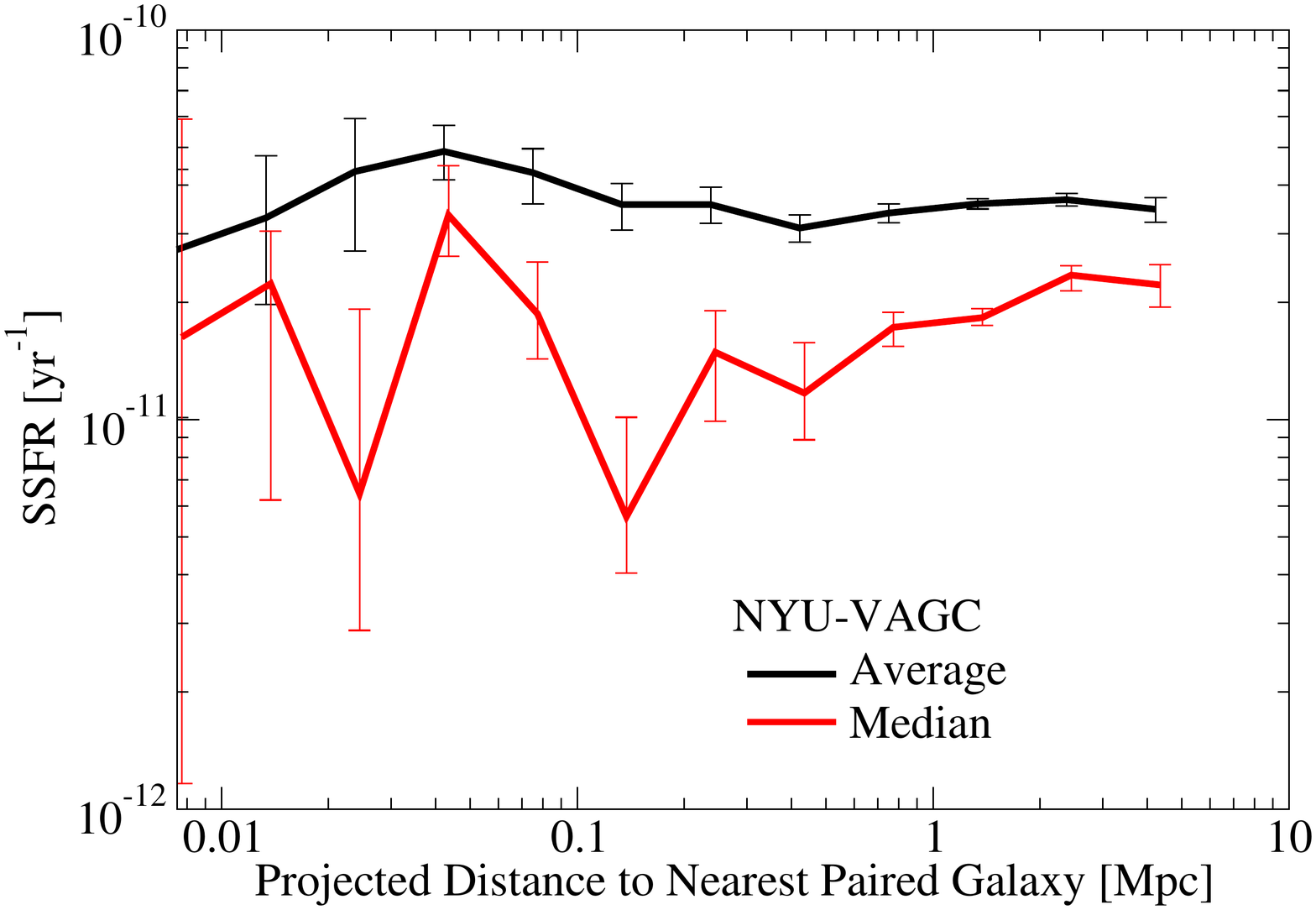}\\[-3ex]
\plotgrace{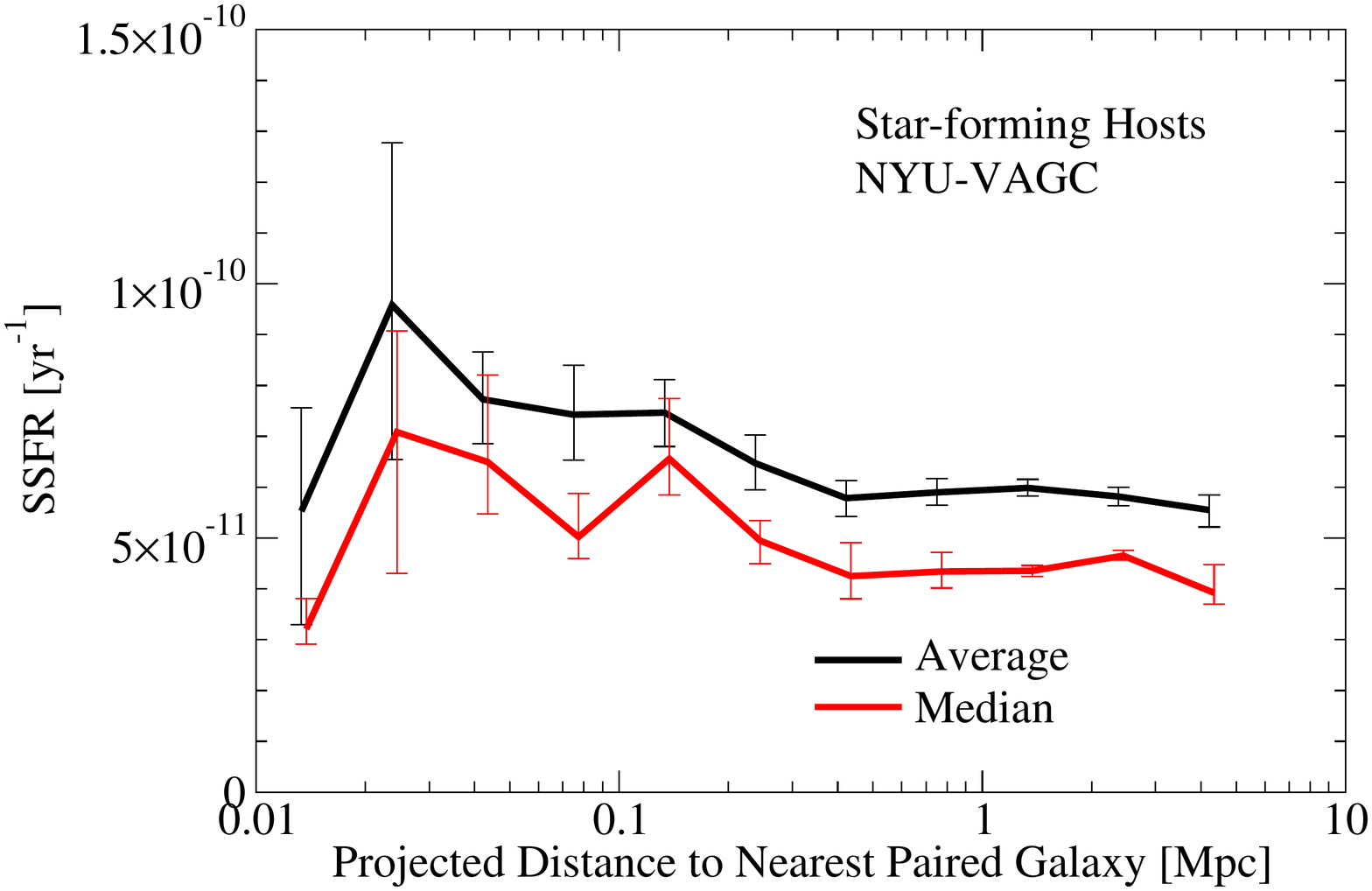}\plotgrace{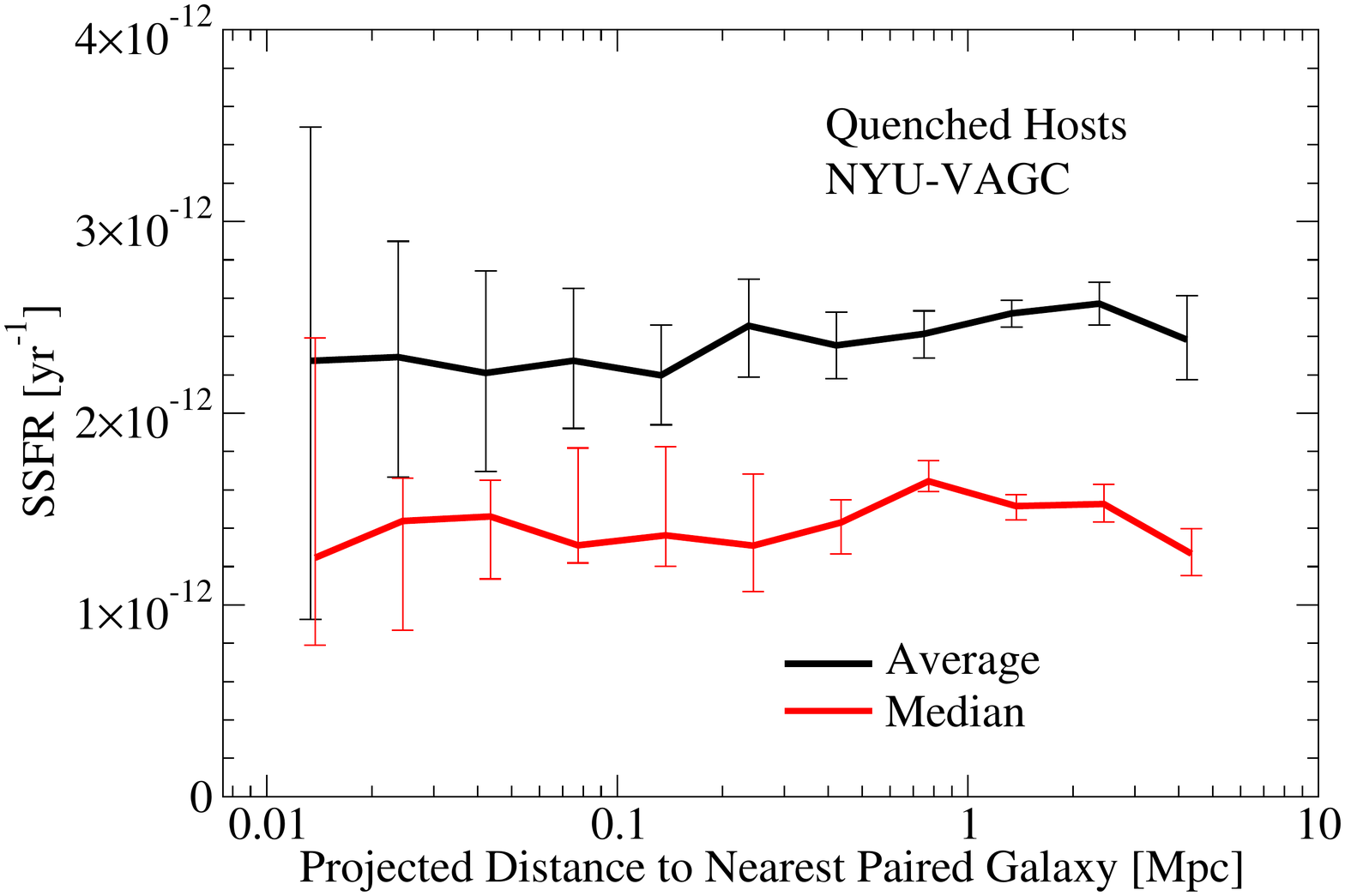}\\[-3ex]
\caption{Top-left panel: the star-forming fraction of host galaxies as a function of the distance to the nearest paired galaxy, using stellar masses and redshifts from the NYU-VAGC \citep{Blanton05}.  In this catalog, galaxies missing spectroscopic redshifts are assigned the redshift of the nearest neighbour galaxy, which results in a $\sim$ 3 times larger sample size for close galaxy pairs within 55".  This panel is analogous to Fig.\ \ref{f:sdss_sfr}, top-right panel.  Top-right panel: SSFRs for host galaxies (still from \citealt{Brinchmann04}) as a function of distance to the nearest paired galaxy, analogous to the bottom-right panel of Fig.\ \ref{f:sdss_sfr}, using NYU-VAGC stellar masses and redshifts.  Bottom-left panel: SSFRs for star-forming host galaxies as a function of distance to the nearest paired galaxy, analogous to the top-right panel of Fig.\ \ref{f:sdss_sf_q}, using NYU-VAGC stellar masses and redshifts.  Bottom-right panel: SSFRs for quenched host galaxies as a function of distance to the nearest paired galaxy, analogous to the bottom-right panel of Fig.\ \ref{f:sdss_sf_q}, using NYU-VAGC stellar masses and redshifts.}
\label{f:blanton}
\end{figure*}

We have tested varying the isolation criteria for host galaxies, the definition of a paired galaxy, as well as the separation distance considered ``close.''  In Figs.\ \ref{f:ssfr_var} and \ref{f:sf_frac_var}, we show the impact on the specific star formation rate distributions as well as the overall star-forming fraction of host galaxies.  We find no impact large enough to affect our main conclusions; e.g., that major halo mergers do not result in significant changes to star formation activity.

We have also explored two spectroscopic-only indicators of galaxy formation.  These have not been included in the main discussion because the SDSS fibre size (3" diameter) only covers a small fraction of the host galaxy, which may not be representative of the overall star-formation activity \citep{Salim07}.  The fibre-only SSFRs \citep{Brinchmann04} shown in Fig.\ \ref{f:fibre_ssfrs} do not provide any different picture than total galaxy SSFRs (Figs.\ \ref{f:ssfr_var} and \ref{f:sf_frac_var}).  The 4000\AA\ break strength (D$_n$(4000)) is more interesting.  As with galaxy total SSFRs, a larger fraction of host galaxies with close pairs seem to have older stellar populations (D$_n$(4000)$ > 1.7$).  However, star-forming host galaxies in the close pairs sample appear to have slightly younger ages than star-forming galaxies in the host sample.  This difference is statistically significant, but small (median D$_n$(4000) of $1.359^{+0.016}_{-0.008}$ for host galaxies with close pairs, versus $1.397^{+0.003}_{-0.002}$ for the host sample).  While this could indicate a younger stellar population, changes of this magnitude are also possible with dust and metallicity differences \citep{SHARDS13}.  Future IFU spectroscopy of close pairs \citep[e.g.,][]{Bundy15} or deeper spectroscopy at higher redshifts may reveal larger differences in galaxy discs.

Many previous works have excluded galaxies classified as AGNs (based on the BPT diagram; \citealt{BPT}) due to the difficulty of estimating star formation rates from emission lines.  The \cite{Brinchmann04} SFRs avoid this issue by basing AGN and composite fibre SFRs on D$_n$(4000), using the D$_n$(4000)--SSFR distribution for non-AGN-contaminated galaxies.  While this could introduce small biases for AGN and composite galaxies, the low redshifts of our sample ($z<0.06$) mean that SDSS fibre sizes (3'') cover a minority of the galaxy light.  As a result, the majority of the SFR estimate is based on photometry, which is calculated in the same way for all galaxies.  Nonetheless, for completeness, we show galaxy SSFR distributions for the host and close pairs samples in Fig.\ \ref{f:agn} excluding AGNs, LINERs, and composite host galaxies.

Finally, we have tested using stellar masses and redshifts from the NYU Value-Added Galaxy Catalog (NYU-VAGC; \citealt{Blanton05}).  This catalog's main advantage is that redshifts for fibre-collided galaxies are taken from the nearest available galaxy.  This more than triples the available statistics for close pairs separated by $<$55".  However, as in Fig.\ \ref{f:sdss_sfr}, there remains no evidence for extremely close pairs having a larger star-forming fraction (Fig.\ \ref{f:blanton}, top panels).  Similarly, boosted specific star formation rates are seen in star-forming host galaxies in close pairs, whereas no boost is seen for quenched galaxies, regardless of pair separation (Fig.\ \ref{f:blanton}, lower panels).

\section{Alternate Mock Catalog Construction Methods}
\label{a:mocks}

\begin{figure*}
\vspace{-5ex}
\plotgrace{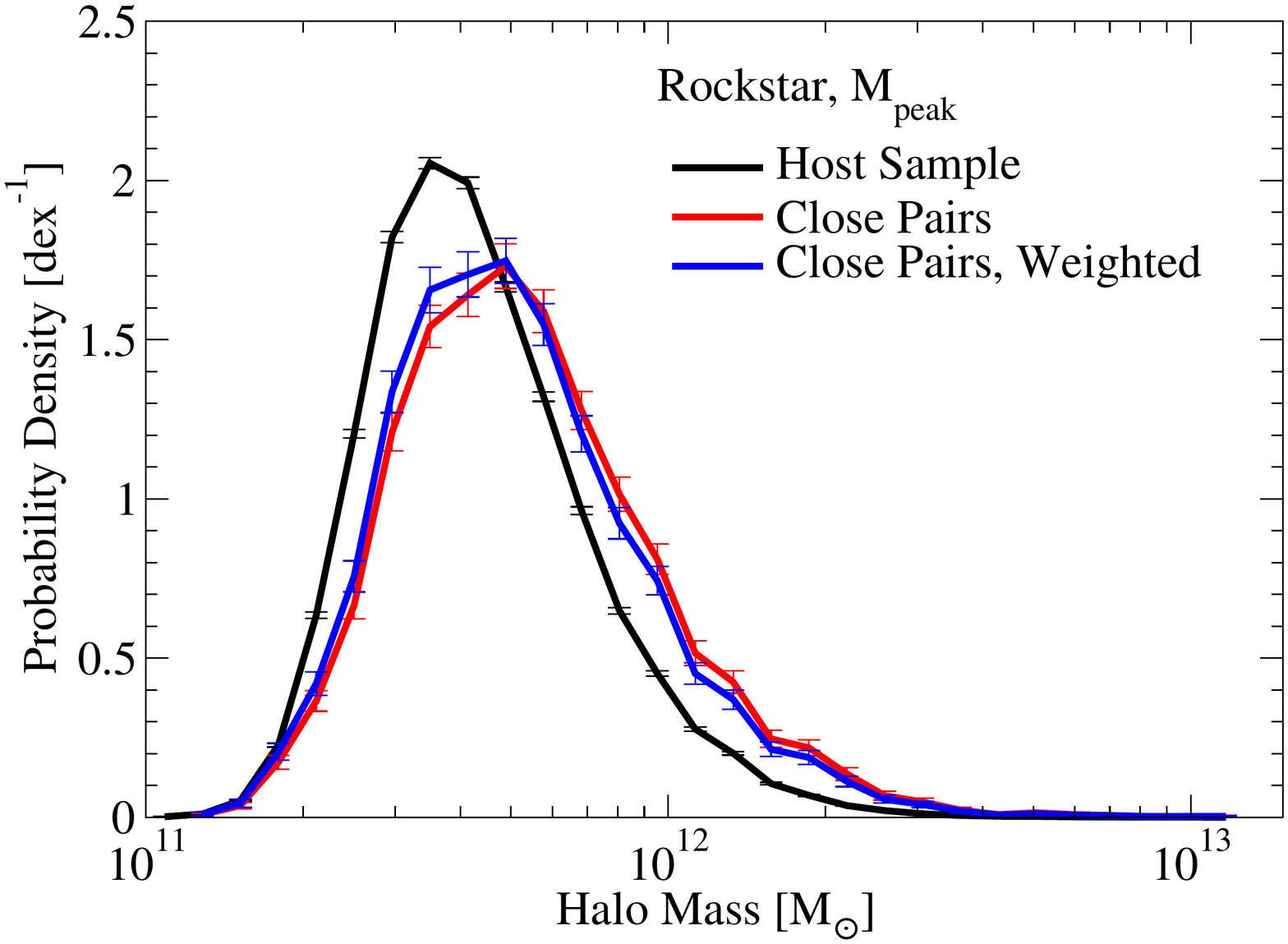}\plotgrace{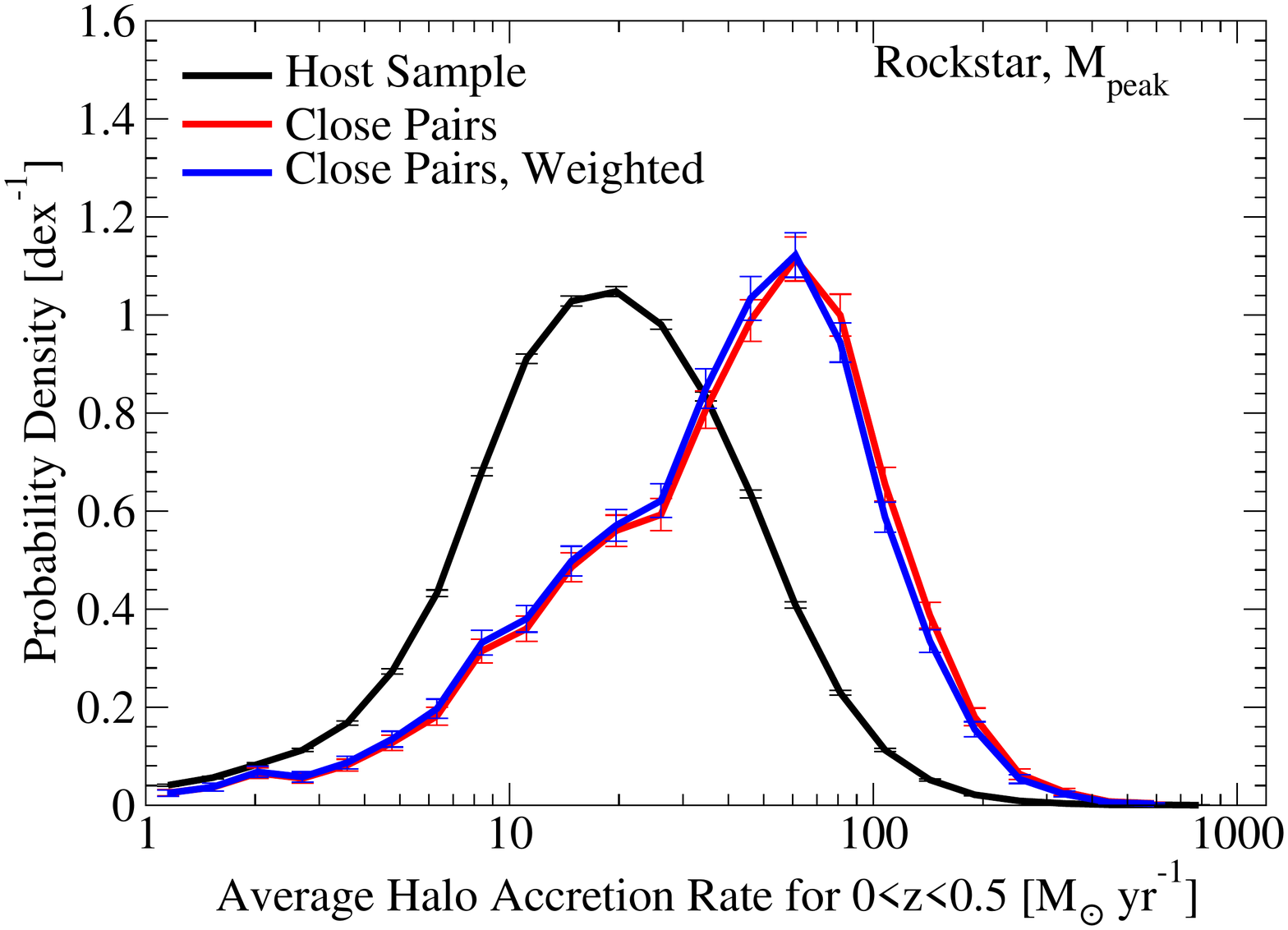}\\[-3ex]
\plotgrace{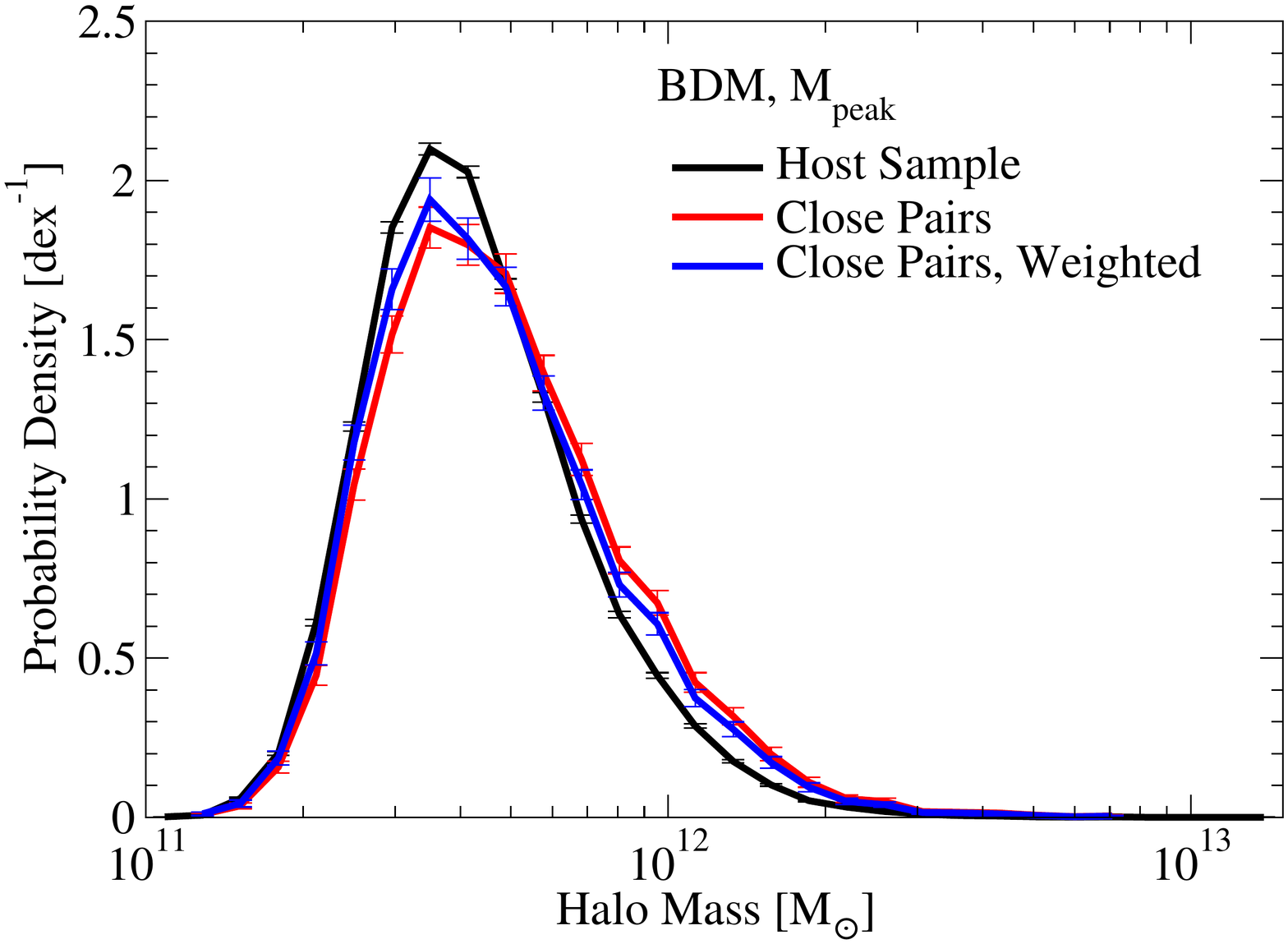}\plotgrace{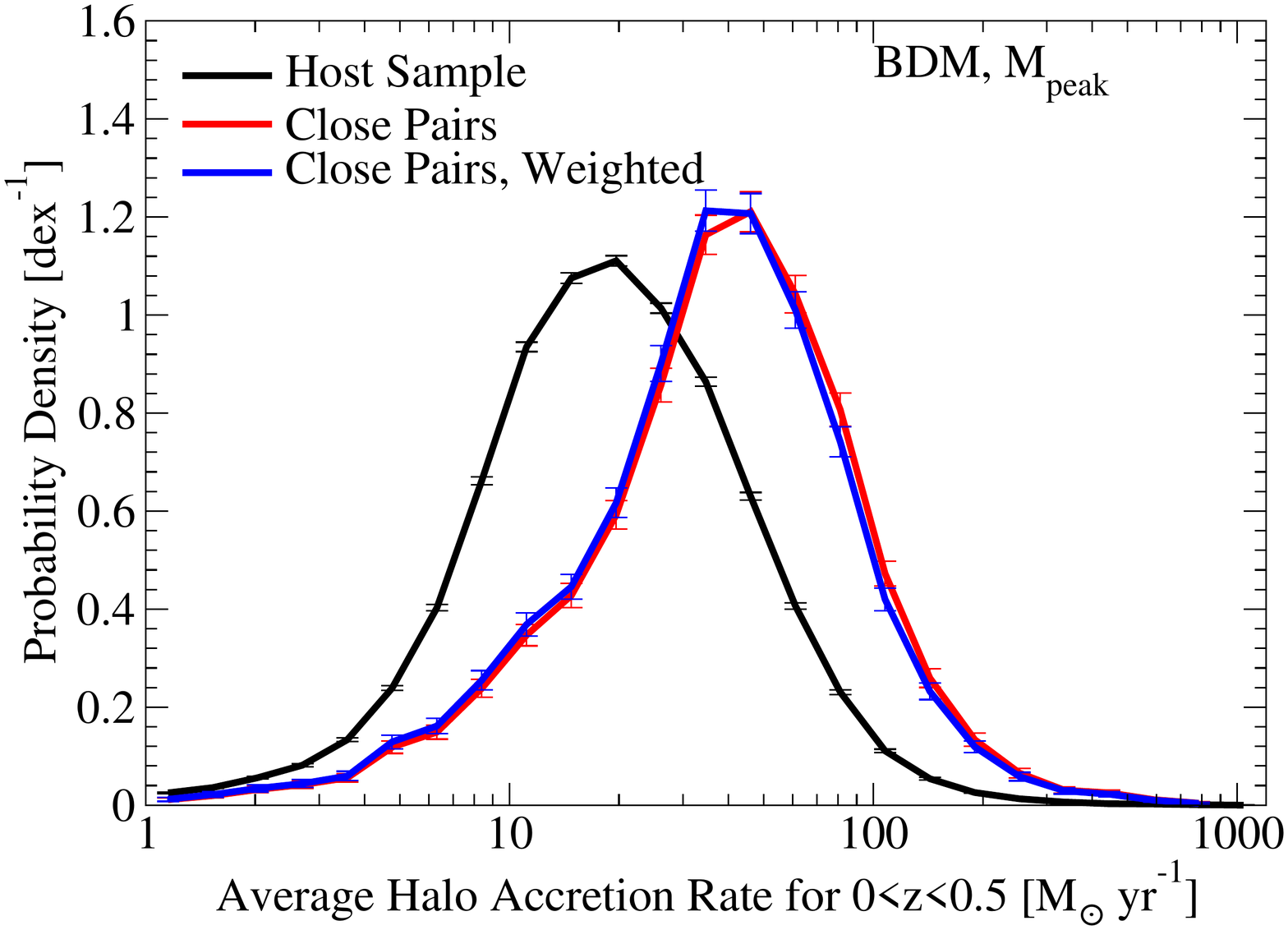}\\[-3ex]
\plotgrace{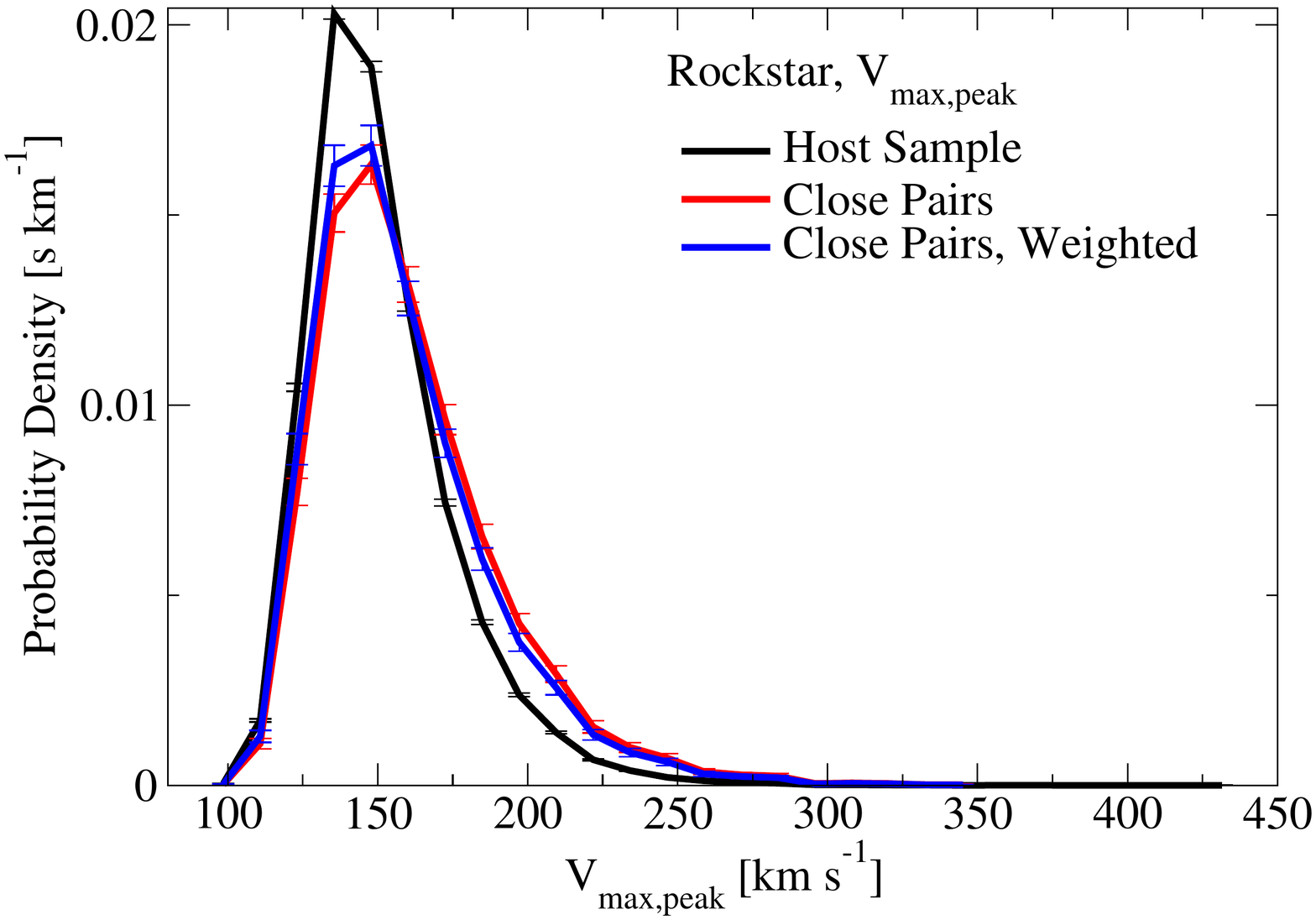}\plotgrace{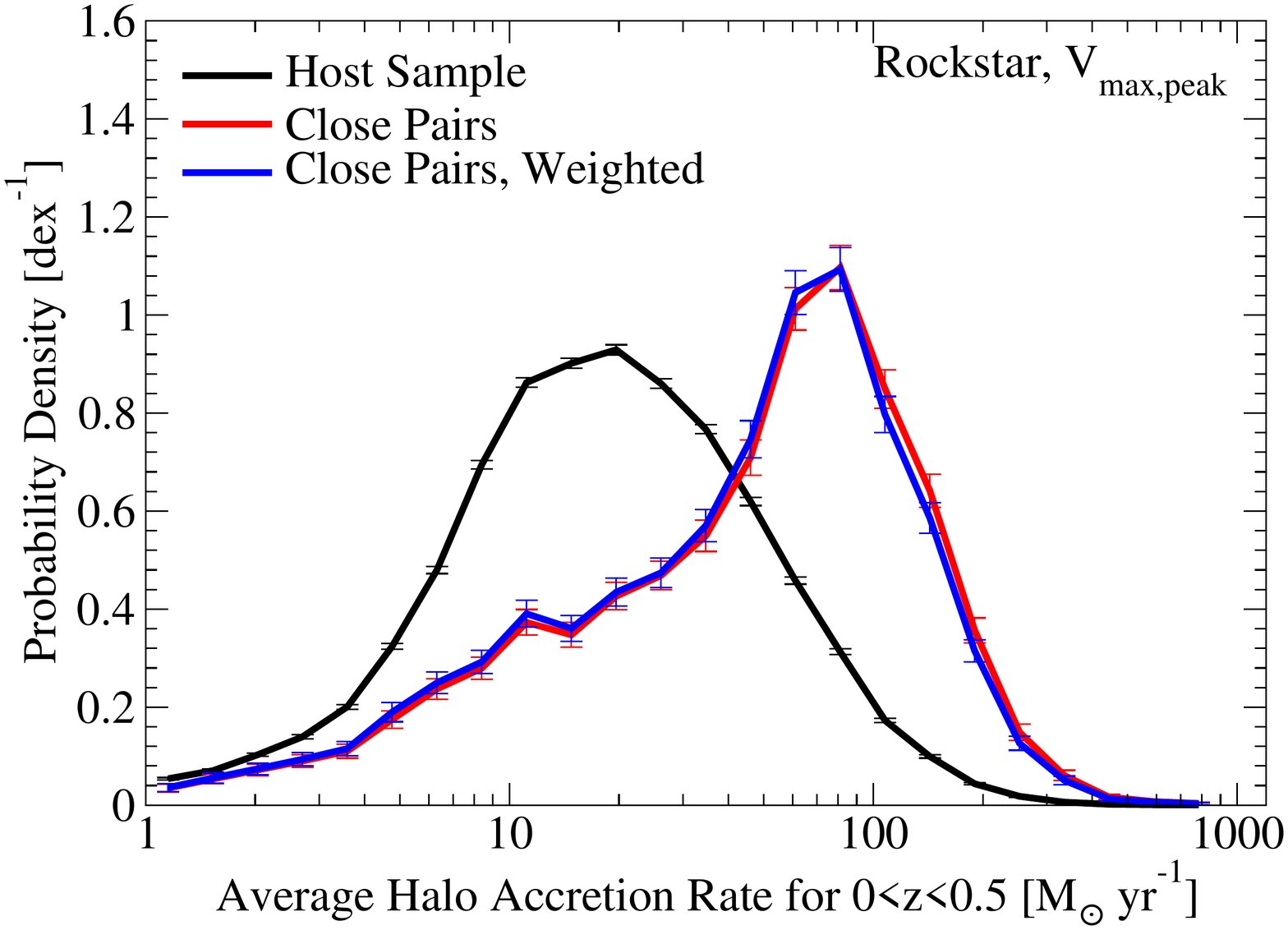}\\[-3ex]
\caption{Left panels: Peak halo mass (or peak $\vmax$, as appropriate) for haloes hosting galaxies in the host sample as well as the close pairs sample for three alternate mock catalogs.  Right panels: averaged halo accretion rates for $0<z<0.5$ for the same alternate mock catalogs.}
\label{f:mocks}
\end{figure*}

As noted in \cite{Reddick12}, many different ways exist to abundance match galaxies to haloes; additionally, many different halo finders exist \citep[see ][for comparisons and reviews]{Knebe11,Knebe13}.  We have therefore explored three alternate ways of assigning galaxies to haloes in mock catalogs.  These include abundance matching on $M_p$ (peak historical mass) and $v_\mathrm{max,peak}$ (peak historical $\vmax$) with the \textsc{Rockstar} halo finder, as well as abundance matching on $M_p$ with the \textsc{BDM} halo finder.

As shown in Fig.\ \ref{f:mocks}, the choices of abundance matching proxy and halo finder both affect quantitative details for galaxy host halo masses (or $\vmax$) and accretion history.  Several qualitative facts remain, however.  Regardless of the catalog, our selection criteria for close pairs does not significantly bias present-day host halo properties relative to those for the host sample.  The major merger fractions for the close pairs samples are 47\%, 52\%, and 44\% for the \textsc{Rockstar} $M_p$, \textsc{Rockstar} $v_\mathrm{max,peak}$, and \textsc{BDM} $M_p$ catalogs, respectively.  Additionally, regardless of the catalog, host haloes of galaxies in close pairs have significantly larger recent total accretion rates as compared to the host sample.  In combination, these suggest that the selection criteria we have chosen are a robust way to preferentially identify galaxies whose host haloes are undergoing major mergers and have had more recent formation times without imposing a strong bias on the host halo mass or $\vmax$.

\section{Stellar Mass Function Calculation}
\label{a:smf}

\begin{figure*}
\vspace{-5ex}
\plotgrace{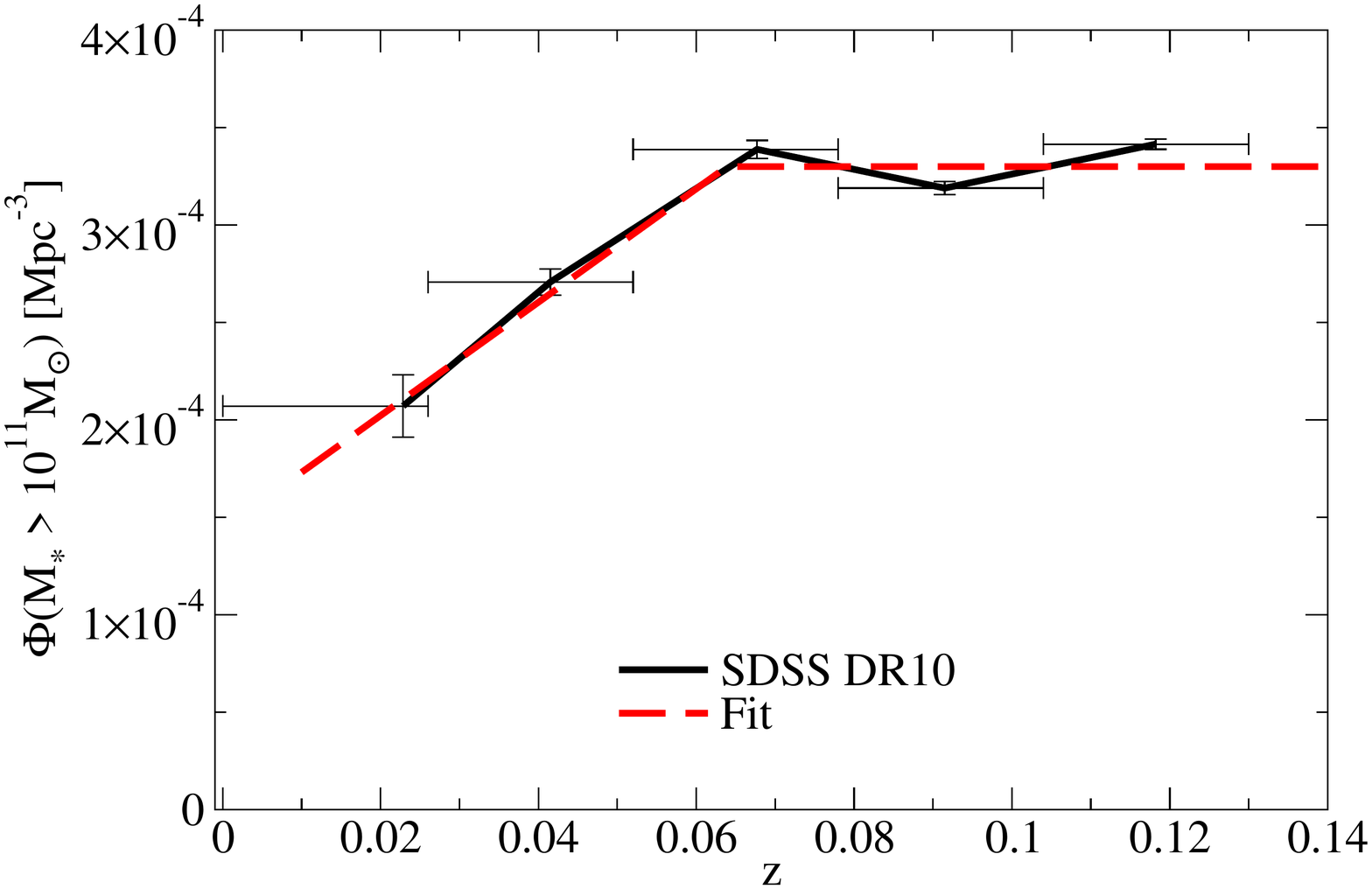}\plotgrace{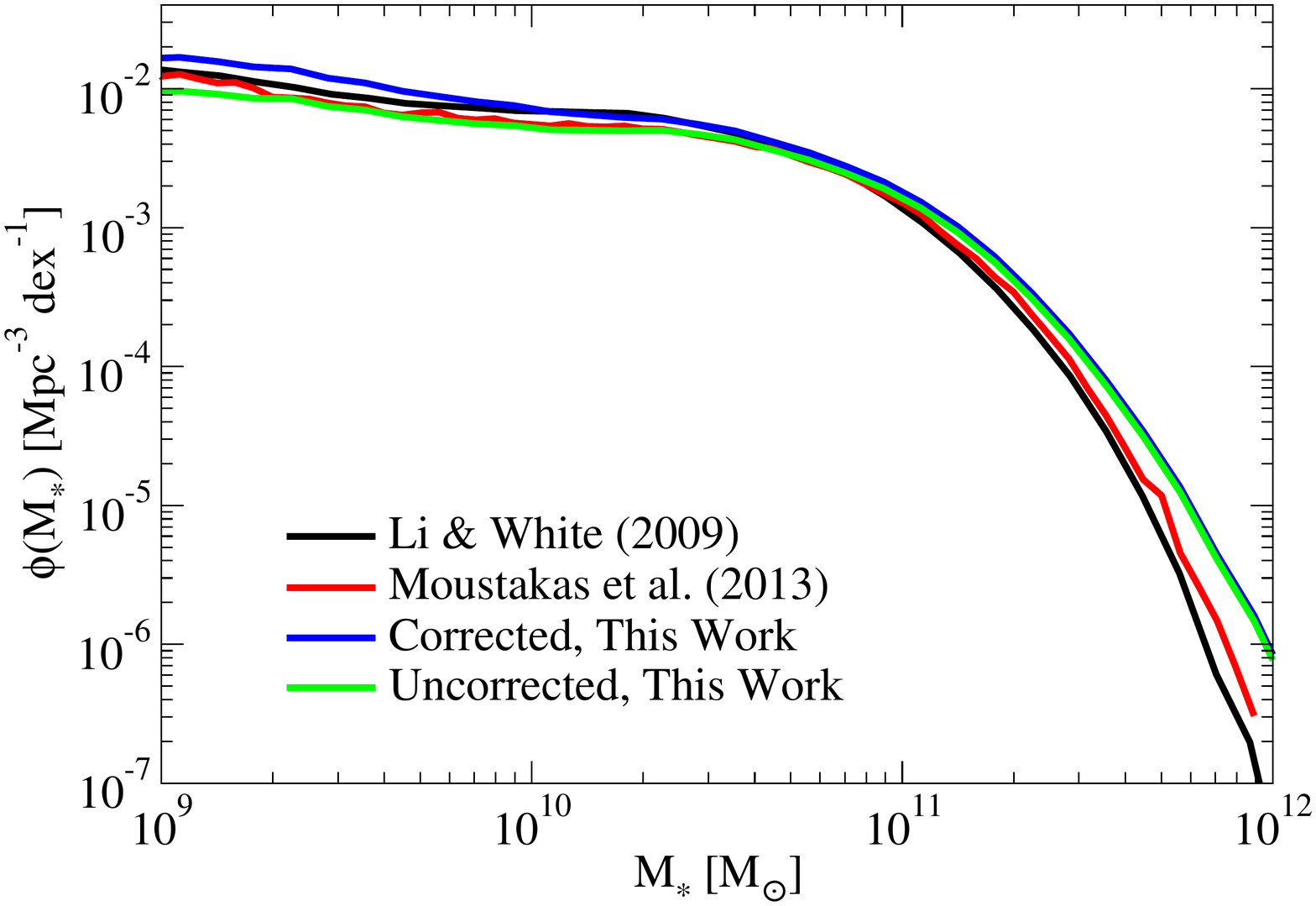}\\[-3ex]
\caption{Left panel: number density of $M_* > 10^{11}\,\Msun$ galaxies in the SDSS as a function of redshift.  Horizontal error bars show bin widths; data point centres are located at the median bin redshift.  Vertical error bars show Poisson uncertainties.  Right panel: corrected and uncorrected stellar mass functions compared to previous literature results.}
\label{f:a_smf}
\end{figure*}

The local volume ($z<0.07$) is underdense compared with the nearby ($0.07<z<0.2$) universe \citep[see][and references therein]{Baldry08,Keenan13}.  When abundance-matching to a simulation at the universe's typical density, it is necessary to correct for this effect; otherwise the stellar mass--halo mass relation and satellite fractions will be underestimated.  The local cosmological underdensity results in a relatively uniform reduction in galaxy counts at all masses \citep{Baldry08}, so massive galaxy counts can be used to trace the underdensity as a function of redshift.  Number densities for $M_* > 10^{11}\,\Msun$ galaxies as a function of redshift in our SDSS sample are therefore shown in Fig.\ \ref{f:a_smf}.  We model the local underdensity as linearly dependent on redshift out to $z=0.0644$, where we assume that it reaches the cosmological mean (see fit in Fig.\ \ref{f:a_smf}).  The corresponding correction factor for galaxy number counts is then:
\begin{equation}
f_\mathrm{corr}(z) = \begin{cases}
\frac{0.00033}{0.00291 z + 0.000144}, & \text{if }z<0.0644\\ 
1, & \text{otherwise }\\ 
\end{cases}
\end{equation}
In the absence of detailed completeness information as a function of redshift and \cite{Kauffmann03} stellar mass, we have also scaled all number counts to account for the average SDSS spectroscopic completeness fraction of 92\%.  The resulting stellar mass function (both before and after corrections) is shown in Fig.\ \ref{f:a_smf}, with comparisons to previous literature results \citep{li-2009,Moustakas13}.

\end{document}